\documentclass{article}
\usepackage{graphicx}
\usepackage{amsmath}
\usepackage{amsfonts}
\usepackage{amssymb}

\newtheorem{theorem}{Theorem}

\newtheorem{lemma}[theorem]{Lemma}

\newtheorem{proposition}[theorem]{Proposition}
\newtheorem{remark}[theorem]{Remark}

\newenvironment{proof}[1][Proof]{\textbf{#1.} }{\ \rule{0.5em}{0.5em}}

\begin{document}

\title{Invariants of nine dimensional real Lie algebras with nontrivial Levi decomposition}
\author{R. Campoamor-Stursberg\\Dpto. Geometr\'{\i}a y Topolog\'{\i}a,\\ Fac. CC.
Matem\'aticas, Universidad Complutense de Madrid,\\ Plaza de
Ciencias 3, E-28040 Madrid.\\ rutwig@mat.ucm.es}
\date{}

\maketitle

\begin{abstract}
The generalized Casimir invariants of real indecomposable Lie
algebras admitting a nontrivial Levi decomposition are determined.
\end{abstract}

\section{Introduction}

Invariant functions of the coadjoint representation of Lie
algebras constitute an important tool in representation theory and
many applications, both mathematical and physical, such as
completely integrable systems, differential equations or symmetry
analysis in physical problems. In Representation theory, the
eigenvalues of Casimir invariants are used to label irreducible
representations of groups, as well as for the study of reduction
chains with respect to subgroups. The semisimple case has been
completely solved, mainly by the work of Casimir, Van der Waerden,
Chevalley or Racah \cite{Cas,Che,Ra1}. In this mathematical
approach, polynomial invariants are proven to be elements in the
centre of the universal enveloping algebra, and the number of such
operators is given by the rank of the algebra. This classical
result allows an extension to algebras having rational invariants,
where these are obtained as ratios of semi-invariants of the
coadjoint representation \cite{Che}. Extensive work has been done
on the eigenvalues of Casimir operators of classical groups and
their generating functions \cite{Pe,Po}. For the case of non
semisimple Lie algebras no general criteria for the number and
structure of independent invariants exist, up to certain specific
classes \cite{AA,Ca,C35,Tro1}. Various methods have been developed
to compute the invariants of Lie algebras. The classical study of
the universal enveloping algebra is useful for the semisimple and
reductive case, as well as some other classes of algebras
\cite{Cas,Za}. The Kirillov approach has been proven useful for
certain types of semidirect products, although the most employed
technique is the analytical method, i.e., the formulation of the
problem by means of differential equations \cite{Di}. For specific
classes of algebras some purely algebraic methods based on
representation theory have been developed \cite{C33,Pe94}, such as
semidirect products of Heisenberg and semisimple algebras, the
special affine algebras $\frak{sa}(n,\mathbb{R})$ used in the
metric-affine gravity theories or Borel subalgebras of simple Lie
algebras \cite{Ahn,C35,H1,Pe94,Que,Tro1}. Recently a new method
based on the moving frames of E. Cartan has been proposed
\cite{Bo}. This approach, which has proven to be of interest in
the symmetry analysis of differential equations (see \cite{Ol,Ste}
and references therein) has the advantage of avoiding completely
the integration of differential equations, and uses the inner
automorphism group of Lie algebras. This new algorithm provides
considerable simplification in the description of invariants, and
constitutes a powerful tool to analyze solvable Lie algebras in
any dimension.

\medskip

Invariants of Lie algebras have been computed for solvable Lie
algebras up to dimension six (\cite{Bo,Nd,PWZ} and references
therein) and in dimension 7 for nilpotent Lie algebras \cite{Ro}.
Some large classes of solvable algebras in any dimension have also
been studied, such as triangular Lie algebras, algebras with
abelian or Heisenberg radicals, or solvable algebras with graded
nilradical of maximal nilpotence index \cite{C56,Tre,SW}. For
non-solvable Lie algebras various references on the structure and
properties of invariants exist, and for important classes of
algebras, such as the inhomogeneous and the special affine Lie
algebras, the invariants have been computed explicitly
\cite{C33,C35,De,Pe94,Pe06,Que}. Many particular algebras of
physical importance have also been considered, including the
classical kinematical Lie algebras and their subalgebras
\cite{PW2} and algebras related to Cayley-Klein geometries
\cite{H1}.

\medskip

In this work we determine a fundamental system of invariants for
nine dimensional real indecomposable Lie algebras having a
nontrivial Levi decomposition. The classification of these
algebras was obtained in \cite{Tu3}. Although we use the classical
analytical approach, direct integration of the invariants is a
cumbersome task. In order to obtain the invariants, we consider
the theory of semi-invariants for the coadjoint representation
\cite{Tro1,Tro126} and the reduction to subalgebras the invariants
of which can be computed easily or are already known. The
procedure is based in the labelling of representations using
subgroups chains (so called missing label operator problem)
developed for arbitrary Lie algebras in \cite{Sh}.

\medskip

Unless otherwise stated, any Lie algebra $\frak{g}$ considered
here is of finite dimension over the field $\mathbb{K}=\mathbb{R}$
and indecomposable, i.e., not splittable into a direct sum of
ideals.

\section{Semidirect sums of Lie algebras}

The classification of Lie algebras is reduced to specific classes
due to the Levi decomposition theorem, which states that any Lie
algebra is formed from a semisimple Lie algebra $\frak{s}$ (called
the Levi factor of $\frak{g}$) and a maximal solvable ideal
$\frak{r}$, called the radical. It follows that the Levi factor
$\frak{s}$ acts on $\frak{r}$, in either one of the two following
forms:%
\begin{align}
\left[  \frak{s},\frak{r}\right]   &  =0\\
\left[  \frak{s},\frak{r}\right]   &  \neq0\nonumber
\end{align}
In the first case we obtain a decomposable algebra
$\frak{s}\oplus\frak{r}$, whereas the second possibility implies
the existence of a representation $R$ of $\frak{s}$ which
describes the action, i.e., such that
\begin{equation}
\left[  x,y\right]  =R\left(  x\right)  .y,\;\forall x\in\frak{s}%
,y\in\frak{r.} \label{eq1}%
\end{equation}
We will use the notation $\overrightarrow{\oplus}_{R}$ to describe
semidirect products. Since (\ref{eq1}) implies that the radical is
a module over $\frak{s}$, we have to expect structural
restrictions on the radical (for direct sums any solvable Lie
algebra being suitable). Two important structural properties on
semidirect products were obtained in \cite{Tu3}:

\begin{proposition}
Let $\frak{s}\overrightarrow{\oplus}_{R}\frak{r}$ be a Levi
decomposition of a Lie algebra $\frak{g}$.

\begin{enumerate}
\item  If $R$ is irreducible, then the radical $\frak{r}$ is
abelian.

\item  If the representation $R$ does not posses a copy of the
trivial representation, then the radical $\frak{r}$ is a nilpotent
Lie algebra.
\end{enumerate}
\end{proposition}

Indecomposable Lie algebras (i.e., that do not decompose as a
direct sum of ideals) of dimension $n\leq8$ and having non-trivial
Levi decomposition were completely classified by Turkowski in
\cite{Tu2}, while the nine dimensional case was analyzed in
\cite{Tu3}. All these algebras have a Levi part of rank one,
isomorphic to the compact form $\frak{so}\left(  3\right)  $ or
the real normal form $\frak{sl}\left(  2,\mathbb{R}\right)
$.\footnote{Algebras having a Levi part of higher rank are
necessarily decomposable.
\par
{}} In dimension nine, 63 isomorphism classes of indecomposable
algebras, some depending on one or more parameters, were found.
Additionally we define the
Lie algebra $L_{9,7}^{\ast}$ defined by the nontrivial brackets%
\[%
\begin{array}
[c]{llll}%
\left[  X_{1},X_{2}\right]  =X_{3}, & \left[  X_{1},X_{3}\right]
=-X_{2}, &
\left[  X_{2},X_{3}\right]  =X_{1}, & \left[  X_{1},X_{4}\right]  =\frac{1}%
{2}X_{7},\\
\left[  X_{1},X_{5}\right]  =\frac{1}{2}X_{5}, & \left[
X_{1},X_{6}\right] =-\frac{1}{2}X_{5}, & \left[
X_{1},X_{7}\right]  =-\frac{1}{2}X_{4}, &
\left[  X_{2},X_{4}\right]  =\frac{1}{2}X_{5},\\
\left[  X_{2},X_{5}\right]  =-\frac{1}{2}X_{4}, & \left[
X_{2},X_{6}\right] =\frac{1}{2}X_{7}, & \left[  X_{2},X_{7}\right]
=-\frac{1}{2}X_{6}, & \left[
X_{3},X_{4}\right]  =\frac{1}{2}X_{6},\\
\left[  X_{3},X_{5}\right]  =-\frac{1}{2}X_{7}, & \left[
X_{3},X_{6}\right] =-\frac{1}{2}X_{4}, & \left[
X_{3},X_{7}\right]  =\frac{1}{2}X_{5}, & \left[
X_{4},X_{6}\right]  =X_{8},\\
\left[  X_{5},X_{7}\right]  =X_{8}, & \left[  X_{4},X_{9}\right]
=X_{4}, &
\left[  X_{5},X_{9}\right]  =X_{5}, & \left[  X_{6},X_{9}\right]  =X_{6},\\
\left[  X_{7},X_{9}\right]  =X_{7}, & \left[  X_{8},X_{9}\right]
=2X_{8}. & &
\end{array}
\]
This algebra has the Levi decomposition
\begin{equation}
L_{9,7}^{\ast}=\frak{so}\left(  3\right)  \overrightarrow{\oplus}_{R_{4}%
\oplus2D_{0}}\frak{g}_{6,82}^{2,0,0}, \label{MA}%
\end{equation}
where the notation for the radical $\frak{g}_{6,82}^{2,0,0}$ is
that used in \cite{Mu}. Since the real solvable Lie algebras
$\frak{g}_{6,82}^{2,0,0}$ and $\frak{g}_{6,92}^{\ast}$ are
non-isomorphic, the Lie algebras $L_{9,7}^{\ast}$ and
$L_{9,7}^{p}$ of \cite{Tu3} are non-isomorphic for any $p$,
showing that this algebra is missing in the list of \cite{Tu2}.

\section{Method for computing invariants}

Among the possible approaches to compute the invariants of Lie
algebras, we choose the analytical one. Let $G$ be a connected Lie
group, $\frak{g}$ its corresponding Lie algebra and
$Ad:G\rightarrow GL\left(  \frak{g}\right)  $ the adjoint
representation. The coadjoint representation of $G$ is given by
the mapping:
\[
Ad^{\ast}:G\rightarrow GL\left(  \frak{g}^{\ast}\right)  :\;\left(
Ad_{g}^{\ast}F\right)  \left(  x\right)  =F\left(
ad_{g^{-1}}x\right)  ,\;
\]
where $g\in G,F\in\frak{g}^{\ast},x\in\frak{g}$. Denoting the
space of
analytical functions on $\frak{g}^{\ast}$ by $C^{\infty}\left(  \frak{g}%
^{\ast}\right)  $, we say that a function $F\in C^{\infty}\left(
\frak{g}^{\ast}\right)  $ is a semi-invariant for the coadjoint
representation if it verifies the condition
\[
F\left(  ad_{g^{-1}}x\right)  =\chi\left(  g\right)  F\left(
x\right)
\]
for any $g\in G$, $\chi$ being a character of $G$. In particular,
if $\chi\left(  g\right)  =1$, then $F$ is called an invariant of
$\frak{g}$. It follows at once that $F$ is an invariant if and
only if $F$ is constant on each coadjoint orbit. The usual method
to compute the (semi-)invariants of a Lie algebra is making use of
the theory of linear partial differential equations \cite{Fo}. Let
$\left\{  X_{1},..,X_{n}\right\}  $ be a basis of $\frak{g}$ and
let $\left\{  C_{ij}^{k}\right\}  $ be its structure constants
over this basis\ and $\left\{  x_{1},..,x_{n}\right\}  $ the
corresponding dual basis. We define the following differential
operators in the space $C^{\infty}\left(  \frak{g}^{\ast}\right)
$:
\begin{equation}
\varphi\left(  X_{i}\right)
=\widehat{X}_{i}:=C_{ij}^{k}x_{k}\frac{\partial }{\partial x_{j}},
\end{equation}
where $\left[  X_{i},X_{j}\right]  =C_{ij}^{k}X_{k}$ \ $\left(
1\leq i<j\leq
n\right)  $. It is not difficult to verify that the operators $\widehat{X}%
_{i}$ satisfy the brackets $\left[
\widehat{X}_{i},\widehat{X}_{j}\right]
=C_{ij}^{k}\widehat{X}_{k}$, therefore $\varphi$ defines a
representation of $\frak{g}$. Using this fact, the invariance
condition is then translated to an analytical condition
\cite{Tro1}: Let $f=\sum_{i=1}^{n}f_{i}x_{i}$ be a linear function
in $\frak{g}^{\ast}$, $\xi\in\frak{g}$ an arbitrary element and
Exp$:\frak{g\rightarrow} G$ the exponential function$.$

\begin{proposition}
For an analytical function $F\in C^{\infty}\left(
\frak{g}^{\ast}\right)  $
the following identity holds:%
\begin{equation}
\left.  \frac{d^{n}}{dt^{n}}\right|  _{t=0}F\left(  Ad_{Exp\left(
t\xi\right)  }^{\ast}f\right)  =\left(  -\varphi^{n}\left(
\xi\right)
F\right)  \left(  f\right)  .\label{DG1}%
\end{equation}
\end{proposition}

\begin{proof}
For any $\left\{  x_{1},..,x_{n}\right\}  $ we have the identity%
\[
x_{i}\left(  Ad_{Exp\left(  t\xi\right)  }^{\ast}f\right)  =\left(
Ad_{Exp\left(  t\xi\right)  }^{\ast}f\right)  \left(  X_{i}\right)
=f\left( Ad_{Exp\left(  -t\xi\right)  }X_{i}\right)  .
\]
This implies that
\begin{align*}
\left.  \frac{d}{dt}\right|  _{t=0}F\left(  Ad_{Exp\left(
t\xi\right) }^{\ast}f\right)   &  =\frac{\partial F}{\partial
x_{i}}\left(  f\right) \left.  \frac{d}{dt}\right|
_{t=0}x_{i}\left(  Ad_{Exp\left(  t\xi\right)
}^{\ast}f\right)  \\
&  =\frac{\partial F}{\partial x_{i}}\left(  f\right)  \left.  \frac{d}%
{dt}\right|  _{t=0}f\left(  Ad_{Exp\left(  -t\xi\right)
}X_{i}\right)  ,
\end{align*}
and $f$ being a linear function:%
\begin{equation}
\left.  \frac{d}{dt}\right|  _{t=0}f\left(  Ad_{Exp\left(
-t\xi\right) }X_{i}\right)  =f\left(  -ad\left(  \xi\right)
\left(  X_{i}\right)  \right) =-C_{ji}^{k}\xi^{j}f_{k}.
\end{equation}
Therefore
\[
\left.  \frac{d}{dt}\right|  _{t=0}F\left(  Ad_{Exp\left(
t\xi\right)
}^{\ast}f\right)  =-C_{ji}^{k}\xi^{j}f_{k}\frac{\partial F}{\partial x_{i}%
}\left(  \xi\right)  =-\left(  \varphi\left(  \xi\right)  F\right)
\left( f\right)  .
\]
Using the fact that
\[
Ad_{Exp\left(  t\xi\right)  }^{\ast}Ad_{Exp\left(
t^{\prime}\xi\right) }^{\ast}f=Ad_{Exp\left(  \left(
t+t^{\prime}\right)  \xi\right)  }^{\ast}f,
\]
the general case $n>1$ is proved recursively.
\end{proof}

\begin{theorem}
A function $F\in C^{\infty}\left(  \frak{g}^{\ast}\right)  $ is a
semi-invariant if and only if it is a solution of the following
system:
\begin{equation}
\widehat{X}_{i}F(x_{1},..,x_{n})=C_{ij}^{k}x_{k}\frac{\partial}{\partial
x_{j}}F\left(  x_{1},..,x_{n}\right)  =-d\chi\left(  X_{i}\right)
F,\quad
\end{equation}
where $1\leq i\leq n$ and $d\chi$ is the derivative of the
character $\chi$ at the identity element of $G$. In particular,
$F$ is an invariant if and only if
\begin{equation}
\widehat{X}_{i}F=0,\;1\leq i\leq n.\label{sys}%
\end{equation}
\end{theorem}

\begin{proof}
From the semi-invariance condition
\[
F\left(  Ad_{Exp\left(  t\xi\right)  }^{\ast}f\right)  =\chi\left(
Exp\left( t\xi\right)  \right)  F\left(  f\right)
\]
it follows by (\ref{DG1}) that
\begin{equation}
\left.  \frac{d}{dt}\right|  _{t=0}F\left(  Ad_{Exp\left(
t\xi\right) }^{\ast}f\right)  =\left.  \frac{d}{dt}\right|
_{t=0}\chi\left(  Exp\left( t\xi\right)  \right)  F\left(
f\right)  =-d\chi\left(  \xi\right)  F\left(
f\right)  .\label{DG2}%
\end{equation}
In particular, if $\chi=1$, equation (\ref{DG2}) implies that
\[
\widehat{X}_{i}\left(  F\right)  =0,\;1\leq i\leq n.
\]
Conversely, if $\left(  \left(  -\varphi\left(  \xi\right)
\right) ^{n}F\right)  \left(  f\right)  =d\chi\left(  \xi\right)
F\left(  f\right) $, expanding $F\left(  Ad_{Exp\left(
t\xi\right)  }^{\ast}f\right)  $ as a
Taylor series we obtain%
\[
F\left(  Ad_{Exp\left(  t\xi\right)  }^{\ast}f\right)  =F\left(
f\right)
+\sum_{n=1}^{\infty}\frac{\left(  -\varphi\left(  \xi\right)  \right)  ^{n}%
F}{n!}\left(  f\right)  t^{n}.
\]
Inserting $d\chi\left(  \xi\right)  F\left(  f\right)  $ into the
previous equation we obtain
\[
F\left(  Ad_{Exp\left(  t\xi\right)  }^{\ast}f\right)  =F\left(
f\right)
+\sum_{n=1}^{\infty}\frac{\left(  d\chi\left(  \xi\right)  \right)  F}%
{n!}\left(  f\right)  t^{n},
\]
and since $\chi\left(  Exp\left(  t\xi\right)  \right)  =Exp\left(
t\,d\chi\left(  \xi\right)  \right)  $ we deduce
\[
F\left(  Ad_{Exp\left(  t\xi\right)  }^{\ast}f\right)  =\chi\left(
Exp\left( t\xi\right)  \right)  F\left(  f\right)  .
\]
For the special case $\chi=1$ we recover the invariance condition.
\end{proof}

\medskip

This theorem reduces the determination of the invariants to a
system of linear first-order partial differential equations. A
maximal set of functionally independent invariants of $\frak{g}$
is called a fundamental set of invariants. For any given Lie
algebra $\frak{g}$, the number of functionally independent
invariants can be computed from the brackets. More specifically,
let $\left(  C_{ij}^{k}x_{k}\right)  $ be the matrix which
represents the commutator table over the basis $\left\{  X_{1},..,X_{n}%
\right\}  $, the $C_{ij}^{k}$ being the structure constants over
this basis. It follows from system (\ref{sys}) that the cardinal
$\mathcal{N}\left( \frak{g}\right)  $ of a fundamental set of
invariants of $\frak{g}$ is given by \cite{Be}:
\begin{equation}
\mathcal{N}\left(  \frak{g}\right)
=\dim\,\frak{g}-\mathrm{rank}\left(
C_{ij}^{k}x_{k}\right)  . \label{BB}%
\end{equation}
In particular, the number of functionally independent invariants
has the same parity as the dimension of the algebra, the
commutator matrix being skew-symmetric. This number can also be
derived using the equivalent approach of differential forms
\cite{C43}. However, the number of polynomial solutions of $\left(
2\right) $ will be strictly lower than
$\dim\,\frak{g}-\mathrm{rank}\left( C_{ij}^{k}x_{k}\right) $, and
only for particular classes of Lie algebras, such as semi-simple
or nilpotent Lie algebras we will obtain an equality \cite{AA,Ra}.
As known, if $\frak{A}$ denotes the universal enveloping algebra
of $\frak{g}$ and $Z\left( \frak{A}\right)  $ its center, then the
elements in $Z\left( \frak{A}\right)  $ correspond to the Casimir
operators \cite{Cas}. This set indeed coincides with the set of
polynomial invariants of $Ad^{\ast}$, while the rational
invariants correspond to ratios of polynomials contained in
$\frak{A}$. The Casimir operators are recovered from system
(\ref{sys}) using the symmetrization map%
\[
Sym\left(  x_{1}^{a_{1}},..,x_{k}^{a_{k}}\right)
=\frac{1}{k!}\sum_{\sigma\in S_{n}}x_{\sigma\left(  1\right)
}^{a_{1}}..x_{\sigma\left(  k\right) }^{a_{k}},
\]
$S_{k}$ being the symmetric group on $k$ letters \cite{AA}.
However, the system (\ref{sys}) can also have non-rational
solutions, which do not have such an interpretation in terms of
the enveloping algebra. Such solutions are usually called
generalized Casimir invariants.

\bigskip

Unfortunately, there is no general method to solve the system
(\ref{sys}), and only for certain types of algebras fundamental
systems of solutions can be determined. Recently an alternative
method was proposed in \cite{Pe94}, which turns out to be of
interest for Lie algebras admitting only one (independent)
invariant. This method reduces the problem of system (\ref{sys})
to the integration of a total differential equation whose
coefficients can be computed by determinants. The system can thus
be reduced to an equation:
\begin{equation}
dF=dx_{1}+U_{12}dx_{2}+...+U_{1n}dx_{n}=0,
\end{equation}
where the $U_{1i}$ are functions of the generators of $\frak{g}$
obtained from the ratios $\frac{\partial F/\partial
x_{i}}{\partial F/\partial x_{1}}$ by the rule of Cramer applied
to the commutator matrix $A(\frak{g})$ after having deleted the
dependent row \cite{Pe94}. That is,
\begin{equation}
U_{1i}=\frac{\Delta_{i}}{\Delta_{1}}=\frac{f_{i}}{f_{1}}%
\end{equation}
where the $\Delta_{i}$ are simply the determinants obtained from
the equation
\begin{equation}
A(\frak{g})(x_{i})^{T}=B,
\end{equation}
the latter $(n\times1)$-matrix expressing the coefficients of the
dependent row. Therefore the solution of (2) is
$F=\sum_{i=1}^{n}f_{i}x_{i}$, where $f_{i}$ is the result of (5)
after deleting the common factors of the determinants. Thus after
evaluating $\dim(\frak{g})-1$ determinants and integrating
equation (4), the Casimir operator is found. For some specific
types of Lie algebras having one invariant other direct methods to
integrate the system \index{(sys)} have been developed \cite{C33}.

\bigskip If the Lie algebra is decomposable, i.e., $\frak{g}=\frak{s\oplus r}%
$, then it is a direct consequence of \index{(BB)} that
$\mathcal{N}\left(  \frak{g}\right)  =\mathcal{N}\left(
\frak{s}\right)  +\mathcal{N}\left(  \frak{r}\right)  $. Since the
sum is direct, we have that $\left[  \frak{s},\frak{r}\right]  =0$
and therefore the rank of the matrix $A\left(  \frak{g}\right)  $
is the sum of the ranks of $A\left(  \frak{s}\right)  $ and
$A\left(  \frak{r}\right)  $. For algebras having a non-trivial
Levi decomposition, no apparent relation between the number of
invariants of the Levi factor and the radical and the number of
invariants of the semidirect sum exists, as shown by the special
affine Lie algebras \cite{C33,Pe94}. In particular, it is not
sufficient to determine the invariants of solvable Lie algebras to
have an overview of invariants of Lie algebras. At the present
time there is no hope to construct a theory of invariants that
covers all types of Lie algebras.

\section{Invariants of nine dimensional real algebras}

In this section we recall some specific results and properties
that have been used to solve system (\ref{sys}) for the nine
dimensional Lie algebras with nontrivial Levi decomposition. To
this extent, we recall the notions of contraction of Lie algebras
(for general properties see e.g. \cite{We}). Let $\frak{g}$ be a
Lie algebra and $\Phi_{t}\in Aut(\frak{g})$ a family of
automorphisms of $\frak{g}$, where $t\in\mathbb{N}$. For any
$X,Y\in\frak{g}$ define
\begin{equation}
\left[X,Y\right]_{\Phi_{t}}:=\left[\Phi_{t}(X),\Phi_{t}(Y)\right]=\Phi_{t}(\left[X,Y\right]).
\end{equation}
Obviously $\left[X,Y\right]_{\Phi_{t}}$ are the brackets of the
Lie algebra over the transformed basis. If the limit
\begin{equation}
\left[X,Y\right]_{\infty}:=\lim_{t\rightarrow
\infty}\Phi_{t}^{-1}\left[\Phi_{t}(X),\Phi_{t}(Y)\right]
\label{Ko}
\end{equation}
exists for any $X,Y\in\frak{g}$, equation (\ref{Ko}) defines a Lie
algebra $\frak{g}^{\prime}$ called contraction of $\frak{g}$,
nontrivial if  $\frak{g}$ and $\frak{g}^{\prime}$ are
nonisomorphic.

\begin{lemma}
Let $\frak{g}$ be a nine dimensional indecomposable Lie algebra
with nontrivial Levi decomposition. Then $\mathcal{N}\left(
\frak{g}\right)  =1,3.$
\end{lemma}

\begin{proof}
Let $\left\{  X_{1},..,X_{n},X_{n+1},..,X_{n+m}\right\}  $ be a
basis of $\frak{g}$ such that $\left\{  X_{1},..,X_{n}\right\}  $
is a basis of $\frak{s}$ and $\left\{  X_{n+1},..,X_{n+m}\right\}
$ is a basis of $\frak{r}.$ Let $\left\{  C_{ij}^{k}\right\}
_{1\leq i,j,k\leq n+m}$ be the structure constants of $\frak{g}$
over this basis. If we consider the change
of basis defined by:%
\begin{equation}%
\begin{array}
[c]{l}%
X_{i}^{\prime}:=X_{i},\;1\leq i\leq n,\\
X_{n+j}^{\prime}:=\frac{1}{n}X_{n+j},\;1\leq j\leq m,
\end{array}
\end{equation}
then, over the new basis the brackets are:%
\begin{equation}%
\begin{array}
[c]{l}%
\left[  X_{i}^{\prime},X_{j}^{\prime}\right]  =\left[
X_{i},X_{j}\right]
,\;1\leq i,j\leq n\\
\left[  X_{i}^{\prime},X_{n+j}^{\prime}\right]  =\frac{1}{n}\left[
X_{i},X_{n+j}\right]  ,\;1\leq i\leq n,1\leq j\leq m\\
\left[  X_{n+i}^{\prime},X_{n+j}^{\prime}\right]
=\frac{1}{n^{2}}\left[ X_{n+i},X_{n+j}\right]  ,\;1\leq i,j\leq m.
\end{array}
\end{equation}
Since $\left[  X_{i}^{\prime},X_{n+j}^{\prime}\right]  =\frac{1}{n}%
C_{i,j+n}^{n+k}X_{n+k}=C_{i,j+n}^{n+k}X_{n+k}^{\prime}$, this
 shows that the Levi part $\frak{s}$ and the
representation $R$ of $\frak{s}$ on the radical remain unchanged,
while the brackets of the radical $\frak{r}$
adopt the form:%
\begin{equation}
\left[  X_{n+i}^{\prime},X_{n+j}^{\prime}\right]  =\frac{1}{n}C_{n+i,n+j}%
^{n+k}X_{n+k}^{\prime},\;1\leq i,j,k\leq m.
\end{equation}
If we consider the limit $n\rightarrow\infty,$ we obtain a Lie
algebra with the same Levi subalgebra and describing
representation, but where the radical satisfies the brackets
\[
\left[  X_{n+i}^{\prime},X_{n+j}^{\prime}\right]  =0.
\]
This shows that the affine Lie algebra $\frak{s}\overrightarrow{\oplus}%
_{R}\left(  \deg R\right)  L_{1}$ is a In\"{o}n\"{u}-Wigner
contraction of $\frak{g}$. Using either the fact that contractions
increase or leave invariant the number of independent solutions of
(\ref{sys}), it follows that
\begin{equation}
\mathcal{N}\left(  \frak{g}\right)  \leq\mathcal{N}\left(  \frak{s}%
\overrightarrow{\oplus}_{R}\left(  \deg R\right)  L_{1}\right)  .
\end{equation}
Table 1 indicates the number of invariants of nine dimensional
affine algebras obtained by contraction of indecomposable algebras
$\frak{g}$ having a nontrivial Levi decomposition. The problem is
therefore reduced to prove that any indecomposable Lie algebra
contracting onto $\frak{so}\left(  3\right)
\overrightarrow{\oplus}_{ad\frak{so}\left(  3\right)
\oplus3D_{0}}6L_{1}$ or
$\frak{sl}\left(  2,\mathbb{R}\right)  \overrightarrow{\oplus}_{D_{\frac{1}%
{2}}\oplus4D_{0}}6L_{1}$ has at most three
invariants\footnote{Since $\left(
\frak{sl}\left(  2,\mathbb{R}\right)  \overrightarrow{\oplus}_{D_{1}%
\oplus3D_{0}}6L_{1}\right)  \otimes_{\mathbb{R}}\mathbb{C}$ is
isomorphic to
$\left(  \frak{so}\left(  3\right)  \overrightarrow{\oplus}_{ad\frak{so}%
\left(  3\right)  \oplus3D_{0}}6L_{1}\right)
\otimes_{\mathbb{R}}\mathbb{C}$, it suffices to prove the
assertion for one of the algebras, the property being preserved by
the real forms.}. We prove it for the first algebra, the remaining
case being analogous.

Let $\left\{  X_{4},..,X_{9}\right\}  $ be a basis of the radical
$\frak{r}$ such that $\left\{  X_{4},..,X_{6}\right\}  $
transforms according to the adjoint representation of
$\frak{so}\left(  3\right)  $ and the remaining elements by the
trivial representation. The Jacobi conditions imply that
\begin{align*}
\left[  X_{4},X_{5}\right]   &  =\left[  X_{4},X_{6}\right]
=\left[
X_{5},X_{6}\right]  =0,\\
\left[  X_{i},X_{7+j}\right]   &  =\lambda_{j}X_{i},\;i=4,5,6;\;j=0,1,2,\\
\left[  X_{7+i},X_{7+j}\right]   &  \in\mathbb{R}\left\langle X_{7}%
,X_{8},X_{9}\right\rangle ,\;0\leq i<j\leq2.\;
\end{align*}
Since the algebra is indecomposable, at least one $\lambda_{i}$
must be nonzero. Without loss of generality we can suppose that
$\lambda_{0}\neq0$. A change of basis allows to put
$\lambda_{1}=\lambda_{2}=0$. In order that $\frak{g}$ is
indecomposable, the following case holds:
\begin{align*}
\left[  X_{7},X_{8}\right]   &  =\alpha_{1}X_{8}+\beta X_{9},\\
\left[  X_{7},X_{9}\right]   &  =\gamma X_{8}+\alpha_{2}X_{9}%
\end{align*}
with rank$\left(
\begin{array}
[c]{cc}%
\alpha_{1} & \beta\\
\gamma & \alpha_{2}%
\end{array}
\right)  \geq1$. Further we obtain that
\[
\left[  X_{8},X_{9}\right]  =\delta_{1}X_{8}+\delta_{2}X_{9}%
\]
with
\[
\alpha_{2}\delta_{1}-\delta_{2}\gamma=\alpha_{1}\delta_{2}-\delta_{1}\beta=0.
\]
In any case it follows that%
\[
\left(
\begin{array}
[c]{ccc}%
0 & \alpha_{1}x_{8}+\beta x_{9} & \gamma x_{8}+\alpha_{2}x_{9}\\
-\alpha_{1}x_{8}-\beta x_{9} & 0 & \delta_{1}x_{8}+\delta_{2}x_{9}\\
-\gamma x_{8}-\alpha_{2}x_{9} & -\delta_{1}x_{8}-\delta_{2}x_{9} &
0
\end{array}
\right)  \geq2,
\]
showing that
\[
\mathrm{rank}\,A\left(  \frak{g}\right)  \geq6,
\]
and therefore $\mathcal{N}\left(  \frak{g}\right)  \leq3.$
\end{proof}%

\begin{table}
\caption{\label{Tab1} Affine Lie algebras obtained by contraction
from\newline nine dimensional algebras with Levi part
$\frak{s}\neq 0$.}
\begin{tabular}
[c]{|c|c|c|}\hline Affine Lie algebra $\frak{g}$ &
$\mathcal{N}\left(  \frak{g}\right)  $ & \\\hline $\frak{so}\left(
3\right) \overrightarrow{\oplus}_{ad\frak{so}\left( 3\right)
\oplus3D_{0}}6L_{1}$ & $5$ & decomposable\\
$\frak{so}\left(  3\right)  \overrightarrow{\oplus}_{R_{4}\oplus2D_{0}}6L_{1}$%
& $3$ & decomposable\\
$\frak{so}\left(  3\right)  \overrightarrow{\oplus}_{R_{5}\oplus D_{0}}6L_{1}$%
& $3$ & decomposable\\
$\frak{so}\left(  3\right)
\overrightarrow{\oplus}_{2ad\frak{so}\left(
3\right)  }6L_{1}$ & $3$ & indecomposable\\
$\frak{sl}\left(  2,\mathbb{R}\right)  \overrightarrow{\oplus}_{D_{\frac{1}%
{2}}\oplus4D_{0}}6L_{1}$ & $5$ & decomposable\\
$\frak{sl}\left(  2,\mathbb{R}\right)  \overrightarrow{\oplus}_{D_{1}%
\oplus3D_{0}}6L_{1}$ & $5$ & decomposable\\
$\frak{sl}\left(  2,\mathbb{R}\right)  \overrightarrow{\oplus}_{D_{\frac{3}%
{2}}\oplus2D_{0}}6L_{1}$ & $3$ & decomposable\\
$\frak{sl}\left(  2,\mathbb{R}\right)
\overrightarrow{\oplus}_{2D_{\frac
{1}{2}}\oplus2D_{0}}6L_{1}$ & $3$ & decomposable\\
$\frak{sl}\left(  2,\mathbb{R}\right)
\overrightarrow{\oplus}_{D_{1}\oplus
D_{\frac{1}{2}}\oplus D_{0}}6L_{1}$ & $3$ & decomposable\\
$\frak{sl}\left(  2,\mathbb{R}\right)
\overrightarrow{\oplus}_{D_{2}\oplus
D_{0}}6L_{1}$ & $3$ & decomposable\\
$\frak{sl}\left(  2,\mathbb{R}\right)  \overrightarrow{\oplus}_{D_{\frac{5}%
{2}}}6L_{1}$ & $3$ & indecomposable\\
$\frak{sl}\left(  2,\mathbb{R}\right)  \overrightarrow{\oplus}_{D_{\frac{3}%
{2}}\oplus D\frac{1}{2}}6L_{1}$ & $3$ & indecomposable\\
$\frak{sl}\left(  2,\mathbb{R}\right)  \overrightarrow{\oplus}_{2D_{1}}6L_{1}$%
& $3$ & indecomposable\\
$\frak{sl}\left(  2,\mathbb{R}\right)
\overrightarrow{\oplus}_{3D_{\frac {1}{2}}}6L_{1}$ & $3$ &
indecomposable\\\hline
\end{tabular}
\end{table}

\medskip

We now recall some properties that have been used in the
integration of the system (\ref{sys}) for the nine dimensional
algebras with Levi decomposition.

\bigskip

\begin{proposition}
If $\frak{g}$ admits a codimension one subalgebra $\frak{k}$, then
$\frak{g}$
and $\frak{k}$ have $\left[  \frac{\mathcal{N}\left(  \frak{g}\right)  }%
{2}\right]  $ invariants in common.
\end{proposition}

\begin{proof}
In \cite{Sh} it was shown that the number of invariants of a Lie
algebra $\frak{g}$ is related with that of subalgebras by means of
the following
inequality:%
\[
\dim\frak{g}-\mathcal{N}\left(  \frak{g}\right)  -\dim\frak{k}-\mathcal{N}%
\left(  k\right)  +2l^{\prime}\geq0,
\]
where $l^{\prime}$ is the number of invariants of $\frak{g}$
depending only on variables of the subalgebra $\frak{k}$. For the
case of codimension one
subalgebras we therefore obtain%
\[
1-\mathcal{N}\left(  \frak{g}\right)  -\mathcal{N}\left(  k\right)
+2l^{\prime}\geq0.
\]
By the preceding lemma, two cases must be distinguished:

\begin{enumerate}
\item  If $\ \mathcal{N}\left(  \frak{g}\right)  =3$,we obtain the inequality%
\[
2l^{\prime}-\mathcal{N}\left(  k\right)  -2\geq0,
\]
thus
\[
l^{\prime}\geq1+\frac{\mathcal{N}\left(  \frak{k}\right)
}{2}\geq\left[ \frac{\mathcal{N}\left(  \frak{g}\right)
}{2}\right]  .
\]
In particular, in this case the codimension subalgebra can possess
up to four independent invariants.

\item  If $\mathcal{N}\left(  \frak{g}\right)  =1$, then
\[
2l^{\prime}-\mathcal{N}\left(  \frak{k}\right)  \geq0,
\]
and since
\[
2\geq2l^{\prime}\geq\mathcal{N}\left(  k\right)  ,
\]
either $l^{\prime}=0$ or $l^{\prime}=1$. The subalgebra $\frak{k}$
has at most two invariants.
\end{enumerate}
\end{proof}

\bigskip The preceding result is of interest for the computation of invariants
using codimension one subalgebras of $\frak{g}$ (if they exist).
More specifically, suppose that $\frak{k}$ is a codimension one
subalgebra of $\frak{g}$ and $\mathcal{F}=\left\{
J_{1},..,J_{l}\right\}  $ a fundamental system of invariants of
$\frak{k}$. Suppose further that there exists an element
$X\in\frak{g}$ with $X\notin\left[  \frak{g},\frak{g}\right]  $
such that for any invariant $F$ of $\frak{g}$ we have
$\frac{\partial F}{\partial x}=0$. Then the system (\ref{sys})
associated to $\frak{g}$ is the union of the subsystem
corresponding to the subalgebra $\frak{k}$ and the equation
\begin{equation}
\widehat{X}\left(  F\right)  =C_{iX}^{j}x_{j}\frac{\partial F}{\partial x_{i}%
}=0. \label{Eq}%
\end{equation}
If the set $\mathcal{F}=\left\{  J_{1},..,J_{l}\right\}  $
transforms linearly by the differential operator $\widehat{X}$,
i.e., if
\[
\widehat{X}\left(  J_{i}\right)  =\alpha^{j}J_{j},\;1\leq i\leq l,
\]
then we can always find a function $\Phi\left(
J_{1},..,J_{l}\right)  $ that is a common solution to the
subalgebra and the equation (\ref{Eq}) \cite{Di}. This procedure
always works when we have a Lie algebra that is an extension by a
derivation of a codimension one subalgebra \cite{C33}, and
constitutes the main technique employed in this work to compute
the invariants of nine dimensional Lie algebras.

\medskip

As example, consider the Lie algebra $L_{9,7}^{\ast}$ given by the
brackets (\ref{MA}). In this case, the system to be solved is
\[
\left(
\begin{array}
[c]{rrrrrrrrr}%
0 & x_{3} & -x_{2} & \frac{x_{7}}{2} & \frac{x_{6}}{2} &
-\frac{x_{5}}{2} &
-\frac{x_{4}}{2} & 0 & 0\\
-x_{3} & 0 & x_{1} & \frac{x_{5}}{2} & -\frac{x_{4}}{2} &
\frac{x_{7}}{2} &
-\frac{x_{6}}{2} & 0 & 0\\
x_{2} & -x_{1} & 0 & \frac{x_{6}}{2} & -\frac{x_{7}}{2} &
-\frac{x_{4}}{2} &
\frac{x_{5}}{2} & 0 & 0\\
-\frac{x_{7}}{2} & -\frac{x_{5}}{2} & -\frac{x_{6}}{2} & 0 & 0 &
x_{8} & 0 &
0 & x_{4}\\
-\frac{x_{4}}{2} & \frac{x_{4}}{2} & \frac{x_{7}}{2} & 0 & 0 & 0 &
x_{8} & 0 &
x_{5}\\
\frac{x_{5}}{2} & -\frac{x_{7}}{2} & \frac{x_{4}}{2} & -x_{8} & 0
& 0 & 0 &
0 & x_{6}\\
\frac{x_{4}}{2} & \frac{x_{6}}{2} & -\frac{x_{5}}{2} & 0 & -x_{8}
& 0 & 0 &
0 & x_{7}\\
0 & 0 & 0 & 0 & 0 & 0 & 0 & 0 & 2x_{8}\\
0 & 0 & 0 & -x_{4} & -x_{5} & -x_{6} & -x_{7} & -2x_{8} & 0
\end{array}
\right)  \left(
\begin{array}
[c]{c}%
\partial_{x_{1}}F\\
\partial_{x_{2}}F\\
\partial_{x_{3}}F\\
\partial_{x_{4}}F\\
\partial_{x_{5}}F\\
\partial_{x_{6}}F\\
\partial_{x_{7}}F\\
\partial_{x_{8}}F\\
\partial_{x_{9}}F
\end{array}
\right)  =0
\]
Since the rank of the matrix is $8$, there is only one invariant.
It follows from
\[
\widehat{X}_{8}\left(  F\right)  =2x_{8}\frac{\partial F}{\partial
x_{9}}=0
\]
that the solutions do not depend on $x_{9}$. We can therefore
consider the eight dimensional subalgebra of $L_{9,7}^{\ast}$
spanned by $\left\langle X_{1},..,X_{8}\right\rangle $. This
subalgebra is isomorphic to the semidirect product of
$\frak{so}\left(  3\right)  $ with a five dimensional Heisenberg
Lie algebra $\frak{h}_{2}$, and a fundamental system of invariants
can be computed by by means of determinantal methods \cite{C40}.
By this procedure we
obtain%
\begin{align*}
I_{1} &  =16\left(  x_{1}^{2}+x_{2}^{2}+x_{3}^{2}\right)
x_{8}^{2}+\left( x_{4}^{2}+x_{5}^{2}+x_{6}^{2}+x_{7}^{2}\right)
^{2}+16x_{1}x_{8}\left(
x_{4}x_{5}+x_{6}x_{7}\right)  +\\
&  -16x_{2}x_{8}\left(  x_{5}x_{6}-x_{4}x_{7}\right)
+8x_{3}x_{8}\left(
x_{4}^{2}-x_{5}^{2}+x_{6}^{2}-x_{7}^{2}\right)  ,\\
I_{2} &  =x_{8}.
\end{align*}
If we evaluate these functions in the differential operator
$\widehat{X}_{9}$, it is straightforward to verify that \
\begin{align}
\widehat{X}_{9}\left(  I_{1}\right)   &  =-4I_{1},\nonumber\\
\widehat{X}_{9}\left(  I_{2}\right)   &  =-2I_{2}.\label{SI}%
\end{align}
We will call (\ref{SI}) the semi-invariance conditions with
respect to the operator $\widehat{X}_{9}$. Since $I_{1}$ and
$I_{2}$ are semi-invariants of $L_{9,7}^{\ast}$, in order to find
an invariant, we consider the new variables
$u=I_{1}$ and $v=I_{2}$ and the differential equation%
\begin{equation}
\frac{\partial}{\partial u}F\left(  u,v\right)
+\frac{\widehat{X}_{9}\left( v\right)  }{\widehat{X}_{9}\left(
u\right)  }\frac{\partial}{\partial v}F\left(  u,v\right)  =0,
\end{equation}
with solution $J=uv^{-2}$. Therefore the function
$J=I_{1}I_{2}^{-2}$ provides the invariant of $L_{9,7}^{\ast}$. It
can actually be shown that $L_{9,7}^{\ast}$ is an extension of the
subalgebra $L\left\langle X_{1},..,X_{8}\right\rangle $ by a
derivation of it radical (isomorphic to $\frak{h}_{2}$).

\medskip

Indecomposable Lie algebras with a Levi subalgebra of rank one
have some special properties that do not hold for higher ranks,
and that are useful for the computation of invariants \cite{Ca3}.
These often allow to find adequate subalgebras to obtain the
semi-invariants of the algebra, or even to obtain alternative
criteria to integrate system (\ref{sys}):

\begin{proposition}
Let $\frak{s}=\frak{so}\left(  3\right)  ,\frak{sl}\left(  2,\mathbb{R}%
\right)  $. If the radical $\frak{r}$ of $\frak{s}\overrightarrow{\oplus}%
_{R}\frak{r}$ has a one dimensional centre, then the
representation $R$ describing the semidirect sum contains a copy
of the trivial representation $D_{0}$.
\end{proposition}

This result specially applies to nilpotent radicals that contract
onto the Heisenberg algebra or solvable radicals that contain some
zero weight for its action on the nilradical \cite{C56,Tre}.

\medskip

We finally recall a result that is of interest for Lie algebras
with abelian radical.

\begin{proposition}
Let $R$ be a representation of $\frak{s}$, where
$\frak{s}=\frak{so}(3)$ or $\frak{sl}(2,\mathbb{R})$. If ${\rm
dim}R > 3$ and the radical of
$\frak{s}\overrightarrow{\oplus}_{R}\frak{r}$ is abelian, then
there exists a fundamental set of invariants formed by functions
$F_{i}$ depending only on variables associated to elements of
$\frak{r}$.
\end{proposition}

The proof can be found in \cite{Ca3}. This case turns out to be
usually more difficult to integrate, particularly when the
representation $R$ is irreducible. In the case that an algebra has
an abelian radical, the contraction procedure is often employed to
obtain the invariants \cite{C43,H1}. For algebras that cannot be
obtained by a limiting process, we are often led to integrate the
corresponding system (\ref{sys}) directly. Even if we can make
some predictions on the invariants, the effective computation for
some affine Lie algebras (i.e., semidirect product of abelian and
semisimple Lie algebras) can be extremely complicated, due to the
enormous number of terms (see e.g. the Lie algebra $L_{9,59}$).
This seems to indicate that, even if the system (\ref{sys}) is
extremely simple in appearance, no general method leading to a
fundamental system of invariants can be obtained.

\newpage

\section{Tables of Invariants of nine dimensional algebras}

We shortly resume the notation used along the tables. Let
$\frak{g}=\frak{s}\overrightarrow{\oplus}_{R}\frak{r}$ be a nine
dimensional Lie algebra with nontrivial Levi decomposition, where
$\frak{s}$ is the Levi subalgebra, $\frak{r}$ the radical and $R$
the representation of $\frak{s}$ describing the semidirect
product. The structure tensor $\left\{ C_{ij}^{k}\right\}  $ of
$\frak{g}$ is given over a basis $\left\{ X_{1},..,X_{9}\right\}
$, where $\left\{  X_{1},X_{2},X_{3}\right\}  $ is a basis of the
Levi subalgebra and $\left\{  X_{4},..,X_{9}\right\}  $ is the
basis of the radical. The following information is given in the
tables

\begin{itemize}
\item  The Levi decomposition and the structure tensor after the
classification in \cite{Tu3}. The subindex $k$ in the notation $L_{9,k}%
^{p_{1},..,p_{r}}$ indicates the number of the isomorphism class,
while super-indices $p_{1},..,p_{k}$ refer to continuous
parameters (if any).

\item The notation for the radicals are those employed in
references \cite{Mu} and \cite{PWZ}.

\item  The rank of the matrix $A\left(  \frak{g}\right)  =\left(  C_{ij}%
^{k}x_{k}\right)  .$

\item  The codimension of the derived ideal $[\frak{g,g].}$

\item  Conditions on the invariants of the algebra. If the invariants depend
on all variables $\left\{  x_{1},..,x_{9}\right\}  $, then the invariants are
searched either by direct integration or using alternative methods.

\item  The subalgebra $L\left\langle
X_{i_{1}},..,X_{i_{k}}\right\rangle $ of $\frak{g}$ generated by
the elements $\left\{  X_{i_{1}},..,X_{i_{k}}\right\} $ providing
semi-invariants of $\frak{g}$, only in the case that the
invariants of $\frak{g}$ can be expressed in terms of some
subalgebra invariants. The isomorphism class uses the notations of
\cite{Tu2}

\item  A fundamental system of invariants $\left\{  J_{1},..,J_{t}\right\}  $
of $L\left\langle X_{i_{1}},..,X_{i_{k}}\right\rangle $.

\item  The semi-invariance conditions.

\item  Functions $\Phi_{l}\left(  J_{1},..,J_{t}\right)  $ providing a
fundamental system of invariants of $\frak{g}$.
\end{itemize}

\newpage

$\mathbf{L}_{9,1}^{p,q}\;\left[  pq\neq0\right]  $

\begin{itemize}
\item  Levi decomposition: $L_{9,1}^{p,q}=\frak{so}\left(  3\right)
\overrightarrow{\oplus}_{R}\frak{g}_{6,1}$

\item  Describing representation: $R=ad\frak{so}\left(  3\right)  \oplus3D_{0}.$

\item  Structure tensor%
\[%
\begin{array}
[c]{cccccc}%
C_{12}^{3}=1, & C_{13}^{2}=-1, & C_{23}^{1}=1, & C_{15}^{6}=1, & C_{16}%
^{5}=1, & C_{24}^{6}=-1,\\
C_{26}^{4}=1, & C_{34}^{5}=1, & C_{35}^{4}=-1, & C_{49}^{4}=1, & C_{59}%
^{5}=1, & C_{69}^{6}=1,\\
C_{79}^{7}=p, & C_{89}^{8}=q. &  &  &  & \\
&  &  &  &  & \\
&  &  &  &  &
\end{array}
\]

\item $\mathrm{codim}_{\frak{g}}\left[  \frak{g},\frak{g}\right]  =1.$

\item  Rank $A\left(  \frak{g}\right)  =6$

\item  Conditions on invariants:
\[
\frac{\partial F}{\partial x_{9}}=0.
\]

\item  Codimension one subalgebra: $L\left\langle X_{1},..,X_{8}\right\rangle
\simeq L_{6,1}\oplus2L_{1}$

\item  Invariants of subalgebra:%
\begin{align*}
J_{1} &  =x_{4}^{2}+x_{5}^{2}+x_{6}^{2}\\
J_{2} &  =x_{1}x_{4}+x_{2}x_{5}+x_{3}x_{6}\\
J_{3} &  =x_{7},\;J_{4}=x_{8}%
\end{align*}

\item  Semi-invariance conditions:%
\begin{align*}
\widehat{X}_{9}\left(  J_{1}\right)   &  =-2J_{1},\;\widehat{X}_{9}\left(
J_{2}\right)  =-J_{2}\\
\widehat{X}_{9}\left(  J_{3}\right)   &  =-pJ_{3},\;\widehat{X}_{9}\left(
J_{4}\right)  =-qJ_{4}.
\end{align*}

\item  Invariants of $L_{9,1}^{p,q}$%
\[
I_{1}=\frac{J_{1}^{p}}{J_{3}^{2}},\;I_{2}=\frac{J_{2}^{p}}{J_{3}}%
,\;I_{3}=\frac{J_{3}^{q}}{J_{4}^{p}}.
\]
\newpage
\end{itemize}

$\mathbf{L}_{9,2}^{p}$

\bigskip

\begin{itemize}
\item  Levi decomposition: $L_{9,2}^{p}=\frak{so}\left(  3\right)
\overrightarrow{\oplus}_{R}\frak{g}_{6,2}$

\item  Describing representation: $R=ad\frak{so}\left(  3\right)  \oplus3D_{0}.$

\item  Structure tensor
\[%
\begin{array}
[c]{cccccc}%
C_{12}^{3}=1, & C_{13}^{2}=-1, & C_{23}^{1}=1, & C_{15}^{6}=1, & C_{16}%
^{5}=1, & C_{24}^{6}=-1,\\
C_{26}^{4}=1, & C_{34}^{5}=1, & C_{35}^{4}=1, & C_{49}^{4}=1, & C_{59}%
^{5}=1, & C_{69}^{6}=1,\\
C_{79}^{7}=p, & C_{89}^{8}=1, & C_{79}^{8}=p. &  &  & \\
&  &  &  &  & \\
&  &  &  &  &
\end{array}
\]

\item $\mathrm{codim}_{\frak{g}}\left[  \frak{g},\frak{g}\right]  =\left\{
\begin{tabular}
[c]{ll}%
$1,$ & $p\neq0$\\
$2,$ & $p=0$%
\end{tabular}
\right.  .$

\item  Rank $A\left(  \frak{g}\right)  =6$

\item  Conditions on invariants:
\[
\frac{\partial F}{\partial x_{9}}=0.
\]

\item  Codimension one subalgebra: $L\left\langle X_{1},..,X_{8}\right\rangle
\simeq L_{6,1}\oplus2L_{1}$

\item  Invariants of subalgebra:%
\begin{align*}
J_{1} &  =x_{4}^{2}+x_{5}^{2}+x_{6}^{2}\\
J_{2} &  =x_{1}x_{4}+x_{2}x_{5}+x_{3}x_{6}\\
J_{3} &  =x_{7},\;J_{4}=x_{8}%
\end{align*}

\item  Semi-invariance conditions:%
\begin{align*}
\widehat{X}_{9}\left(  J_{1}\right)   &  =-2J_{1},\;\widehat{X}_{9}\left(
J_{2}\right)  =-J_{2}\\
\widehat{X}_{9}\left(  J_{3}\right)   &  =-pJ_{3},\;\widehat{X}_{9}\left(
J_{4}\right)  =-J_{3}-pJ_{4}.
\end{align*}

\item  Invariants of $L_{9,2}^{p}$%
\begin{align*}
I_{1} &  =\frac{J_{1}}{J_{2}^{2}},\;I_{2}=\frac{J_{2}^{p}}{J_{3}}%
,\;I_{3}=-\frac{J_{3}}{J_{4}}+\frac{1}{p}\ln\left(  J_{3}\right)  ,\;p\neq0.\\
I_{1} &  =\frac{J_{1}}{J_{2}^{2}},\;I_{2}=\frac{J_{2}^{p}}{J_{3}}%
,\;I_{3}=J_{1}\exp\left(  -J_{4}/J_{3}\right)  ,\;p=0.
\end{align*}
\newpage
\end{itemize}

$\mathbf{L}_{9,3}^{p,q}\;[p\neq0,\;q\geq0]$

\bigskip

\begin{itemize}
\item  Levi decomposition: $L_{9,3}^{p,q}=\frak{so}\left(  3\right)
\overrightarrow{\oplus}_{R}\frak{g}_{6,8}$

\item  Describing representation: $R=ad\frak{so}\left(  3\right)  \oplus3D_{0}$

\item  Structure tensor:%
\[%
\begin{array}
[c]{cccccc}%
C_{12}^{3}=1, & C_{13}^{2}=-1, & C_{23}^{1}=1, & C_{15}^{6}=1, & C_{16}%
^{5}=1, & C_{24}^{6}=-1,\\
C_{26}^{4}=1, & C_{34}^{5}=1, & C_{35}^{4}=1, & C_{49}^{4}=p, & C_{59}%
^{5}=p, & C_{69}^{6}=p,\\
C_{79}^{7}=q, & C_{79}^{8}=-1, & C_{89}^{7}=1, & C_{89}^{8}=q. &  & \\
&  &  &  &  & \\
&  &  &  &  &
\end{array}
\]

\item $\mathrm{codim}_{\frak{g}}\left[  \frak{g},\frak{g}\right]  =1.$

\item  Rank $A\left(  \frak{g}\right)  =6$

\item  Conditions on invariants:
\[
\frac{\partial F}{\partial x_{9}}=0.
\]

\item  Codimension one subalgebra: $L\left\langle X_{1},..,X_{8}\right\rangle
\simeq L_{6,1}\oplus2L_{1}$

\item  Invariants of subalgebra:%
\begin{align*}
J_{1} &  =x_{4}^{2}+x_{5}^{2}+x_{6}^{2}\\
J_{2} &  =x_{1}x_{4}+x_{2}x_{5}+x_{3}x_{6}\\
J_{3} &  =x_{7},\;J_{4}=x_{8}%
\end{align*}

\item  Semi-invariance conditions:%
\begin{align*}
\widehat{X}_{9}\left(  J_{1}\right)   &  =-2pJ_{1},\;\widehat{X}_{9}\left(
J_{2}\right)  =-pJ_{2}\\
\widehat{X}_{9}\left(  J_{3}\right)   &  =-qJ_{3}+J_{4},\;\widehat{X}%
_{9}\left(  J_{4}\right)  =-J_{3}-qJ_{4}.
\end{align*}

\item  Invariants of $L_{9,3}^{p,q}$%
\[
I_{1}=\frac{J_{1}}{J_{2}^{2}},\;I_{2}=\frac{J_{2}^{q-i}}{\left(  J_{3}%
-iJ_{4}\right)  ^{p}},\;I_{3}=\ln\left(  x_{7}^{2}+x_{8}^{2}\right)
-2q\arctan\left(  \frac{J_{4}}{J_{3}}\right)  .
\]
\newpage
\end{itemize}

$\mathbf{L}_{9,4}^{{}}$

\begin{itemize}
\item  Levi decomposition: $L_{9,4}^{{}}=\frak{so}\left(  3\right)
\overrightarrow{\oplus}_{R}A_{6,5}$

\item  Describing representation: $R=R_{4}\oplus2D_{0}$

\item  Structure tensor:%
\[%
\begin{array}
[c]{cccccc}%
C_{12}^{3}=1, & C_{13}^{2}=-1, & C_{23}^{1}=1, & C_{14}^{7}=\frac{1}{2}, &
C_{15}^{6}=\frac{1}{2}, & C_{16}^{5}=-\frac{1}{2},\\
C_{17}^{4}=-\frac{1}{2}, & C_{24}^{5}=\frac{1}{2}, & C_{25}^{4}=-\frac{1}%
{2}, & C_{26}^{7}=\frac{1}{2}, & C_{27}^{6}=-\frac{1}{2}, & C_{34}^{6}%
=\frac{1}{2},\\
C_{35}^{7}=-\frac{1}{2}, & C_{36}^{4}=-\frac{1}{2}, & C_{37}^{5}=\frac{1}%
{2}, & C_{45}^{9}=-1, & C_{47}^{8}=1, & C_{56}^{8}=-1,\\
C_{67}^{9}=1. &  &  &  &  & \\
&  &  &  &  &
\end{array}
\]

\item $\mathrm{codim}_{\frak{g}}\left[  \frak{g},\frak{g}\right]  =0.$

\item  Rank $A\left(  \frak{g}\right)  =6$

\item  Invariants of $L_{9,4}^{{}}$%
\[
I_{1}=x_{8},\;I_{2}=x_{9},\;I_{3}=\sqrt{D},
\]
where $D$ is the determinant%
\[
D:=\left|
\begin{array}
[c]{rrrrrrrr}%
0 & x_{3} & -x_{2} & \frac{x_{7}}{2} & \frac{x_{6}}{2} & -\frac{x_{5}}{2} &
-\frac{x_{4}}{2} & x_{1}\\
-x_{3} & 0 & x_{1} & \frac{x_{5}}{2} & -\frac{x_{4}}{2} & \frac{x_{7}}{2} &
-\frac{x_{6}}{2} & x_{2}\\
x_{2} & -x_{1} & 0 & \frac{x_{6}}{2} & -\frac{x_{7}}{2} & -\frac{x_{4}}{2} &
\frac{x_{5}}{2} & x_{3}\\
-\frac{x_{7}}{2} & -\frac{x_{5}}{2} & -\frac{x_{6}}{2} & 0 & -x_{9} & 0 &
x_{8} & \frac{x_{4}}{2}\\
-\frac{x_{4}}{2} & \frac{x_{4}}{2} & \frac{x_{7}}{2} & x_{9} & 0 & -x_{8} &
0 & \frac{x_{5}}{2}\\
\frac{x_{5}}{2} & -\frac{x_{7}}{2} & \frac{x_{4}}{2} & 0 & x_{8} & 0 & x_{9} &
\frac{x_{6}}{2}\\
\frac{x_{4}}{2} & \frac{x_{6}}{2} & -\frac{x_{5}}{2} & 0 & 0 & -x_{9} & 0 &
\frac{x_{7}}{2}\\
-x_{1} & -x_{2} & -x_{3} & -\frac{x_{4}}{2} & -\frac{x_{5}}{2} & -\frac{x_{6}%
}{2} & -\frac{x_{7}}{2} & 0
\end{array}
\right|
\]
\newpage
\end{itemize}

$\mathbf{L}_{9,5}^{p}\;[p\neq0]$

\begin{itemize}
\item  Levi decomposition: $L_{9,5}^{p}=\frak{so}\left(  3\right)
\overrightarrow{\oplus}_{R}\frak{g}_{6,1}$

\item  Describing representation: $R=R_{4}\oplus2D_{0}$

\item  Structure tensor:%
\[%
\begin{array}
[c]{cccccc}%
C_{12}^{3}=1, & C_{13}^{2}=-1, & C_{23}^{1}=1, & C_{14}^{7}=\frac{1}{2}, &
C_{15}^{6}=\frac{1}{2}, & C_{16}^{5}=-\frac{1}{2},\\
C_{17}^{4}=-\frac{1}{2}, & C_{24}^{5}=\frac{1}{2}, & C_{25}^{4}=-\frac{1}%
{2}, & C_{26}^{7}=\frac{1}{2}, & C_{27}^{6}=-\frac{1}{2}, & C_{34}^{6}%
=\frac{1}{2},\\
C_{35}^{7}=-\frac{1}{2}, & C_{36}^{4}=-\frac{1}{2}, & C_{37}^{5}=\frac{1}%
{2}, & C_{49}^{4}=1, & C_{59}^{5}=1, & C_{69}^{6}=1,\\
C_{79}^{7}=1, & C_{89}^{8}=p. &  &  &  & \\
&  &  &  &  &
\end{array}
\]

\item $\mathrm{codim}_{\frak{g}}\left[  \frak{g},\frak{g}\right]  =1.$

\item  Rank $A\left(  \frak{g}\right)  =6$

\item  Conditions on invariants:
\[
\frac{\partial F}{\partial x_{j}}=0,\;j=1,2,3,9.
\]

\item  Codimension one subalgebra: $L\left\langle X_{1},..,X_{8}\right\rangle
\simeq L_{7,2}\oplus L_{1}$

\item  Invariants of subalgebra:%
\begin{align*}
J_{1} &  =x_{4}^{2}+x_{5}^{2}+x_{6}^{2}+x_{7}^{2}\\
J_{2} &  =x_{8}%
\end{align*}

\item  Semi-invariance conditions:%
\[
\widehat{X}_{9}\left(  J_{1}\right)  =-2J_{1},\;\widehat{X}_{9}\left(
J_{2}\right)  =-pJ_{2}.
\]

\item  Invariants of $L_{9,5}^{p}$%
\[
I_{1}=\frac{J_{1}^{p}}{J_{2}^{2}}.
\]
\newpage
\end{itemize}

$\mathbf{L}_{9,6}^{p,q}\;[q\neq0]$

\begin{itemize}
\item  Levi decomposition: $L_{9,6}^{p,q}=\frak{so}\left(  3\right)
\overrightarrow{\oplus}_{R}\frak{g}_{6,11}$

\item  Describing representation: $R=R_{4}\oplus2D_{0}$

\item  Structure tensor:%
\[%
\begin{array}
[c]{cccccc}%
C_{12}^{3}=1, & C_{13}^{2}=-1, & C_{23}^{1}=1, & C_{14}^{7}=\frac{1}{2}, &
C_{15}^{6}=\frac{1}{2}, & C_{16}^{5}=-\frac{1}{2},\\
C_{17}^{4}=-\frac{1}{2}, & C_{24}^{5}=\frac{1}{2}, & C_{25}^{4}=-\frac{1}%
{2}, & C_{26}^{7}=\frac{1}{2}, & C_{27}^{6}=-\frac{1}{2}, & C_{34}^{6}%
=\frac{1}{2},\\
C_{35}^{7}=-\frac{1}{2}, & C_{36}^{4}=-\frac{1}{2}, & C_{37}^{5}=\frac{1}%
{2}, & C_{49}^{4}=p, & C_{49}^{6}=-1, & C_{59}^{5}=p,\\
C_{59}^{7}=-1, & C_{69}^{6}=p, & C_{69}^{4}=1, & C_{79}^{5}=1, & C_{79}%
^{7}=p, & C_{89}^{8}=q.\\
&  &  &  &  &
\end{array}
\]

\item $\mathrm{codim}_{\frak{g}}\left[  \frak{g},\frak{g}\right]  =1.$

\item  Rank $A\left(  \frak{g}\right)  =6$

\item  Conditions on invariants:
\[
\frac{\partial F}{\partial x_{j}}=0,\;j=1,2,3,9.
\]

\item  Codimension one subalgebra: $L\left\langle X_{1},..,X_{8}\right\rangle
\simeq L_{7,2}\oplus L_{1}$

\item  Invariants of subalgebra:%
\begin{align*}
J_{1} &  =x_{4}^{2}+x_{5}^{2}+x_{6}^{2}+x_{7}^{2}\\
J_{2} &  =x_{8}%
\end{align*}

\item  Semi-invariance conditions:%
\[
\widehat{X}_{9}\left(  J_{1}\right)  =-2pJ_{1},\;\widehat{X}_{9}\left(
J_{2}\right)  =-qJ_{2}.
\]

\item  Invariants of $L_{9,6}^{p,q}$%
\begin{align*}
I_{1} &  =\frac{J_{1}^{q}}{J_{2}^{2p}}\text{ for }p\neq0,\\
I_{1} &  =J_{1}\text{ for }p=0.
\end{align*}
\newpage
\end{itemize}

$\mathbf{L}_{9,7}^{p}\;\left(  p\neq0\right)  $

\begin{itemize}
\item  Levi decomposition: $L_{9,7}^{p}=\frak{so}\left(  3\right)
\overrightarrow{\oplus}_{R}\frak{g}_{6,82}^{\ast}$

\item  Describing representation: $R=R_{4}\oplus2D_{0}$

\item  Structure tensor:%
\[%
\begin{array}
[c]{llllll}%
C_{12}^{3}=1, & C_{13}^{2}=-1, & C_{23}^{1}=1, & C_{14}^{7}=\frac{1}{2}, &
C_{15}^{6}=\frac{1}{2}, & C_{16}^{5}=-\frac{1}{2},\\
C_{17}^{4}=-\frac{1}{2}, & C_{24}^{5}=\frac{1}{2}, & C_{25}^{4}=-\frac{1}%
{2}, & C_{26}^{7}=\frac{1}{2}, & C_{27}^{6}=-\frac{1}{2}, & C_{34}^{6}%
=\frac{1}{2},\\
C_{35}^{7}=-\frac{1}{2}, & C_{36}^{4}=-\frac{1}{2}, & C_{37}^{5}=\frac{1}%
{2}, & C_{46}^{8}=1, & C_{49}^{4}=p, & C_{49}^{6}=1,\\
C_{57}^{8}=1, & C_{59}^{5}=p, & C_{59}^{7}=1, & C_{69}^{4}=-1, & C_{69}%
^{6}=p, & C_{79}^{5}=-1,\\
C_{79}^{7}=p, & C_{89}^{8}=2p. &  &  &  &
\end{array}
\]

\item $\mathrm{codim}_{\frak{g}}\left[  \frak{g},\frak{g}\right]  =1.$

\item  Rank $A\left(  \frak{g}\right)  =8$

\item  Conditions on invariants:
\[
\frac{\partial F}{\partial x_{9}}=0.
\]

\item  Codimension one subalgebra: $L\left\langle X_{1},..,X_{8}\right\rangle
\simeq L_{8,2}$

\item  Invariants of subalgebra:%
\begin{align*}
J_{1} &  =x_{8},\\
J_{2} &  =\sqrt{D},
\end{align*}
where
\[
D=\left|
\begin{array}
[c]{rrrrrrrr}%
0 & x_{3} & -x_{2} & \frac{x_{7}}{2} & \frac{x_{6}}{2} & -\frac{x_{5}}{2} &
-\frac{x_{4}}{2} & x_{1}\\
-x_{3} & 0 & x_{1} & \frac{x_{5}}{2} & -\frac{x_{4}}{2} & \frac{x_{7}}{2} &
-\frac{x_{6}}{2} & x_{2}\\
x_{2} & -x_{1} & 0 & \frac{x_{6}}{2} & -\frac{x_{7}}{2} & -\frac{x_{4}}{2} &
\frac{x_{5}}{2} & x_{3}\\
-\frac{x_{7}}{2} & -\frac{x_{5}}{2} & -\frac{x_{6}}{2} & 0 & 0 & x_{8} & 0 &
\frac{x_{4}}{2}\\
-\frac{x_{4}}{2} & \frac{x_{4}}{2} & \frac{x_{7}}{2} & 0 & 0 & 0 & x_{8} &
\frac{x_{5}}{2}\\
\frac{x_{5}}{2} & -\frac{x_{7}}{2} & \frac{x_{4}}{2} & -x_{8} & 0 & 0 & 0 &
\frac{x_{6}}{2}\\
\frac{x_{4}}{2} & \frac{x_{6}}{2} & -\frac{x_{5}}{2} & 0 & -x_{8} & 0 & 0 &
\frac{x_{7}}{2}\\
-x_{1} & -x_{2} & -x_{3} & -\frac{x_{4}}{2} & -\frac{x_{5}}{2} & -\frac{x_{6}%
}{2} & -\frac{x_{7}}{2} & 0
\end{array}
\right|
\]

\item  Semi-invariance conditions:%
\[
\widehat{X}_{9}\left(  J_{1}\right)  =-2pJ_{1},\;\widehat{X}_{9}\left(
J_{2}\right)  =-4pJ_{2}.
\]

\item  Invariants of $L_{9,7}^{p}$%
\[
I_{1}=\frac{J_{2}}{J_{1}^{2}}.
\]
\end{itemize}

\newpage

$\mathbf{L}_{9,7}^{0}\;$

\begin{itemize}
\item  Levi decomposition: $L_{9,7}^{0}=\frak{so}\left(  3\right)
\overrightarrow{\oplus}_{R}\frak{g}_{6,82}^{\ast}$

\item  Describing representation: $R=R_{4}\oplus2D_{0}$

\item  Structure tensor:%
\[%
\begin{array}
[c]{llllll}%
C_{12}^{3}=1, & C_{13}^{2}=-1, & C_{23}^{1}=1, & C_{14}^{7}=\frac{1}{2}, &
C_{15}^{6}=\frac{1}{2}, & C_{16}^{5}=-\frac{1}{2},\\
C_{17}^{4}=-\frac{1}{2}, & C_{24}^{5}=\frac{1}{2}, & C_{25}^{4}=-\frac{1}%
{2}, & C_{26}^{7}=\frac{1}{2}, & C_{27}^{6}=-\frac{1}{2}, & C_{34}^{6}%
=\frac{1}{2},\\
C_{35}^{7}=-\frac{1}{2}, & C_{36}^{4}=-\frac{1}{2}, & C_{37}^{5}=\frac{1}%
{2}, & C_{46}^{8}=1, & C_{49}^{6}=1, & C_{57}^{8}=1,\\
C_{59}^{7}=1, & C_{69}^{4}=-1, & C_{79}^{5}=-1. &  &  & \\
&  &  &  &  &
\end{array}
\]

\item $\mathrm{codim}_{\frak{g}}\left[  \frak{g},\frak{g}\right]  =1.$

\item  Rank $A\left(  \frak{g}\right)  =6$

\item  Codimension one subalgebra: $L\left\langle X_{1},..,X_{8}\right\rangle
\simeq L_{8,2}$

\item  Invariants of subalgebra:%
\begin{align*}
J_{1} &  =x_{8},\\
J_{2} &  =\sqrt{D},
\end{align*}
where
\[
D=\left|
\begin{array}
[c]{rrrrrrrr}%
0 & x_{3} & -x_{2} & \frac{x_{7}}{2} & \frac{x_{6}}{2} & -\frac{x_{5}}{2} &
-\frac{x_{4}}{2} & x_{1}\\
-x_{3} & 0 & x_{1} & \frac{x_{5}}{2} & -\frac{x_{4}}{2} & \frac{x_{7}}{2} &
-\frac{x_{6}}{2} & x_{2}\\
x_{2} & -x_{1} & 0 & \frac{x_{6}}{2} & -\frac{x_{7}}{2} & -\frac{x_{4}}{2} &
\frac{x_{5}}{2} & x_{3}\\
-\frac{x_{7}}{2} & -\frac{x_{5}}{2} & -\frac{x_{6}}{2} & 0 & 0 & x_{8} & 0 &
\frac{x_{4}}{2}\\
-\frac{x_{4}}{2} & \frac{x_{4}}{2} & \frac{x_{7}}{2} & 0 & 0 & 0 & x_{8} &
\frac{x_{5}}{2}\\
\frac{x_{5}}{2} & -\frac{x_{7}}{2} & \frac{x_{4}}{2} & -x_{8} & 0 & 0 & 0 &
\frac{x_{6}}{2}\\
\frac{x_{4}}{2} & \frac{x_{6}}{2} & -\frac{x_{5}}{2} & 0 & -x_{8} & 0 & 0 &
\frac{x_{7}}{2}\\
-x_{1} & -x_{2} & -x_{3} & -\frac{x_{4}}{2} & -\frac{x_{5}}{2} & -\frac{x_{6}%
}{2} & -\frac{x_{7}}{2} & 0
\end{array}
\right|
\]

\item  Semi-invariance conditions:%
\[
\widehat{X}_{9}\left(  J_{1}\right)  =0,\;\widehat{X}_{9}\left(  J_{2}\right)
=0.
\]

\item  Invariants of $L_{9,7}^{0}$%
\begin{align*}
I_{1} &  =J_{1},\;I_{2}=J_{2},\\
I_{3} &  =x_{4}^{5}+x_{5}^{2}+x_{6}^{2}+x_{7}^{2}-2x_{8}x_{9}.
\end{align*}
\end{itemize}

\newpage

$\mathbf{L}_{9,7}^{\ast}$

\begin{itemize}
\item  Levi decomposition: $L_{9,7}^{\ast}=\frak{so}\left(  3\right)
\overrightarrow{\oplus}_{R}\frak{g}_{6,82}^{2,0,0}$

\item  Describing representation: $R=R_{4}\oplus2D_{0}$

\item  Structure tensor:%
\[%
\begin{array}
[c]{llllll}%
C_{12}^{3}=1, & C_{13}^{2}=-1, & C_{23}^{1}=1, & C_{14}^{7}=\frac{1}{2}, &
C_{15}^{6}=\frac{1}{2}, & C_{16}^{5}=-\frac{1}{2},\\
C_{17}^{4}=-\frac{1}{2}, & C_{24}^{5}=\frac{1}{2}, & C_{25}^{4}=-\frac{1}%
{2}, & C_{26}^{7}=\frac{1}{2}, & C_{27}^{6}=-\frac{1}{2}, & C_{34}^{6}%
=\frac{1}{2},\\
C_{35}^{7}=-\frac{1}{2}, & C_{36}^{4}=-\frac{1}{2}, & C_{37}^{5}=\frac{1}%
{2}, & C_{46}^{8}=1, & C_{57}^{8}=1, & C_{49}^{4}=1,\\
C_{59}^{5}=1, & C_{69}^{6}=1, & C_{79}^{7}=1, & C_{89}^{8}=2. &  & \\
&  &  &  &  &
\end{array}
\]

\item  Rank $A\left(  \frak{g}\right)  =8$

\item  Conditions on invariants:
\[
\frac{\partial F}{\partial x_{9}}=0.
\]

\item  Codimension one subalgebra: $L\left\langle X_{1},..,X_{8}\right\rangle
\simeq L_{8,2}$

\item  Invariants of subalgebra:%
\begin{align*}
J_{1} &  =x_{8}\\
J_{2} &  =\sqrt{D},
\end{align*}
where
\[
D=\left|
\begin{array}
[c]{rrrrrrrr}%
0 & x_{3} & -x_{2} & \frac{x_{7}}{2} & \frac{x_{6}}{2} & -\frac{x_{5}}{2} &
-\frac{x_{4}}{2} & x_{1}\\
-x_{3} & 0 & x_{1} & \frac{x_{5}}{2} & -\frac{x_{4}}{2} & \frac{x_{7}}{2} &
-\frac{x_{6}}{2} & x_{2}\\
x_{2} & -x_{1} & 0 & \frac{x_{6}}{2} & -\frac{x_{7}}{2} & -\frac{x_{4}}{2} &
\frac{x_{5}}{2} & x_{3}\\
-\frac{x_{7}}{2} & -\frac{x_{5}}{2} & -\frac{x_{6}}{2} & 0 & 0 & x_{8} & 0 &
\frac{x_{4}}{2}\\
-\frac{x_{4}}{2} & \frac{x_{4}}{2} & \frac{x_{7}}{2} & 0 & 0 & 0 & x_{8} &
\frac{x_{5}}{2}\\
\frac{x_{5}}{2} & -\frac{x_{7}}{2} & \frac{x_{4}}{2} & -x_{8} & 0 & 0 & 0 &
\frac{x_{6}}{2}\\
\frac{x_{4}}{2} & \frac{x_{6}}{2} & -\frac{x_{5}}{2} & 0 & -x_{8} & 0 & 0 &
\frac{x_{7}}{2}\\
-x_{1} & -x_{2} & -x_{3} & -\frac{x_{4}}{2} & -\frac{x_{5}}{2} & -\frac{x_{6}%
}{2} & -\frac{x_{7}}{2} & 0
\end{array}
\right|
\]

\item  Semi-invariance conditions:%
\[
\widehat{X}_{9}\left(  J_{1}\right)  =-2J_{1},\;\widehat{X}_{9}\left(
J_{2}\right)  =-4J_{2}.
\]

\item  Invariants of $L_{9,7}^{\ast}$%
\[
I_{1}=\frac{J_{2}}{J_{1}^{2}}.
\]
\end{itemize}

\begin{remark}
The notation for this algebra corresponds to its natural
localization within the the classification in \cite{Tu3}.
\end{remark}

\newpage

$\mathbf{L}_{9,8}$

\begin{itemize}
\item  Levi decomposition: $L_{9,8}=\frak{so}\left(  3\right)  \overrightarrow
{\oplus}_{R}\mathcal{N}_{6,18}^{0,1,1}$

\item  Describing representation: $R=R_{4}\oplus2D_{0}$

\item  Structure tensor:%
\[%
\begin{array}
[c]{llllll}%
C_{12}^{3}=1, & C_{13}^{2}=-1, & C_{23}^{1}=1, & C_{14}^{7}=\frac{1}{2}, &
C_{15}^{6}=\frac{1}{2}, & C_{16}^{5}=-\frac{1}{2},\\
C_{17}^{4}=-\frac{1}{2}, & C_{24}^{5}=\frac{1}{2}, & C_{25}^{4}=-\frac{1}%
{2}, & C_{26}^{7}=\frac{1}{2}, & C_{27}^{6}=-\frac{1}{2}, & C_{34}^{6}%
=\frac{1}{2},\\
C_{35}^{7}=-\frac{1}{2}, & C_{36}^{4}=-\frac{1}{2}, & C_{37}^{5}=\frac{1}%
{2}, & C_{48}^{4}=1, & C_{49}^{6}=1, & C_{58}^{5}=1,\\
C_{59}^{7}=1, & C_{68}^{6}=1, & C_{69}^{4}=-1, & C_{78}^{7}=1, & C_{79}%
^{5}=-1. & \\
&  &  &  &  &
\end{array}
\]

\item $\mathrm{codim}_{\frak{g}}\left[  \frak{g},\frak{g}\right]  =1.$

\item  Rank $A\left(  \frak{g}\right)  =8$

\item  Condition on invariants.%
\[
\frac{\partial F}{\partial x_{8}}=0.
\]

\item  Codimension one subalgebra: $L\left\langle X_{1},..,X_{8}\right\rangle
\simeq L_{8,4}^{0}$

\item  Invariants of subalgebra:%
\begin{align*}
J_{1} &  =x_{6}^{2}x_{9}+2x_{3}x_{6}^{2}+4x_{2}x_{5}x_{6}+4x_{1}x_{6}%
x_{7}-4x_{2}x_{4}x_{7}+x_{5}^{2}x_{9}\\
&  +2x_{3}x_{4}^{2}+x_{4}^{2}x_{9}+4x_{1}x_{4}x_{5}+x_{7}^{2}x_{9}-2x_{3}%
x_{5}^{2}-2x_{3}x_{7}^{2}\\
J_{2} &  =x_{4}^{2}+x_{5}^{2}+x_{6}^{2}+x_{7}^{2}%
\end{align*}

\item  Semi-invariance conditions:%
\[
\widehat{X}_{9}\left(  J_{1}\right)  =-2J_{1},\;\widehat{X}_{9}\left(
J_{2}\right)  =-2J_{2}.
\]

\item  Invariants of $L_{9,8}$%
\[
I_{1}=\frac{J_{1}}{J_{2}}.
\]
\newpage
\end{itemize}

$\mathbf{L}_{9,9}$

\begin{itemize}
\item  Levi decomposition: $L_{9,9}=\frak{so}\left(  3\right)  \overrightarrow
{\oplus}_{R}\frak{g}_{6,1}^{{}}$

\item  Describing representation: $R=R_{5}\oplus D_{0}$

\item  Structure tensor:%
\[%
\begin{array}
[c]{llllll}%
C_{12}^{3}=1, & C_{13}^{2}=-1, & C_{23}^{1}=1, & C_{14}^{7}=\frac{1}{2}, &
C_{15}^{6}=-\frac{1}{2}, & C_{16}^{5}=2,\\
C_{16}^{8}=-1, & C_{17}^{4}=-2, & C_{18}^{6}=3, & C_{24}^{6}=\frac{1}{2}, &
C_{25}^{7}=\frac{1}{2}, & C_{26}^{4}=-2,\\
C_{27}^{5}=-2, & C_{27}^{8}=-1, & C_{28}^{7}=3, & C_{34}^{58}=2, & C_{35}%
^{4}=-2, & C_{36}^{7}=1,\\
C_{37}^{6}=-1, & C_{49}^{4}=1, & C_{59}^{5}=1, & C_{69}^{6}=1, & C_{79}%
^{7}=1, & C_{89}^{8}=1.\\
&  &  &  &  &
\end{array}
\]

\item $\mathrm{codim}_{\frak{g}}\left[  \frak{g},\frak{g}\right]  =1.$

\item  Rank $A\left(  \frak{g}\right)  =8$

\item  Conditions on invariants:
\[
\frac{\partial F}{\partial x_{j}}=0,\;j=1,2,3,9.
\]

\item  Codimension one subalgebra: $L\left\langle X_{1},..,X_{8}\right\rangle
\simeq L_{8,5}$

\item  Invariants of subalgebra:%
\begin{align*}
J_{1} &  =12\left(  x_{4}^{2}+x_{5}^{2}\right)  +3\left(  x_{6}^{2}+x_{7}%
^{2}\right)  +x_{8}^{2}\\
J_{2} &  =\frac{2}{9}x_{8}^{3}+x_{6}^{2}x_{8}-12x_{4}x_{6}x_{7}-8x_{4}%
^{2}x_{8}+6x_{5}x_{6}^{2}\\
&  -6x_{5}x_{7}^{2}+x_{7}^{2}x_{8}-x_{5}^{2}x_{8}.
\end{align*}

\item  Semi-invariance conditions:%
\[
\widehat{X}_{9}\left(  J_{1}\right)  =-2J_{1},\;\widehat{X}_{9}\left(
J_{2}\right)  =-3J_{2}.
\]

\item  Invariants of $L_{9,9}^{{}}$%
\[
I_{1}=\frac{J_{1}^{3}}{J_{2}^{2}}.
\]
\end{itemize}

\newpage

$\mathbf{L}_{9,10}^{{}}$

\begin{itemize}
\item  Levi decomposition: $L_{9,10}^{{}}=\frak{so}\left(  3\right)
\overrightarrow{\oplus}_{R}6L_{1}$

\item  Describing representation: $R=2ad\frak{so}\left(  3\right)  $

\item  Structure tensor:%
\[%
\begin{array}
[c]{llllll}%
C_{12}^{3}=1, & C_{13}^{2}=-1, & C_{23}^{1}=1, & C_{15}^{6}=1, & C_{16}%
^{5}=-1, & C_{18}^{9}=1,\\
C_{19}^{8}-1, & C_{24}^{6}=-1, & C_{26}^{4}=1, & C_{27}^{9}=-1, & C_{29}%
^{7}=1, & C_{34}^{5}=1,\\
C_{35}^{4}=-1, & C_{37}^{8}=1, & C_{38}^{7}=-1. &  &  & \\
&  &  &  &  & \\
&  &  &  &  &
\end{array}
\]

\item $\mathrm{codim}_{\frak{g}}\left[  \frak{g},\frak{g}\right]  =0.$

\item  Rank $A\left(  \frak{g}\right)  =6$

\item  Conditions on invariants:
\[
\frac{\partial F}{\partial x_{j}}=0,\;j=1,2,3.
\]

\item  Conditions on invariants:
\[
\frac{\partial F}{\partial x_{i}}=0,\;i=1,2,3
\]

\item  Invariants of $L_{9,10}^{{}}$%
\begin{align*}
I_{1}  &  =x_{4}^{2}+x_{5}^{2}+x_{6}^{2},\\
I_{2}  &  =x_{7}^{2}+x_{8}^{2}+x_{9}^{2},\\
I_{3}  &  =x_{4}x_{7}+x_{5}x_{8}+x_{6}x_{9}.
\end{align*}
\end{itemize}

\newpage

$\mathbf{L}_{9,11}^{{}}$

\begin{itemize}
\item  Levi decomposition: $L_{9,11}^{{}}=\frak{so}\left(  3\right)
\overrightarrow{\oplus}_{R}\mathcal{A}_{6,3}^{{}}$

\item  Describing representation: $R=2ad\frak{so}\left(  3\right)  $

\item  Structure tensor:%
\[%
\begin{array}
[c]{llllll}%
C_{12}^{3}=1, & C_{13}^{2}=-1, & C_{23}^{1}=1, & C_{15}^{6}=1, & C_{16}%
^{5}=-1, & C_{18}^{9}=1,\\
C_{19}^{8}-1, & C_{24}^{6}=-1, & C_{26}^{4}=1, & C_{27}^{9}=-1, & C_{29}%
^{7}=1, & C_{34}^{5}=1,\\
C_{35}^{4}=-1, & C_{37}^{8}=1, & C_{38}^{7}=-1, & C_{45}^{9}=1, & C_{46}%
^{8}=-1, & C_{56}^{7}=1.\\
&  &  &  &  &
\end{array}
\]

\item $\mathrm{codim}_{\frak{g}}\left[  \frak{g},\frak{g}\right]  =0.$

\item  Rank $A\left(  \frak{g}\right)  =6$

\item  Invariants of $L_{9,11}^{{}}$%
\begin{align*}
I_{1}  &  =x_{7}^{2}+x_{8}^{2}+x_{9}^{2},\\
I_{2}  &  =x_{4}x_{7}+x_{5}x_{8}+x_{6}x_{9},\\
I_{3}  &  =x_{1}x_{7}+x_{2}x_{8}+x_{3}x_{9}+\frac{1}{2}\left(  x_{4}^{2}%
+x_{5}^{2}+x_{6}^{2}\right)  .
\end{align*}
\end{itemize}

\newpage

$\mathbf{L}_{9,12}$

\begin{itemize}
\item  Levi decomposition: $L_{9,12}=\frak{sl}\left(  2,\mathbb{R}\right)
\overrightarrow{\oplus}_{R}\mathcal{A}_{6,12}$

\item  Describing representation: $R=D_{\frac{1}{2}}\oplus4D_{0}$

\item  Structure tensor:%
\[%
\begin{array}
[c]{llllll}%
C_{12}^{2}=2, & C_{13}^{3}=-2, & C_{23}^{1}=1, & C_{14}^{4}=1, & C_{15}%
^{5}=-1, & C_{25}^{4}=1,\\
C_{34}^{5}=1, & C_{45}^{6}=1, & C_{79}^{6}=1, & C_{89}^{7}=1. &  & \\
&  &  &  &  & \\
&  &  &  &  & \\
&  &  &  &  &
\end{array}
\]

\item $\mathrm{codim}_{\frak{g}}\left[  \frak{g},\frak{g}\right]  =2.$

\item  Rank $A\left(  \frak{g}\right)  =6$

\item  Conditions on invariants:
\[
\frac{\partial F}{\partial x_{9}}=0,\;
\]

\item  Codimension one subalgebra:%
\[
L\left\langle X_{1},..,X_{8}\right\rangle \simeq L_{6,2}\oplus2L_{1}.
\]

\item  Invariants of subalgebra:%
\begin{align*}
J_{1}  &  =4x_{2}x_{3}x_{6}+2x_{2}x_{5}^{2}+2x_{1}x_{4}x_{5}-2x_{3}x_{4}%
^{2}+x_{1}x_{6}^{2},\\
J_{2}  &  =x_{6},\\
J_{3}  &  =x_{7},\\
J_{4}  &  =x_{8}.
\end{align*}

\item  Semi-invariance conditions:%
\begin{align*}
\widehat{X}_{9}\left(  J_{1}\right)   &  =0,\;\;\widehat{X}_{9}\left(
J_{2}\right)  =0,\\
\widehat{X}_{9}\left(  J_{3}\right)   &  =-J_{2},\;\widehat{X}_{9}\left(
J_{4}\right)  =-J_{3}.
\end{align*}

\item  Invariants of $L_{9,12}$%
\begin{align*}
I_{1}  &  =2x_{6}x_{8}-x_{7}^{2},\\
I_{2}  &  =4x_{2}x_{3}x_{6}+2x_{2}x_{5}^{2}+2x_{1}x_{4}x_{5}-2x_{3}x_{4}%
^{2}+x_{1}x_{6}^{2}\\
I_{3}  &  =x_{6}.
\end{align*}
\end{itemize}

\newpage

$\mathbf{L}_{9,13}^{p,q,r}\;[pqr\neq0]$

\begin{itemize}
\item  Levi decomposition: $L_{9,13}^{p,q,r}=\frak{sl}\left(  2,\mathbb{R}%
\right)  \overrightarrow{\oplus}_{R}\frak{g}_{6,1}$

\item  Describing representation: $R=D_{\frac{1}{2}}\oplus4D_{0}$

\item  Structure tensor:%
\[%
\begin{array}
[c]{llllll}%
C_{12}^{2}=2, & C_{13}^{3}=-2, & C_{23}^{1}=1, & C_{14}^{4}=1, & C_{15}%
^{5}=-1, & C_{25}^{4}=1,\\
C_{34}^{5}=1, & C_{49}^{4}=1, & C_{59}^{5}=1, & C_{69}^{6}=p, & C_{79}%
^{7}=q, & C_{89}^{8}=r.\\
&  &  &  &  & \\
&  &  &  &  & \\
&  &  &  &  &
\end{array}
\]

\item $\mathrm{codim}_{\frak{g}}\left[  \frak{g},\frak{g}\right]  =1.$

\item  Rank $A\left(  \frak{g}\right)  =6$

\item  Conditions on invariants:
\[
\frac{\partial F}{\partial x_{9}}=0,
\]

\item  Codimension one subalgebra:%
\[
L\left\langle X_{1},..,X_{8}\right\rangle \simeq\frak{sa}\left(
2,\mathbb{R}\right)  \oplus2L_{1}%
\]

\item  Invariants of subalgebra:%
\begin{align*}
J_{1}  &  =x_{3}x_{4}^{2}-x_{1}x_{4}x_{5}-x_{2}x_{5}^{2},\\
J_{2}  &  =x_{6},\\
J_{3}  &  =x_{7},\\
J_{4}  &  =x_{8}.
\end{align*}

\item  Semi-invariance conditions:%
\begin{align*}
\widehat{X}_{9}\left(  J_{1}\right)   &  =-2J_{1},\;\widehat{X}_{9}\left(
J_{2}\right)  =-pJ_{2},\\
\widehat{X}_{9}\left(  J_{3}\right)   &  =-qJ_{3},\;\widehat{X}_{9}\left(
J_{4}\right)  =-rJ_{4}.
\end{align*}

\item  Invariants of $L_{9,13}^{p,q,r}$%
\[
I_{1}=\frac{J_{1}^{p}}{J_{2}^{2}},\;I_{2}=\frac{J_{1}^{q}}{J_{3}^{2}}%
,\;I_{3}=\frac{J_{1}^{r}}{J_{4}^{2}}.
\]
\end{itemize}

\newpage

$\mathbf{L}_{9,14}^{p,q}\;[p\neq0]$

\begin{itemize}
\item  Levi decomposition: $L_{9,14}^{p,q}=\frak{sl}\left(  2,\mathbb{R}%
\right)  \overrightarrow{\oplus}_{R}\frak{g}_{6,2}$

\item  Describing representation: $R=D_{\frac{1}{2}}\oplus4D_{0}$

\item  Structure tensor:%
\[%
\begin{array}
[c]{llllll}%
C_{12}^{2}=2, & C_{13}^{3}=-2, & C_{23}^{1}=1, & C_{14}^{4}=1, & C_{15}%
^{5}=-1, & C_{25}^{4}=1,\\
C_{34}^{5}=1, & C_{49}^{4}=1, & C_{59}^{5}=1, & C_{69}^{6}=p, & C_{79}%
^{7}=q, & C_{89}^{7}=1,\\
C_{89}^{8}=q. &  &  &  &  & \\
&  &  &  &  & \\
&  &  &  &  &
\end{array}
\]

\item $\mathrm{codim}_{\frak{g}}\left[  \frak{g},\frak{g}\right]  =\left\{
\begin{tabular}
[c]{ll}%
$1,$ & $q\neq0$\\
$2,$ & $q=0$%
\end{tabular}
\right.  .$

\item  Rank $A\left(  \frak{g}\right)  =6$

\item  Conditions on invariants:
\[
\frac{\partial F}{\partial x_{9}}=0,
\]

\item  Codimension one subalgebra:%
\[
L\left\langle X_{1},..,X_{8}\right\rangle \simeq\frak{sa}\left(
2,\mathbb{R}\right)  \oplus2L_{1}%
\]

\item  Invariants of subalgebra:%
\begin{align*}
J_{1}  &  =x_{3}x_{4}^{2}-x_{1}x_{4}x_{5}-x_{2}x_{5}^{2},\\
J_{2}  &  =x_{6},\\
J_{3}  &  =x_{7},\\
J_{4}  &  =x_{8}.
\end{align*}

\item  Semi-invariance conditions:%
\begin{align*}
\widehat{X}_{9}\left(  J_{1}\right)   &  =-2J_{1},\;\widehat{X}_{9}\left(
J_{2}\right)  =-pJ_{2},\\
\widehat{X}_{9}\left(  J_{3}\right)   &  =-qJ_{3},\;\widehat{X}_{9}\left(
J_{4}\right)  =-J_{3}-qJ_{4}.
\end{align*}

\item  Invariants of $L_{9,14}^{p,q}$%
\begin{align*}
I_{1}  &  =\frac{J_{1}^{p}}{J_{2}^{2}},\;I_{2}=\frac{J_{1}^{q}}{J_{3}^{2}%
},\;I_{3}=\frac{J_{4}^{{}}}{J_{3}}-\frac{1}{q}\ln J_{3},\;\left(
q\neq0\right)  ,\\
I_{1}  &  =\frac{J_{1}^{p}}{J_{2}^{2}},\;I_{2}=J_{3},\;I_{3}=\ln\left(
J_{2}J_{3}\right)  -\frac{J_{4}}{J_{3}},\;\left(  q=0\right)  .
\end{align*}
\end{itemize}

\newpage

$\mathbf{L}_{9,15}^{p}$

\begin{itemize}
\item  Levi decomposition: $L_{9,14}^{p,q}=\frak{sl}\left(  2,\mathbb{R}%
\right)  \overrightarrow{\oplus}_{R}\frak{g}_{6,3}$

\item  Describing representation: $R=D_{\frac{1}{2}}\oplus4D_{0}$

\item  Structure tensor:%
\[%
\begin{array}
[c]{llllll}%
C_{12}^{2}=2, & C_{13}^{3}=-2, & C_{23}^{1}=1, & C_{14}^{4}=1, & C_{15}%
^{5}=-1, & C_{25}^{4}=1,\\
C_{34}^{5}=1, & C_{49}^{4}=1, & C_{59}^{5}=1, & C_{69}^{6}=p, & C_{79}%
^{6}=1, & C_{79}^{7}=p,\\
C_{89}^{7}=1, & C_{89}^{8}=p. &  &  &  & \\
&  &  &  &  & \\
&  &  &  &  &
\end{array}
\]

\item $\mathrm{codim}_{\frak{g}}\left[  \frak{g},\frak{g}\right]  =\left\{
\begin{tabular}
[c]{ll}%
$1,$ & $p\neq0$\\
$2,$ & $p=0$%
\end{tabular}
\right.  .$

\item  Rank $A\left(  \frak{g}\right)  =6$

\item  Conditions on invariants:
\[
\frac{\partial F}{\partial x_{9}}=0,
\]

\item  Codimension one subalgebra:%
\[
L\left\langle X_{1},..,X_{8}\right\rangle \simeq\frak{sa}\left(
2,\mathbb{R}\right)  \oplus2L_{1}%
\]

\item  Invariants of subalgebra:%
\begin{align*}
J_{1}  &  =x_{3}x_{4}^{2}-x_{1}x_{4}x_{5}-x_{2}x_{5}^{2},\\
J_{2}  &  =x_{6},\\
J_{3}  &  =x_{7},\\
J_{4}  &  =x_{8}.
\end{align*}

\item  Semi-invariance conditions:%
\begin{align*}
\widehat{X}_{9}\left(  J_{1}\right)   &  =-2J_{1},\;\widehat{X}_{9}\left(
J_{2}\right)  =-pJ_{2},\\
\widehat{X}_{9}\left(  J_{3}\right)   &  =-J_{2}-pJ_{3},\;\widehat{X}%
_{9}\left(  J_{4}\right)  =-J_{3}-pJ_{4}.
\end{align*}

\item  Invariants of $L_{9,15}^{p}$%
\begin{align*}
I_{1}  &  =\frac{J_{1}^{p}}{J_{2}^{2}},\;I_{2}=\frac{2J_{4}^{{}}}{J_{2}^{{}}%
}-\left(  \frac{J_{3}}{J_{2}}\right)  ^{2},\;I_{3}=2\frac{J_{4}}{J_{2}}%
+\frac{1}{p^{2}}\ln^{2}J_{2},\;\left(  p\neq0\right)  ,\\
I_{1}  &  =J_{2},\;I_{2}=\frac{2}{J_{2}J_{4}}-\left(  \frac{J_{3}}{J_{2}J_{4}%
}\right)  ^{2},\;I_{3}=\frac{2J_{3}}{J_{2}}+\ln\left(  \frac{J_{2}}{J_{1}%
}\right)  ,\;\left(  p=0\right)  .
\end{align*}
\end{itemize}

\newpage

$\mathbf{L}_{9,16}^{p,q,r}\;[pq\neq0,\;r\geq0]$

\begin{itemize}
\item  Levi decomposition: $L_{9,16}^{p,q,r}=\frak{sl}\left(  2,\mathbb{R}%
\right)  \overrightarrow{\oplus}_{R}\frak{g}_{6,8}$

\item  Describing representation: $R=D_{\frac{1}{2}}\oplus4D_{0}$

\item  Structure tensor:%
\[%
\begin{array}
[c]{llllll}%
C_{12}^{2}=2, & C_{13}^{3}=-2, & C_{23}^{1}=1, & C_{14}^{4}=1, & C_{15}%
^{5}=-1, & C_{25}^{4}=1,\\
C_{34}^{5}=1, & C_{49}^{4}=p, & C_{59}^{5}=p, & C_{69}^{6}=q, & C_{79}%
^{7}=r, & C_{79}^{8}=-1,\\
C_{89}^{7}=1, & C_{89}^{8}=r. &  &  &  & \\
&  &  &  &  & \\
&  &  &  &  &
\end{array}
\]

\item $\mathrm{codim}_{\frak{g}}\left[  \frak{g},\frak{g}\right]  =1.$

\item  Rank $A\left(  \frak{g}\right)  =6$

\item  Conditions on invariants:
\[
\frac{\partial F}{\partial x_{9}}=0,
\]

\item  Codimension one subalgebra:%
\[
L\left\langle X_{1},..,X_{8}\right\rangle \simeq\frak{sa}\left(
2,\mathbb{R}\right)  \oplus2L_{1}%
\]

\item  Invariants of subalgebra:%
\begin{align*}
J_{1}  &  =x_{3}x_{4}^{2}-x_{1}x_{4}x_{5}-x_{2}x_{5}^{2},\\
J_{2}  &  =x_{6},\\
J_{3}  &  =x_{7},\\
J_{4}  &  =x_{8}.
\end{align*}

\item  Semi-invariance conditions:%
\begin{align*}
\widehat{X}_{9}\left(  J_{1}\right)   &  =-2pJ_{1},\;\widehat{X}_{9}\left(
J_{2}\right)  =-qJ_{2},\\
\widehat{X}_{9}\left(  J_{3}\right)   &  =-rJ_{3}+J_{4},\;\widehat{X}%
_{9}\left(  J_{4}\right)  =-J_{3}-rJ_{4}.
\end{align*}

\item  Invariants of $L_{9,16}^{p,q,r}$%
\[
I_{1}=\frac{J_{1}^{p}}{J_{2}^{2}},\;I_{2}=\left(  J_{3}^{2}+J_{4}^{2}\right)
\left(  \frac{J_{3}+iJ_{4}}{J_{3}-iJ_{4}}\right)  ^{ri},\;I_{3}=J_{2}%
\exp\left(  q\arctan\left(  \frac{J_{3}}{J_{4}}\right)  \right)  .
\]
\end{itemize}

\newpage

$\mathbf{L}_{9,17}^{p,q}\;[pq\neq0]$

\begin{itemize}
\item  Levi decomposition: $L_{9,17}^{p,q}=\frak{sl}\left(  2,\mathbb{R}%
\right)  \overrightarrow{\oplus}_{R}\frak{g}_{6,13}$

\item  Describing representation: $R=D_{\frac{1}{2}}\oplus4D_{0}$

\item  Structure tensor:%
\[%
\begin{array}
[c]{llllll}%
C_{12}^{2}=2, & C_{13}^{3}=-2, & C_{23}^{1}=1, & C_{14}^{4}=1, & C_{15}%
^{5}=-1, & C_{25}^{4}=1,\\
C_{34}^{5}=1, & C_{45}^{6}=1, & C_{49}^{4}=1, & C_{59}^{5}=p, & C_{69}%
^{6}=2, & C_{79}^{7}=p,\\
C_{89}^{8}=q. &  &  &  &  & \\
&  &  &  &  & \\
&  &  &  &  &
\end{array}
\]

\item $\mathrm{codim}_{\frak{g}}\left[  \frak{g},\frak{g}\right]  =1.$

\item  Rank $A\left(  \frak{g}\right)  =6$

\item  Conditions on invariants:
\[
\frac{\partial F}{\partial x_{9}}=0,
\]

\item  Codimension one subalgebra:%
\[
L\left\langle X_{1},..,X_{8}\right\rangle \simeq L_{6,2}\oplus2L_{1}.
\]

\item  Invariants of subalgebras:%
\begin{align*}
J_{1}  &  =2x_{2}x_{3}x_{6}+x_{2}x_{5}^{2}+x_{1}x_{4}x_{5}-x_{3}x_{4}%
^{2}+\frac{1}{2}x_{1}^{2}x_{6},\\
J_{2}  &  =x_{6},\\
J_{3}  &  =x_{7},\\
J_{4}  &  =x_{8}.
\end{align*}

\item  Semi-invariance conditions:%
\begin{align*}
\widehat{X}_{9}\left(  J_{1}\right)   &  =-2J_{1},\;\widehat{X}_{9}\left(
J_{2}\right)  =-2J_{2},\\
\widehat{X}_{9}\left(  J_{3}\right)   &  =-pJ_{3},\;\widehat{X}_{9}\left(
J_{4}\right)  =-qJ_{4}.
\end{align*}

\item  Invariants of $L_{9,17}^{p,q}$%
\[
I_{1}=\frac{J_{1}^{{}}}{J_{2}^{{}}},\;I_{2}=\frac{J_{3}^{2}}{J_{2}^{p}%
},\;I_{3}=\frac{J_{4}^{2}}{J_{3}^{q}}.
\]
\end{itemize}

\newpage

$\mathbf{L}_{9,18}^{p,q}\;[p^{2}+q^{2}\neq0]$

\begin{itemize}
\item  Levi decomposition: $L_{9,18}^{p,q}=\frak{sl}\left(  2,\mathbb{R}%
\right)  \overrightarrow{\oplus}_{R}\frak{g}_{6,13}$

\item  Describing representation: $R=D_{\frac{1}{2}}\oplus4D_{0}$

\item  Structure tensor:%
\[%
\begin{array}
[c]{llllll}%
C_{12}^{2}=2, & C_{13}^{3}=-2, & C_{23}^{1}=1, & C_{14}^{4}=1, & C_{15}%
^{5}=-1, & C_{25}^{4}=1,\\
C_{34}^{5}=1, & C_{49}^{4}=1, & C_{59}^{5}=1, & C_{69}^{6}=p+q, & C_{78}%
^{6}=1, & C_{79}^{7}=p,\\
C_{89}^{8}=q. &  &  &  &  & \\
&  &  &  &  & \\
&  &  &  &  &
\end{array}
\]

\item $\mathrm{codim}_{\frak{g}}\left[  \frak{g},\frak{g}\right]  =\left\{
\begin{tabular}
[c]{ll}%
$1,$ & $pq\neq0$\\
$2,$ & $p=0$ or $q=0$%
\end{tabular}
\right.  .$

\item  Rank $A\left(  \frak{g}\right)  =8$

\item  Conditions on invariants:
\[
\frac{\partial F}{\partial x_{i}}=0,\;i=7,8,9.
\]

\item  Codimension three subalgebra:%
\[
L\left\langle X_{1},..,X_{6}\right\rangle \simeq L_{5,2}\oplus L_{1}%
\]

\item  Invariants of subalgebra:%
\begin{align*}
J_{1}  &  =x_{3}x_{4}^{2}-x_{1}x_{4}x_{5}-x_{2}x_{5}^{2},\\
J_{2}  &  =x_{6},
\end{align*}

\item  Semi-invariance conditions:%
\[
\widehat{X}_{9}\left(  J_{1}\right)  =-2J_{1},\;\widehat{X}_{9}\left(
J_{2}\right)  =-\left(  p+q\right)  J_{2}.
\]

\item  Invariants of $L_{9,18}^{p,q}$%
\begin{align*}
I_{1}  &  =\frac{J_{1}^{p+q}}{J_{2}^{2}},\;p+q\neq0,\\
I_{1}  &  =J_{2},\;p+q=0.
\end{align*}
\end{itemize}

\newpage

$\mathbf{L}_{9,19}^{p}$

\begin{itemize}
\item  Levi decomposition: $L_{9,19}^{p}=\frak{sl}\left(  2,\mathbb{R}\right)
\overrightarrow{\oplus}_{R}\frak{g}_{6,14}$

\item  Describing representation: $R=D_{\frac{1}{2}}\oplus4D_{0}$

\item  Structure tensor:%
\[%
\begin{array}
[c]{llllll}%
C_{12}^{2}=2, & C_{13}^{3}=-2, & C_{23}^{1}=1, & C_{14}^{4}=1, & C_{15}%
^{5}=-1, & C_{25}^{4}=1,\\
C_{34}^{5}=1, & C_{49}^{4}=p, & C_{59}^{5}=p, & C_{69}^{6}=2p, & C_{79}%
^{7}=1, & C_{89}^{6}=1,\\
C_{89}^{8}=2p, & C_{45}^{6}=1. &  &  &  & \\
&  &  &  &  & \\
&  &  &  &  &
\end{array}
\]

\item $\mathrm{codim}_{\frak{g}}\left[  \frak{g},\frak{g}\right]  =\left\{
\begin{tabular}
[c]{ll}%
$1,$ & $p\neq0$\\
$2,$ & $p=0$%
\end{tabular}
\right.  .$

\item  Rank $A\left(  \frak{g}\right)  =6$

\item  Conditions on invariants:
\[
\frac{\partial F}{\partial x_{9}}=0,
\]

\item  Codimension one subalgebra:%
\[
L\left\langle X_{1},..,X_{8}\right\rangle \simeq L_{6,2}\oplus2L_{1}%
\]

\item  Invariants of subalgebra:%
\begin{align*}
J_{1}  &  =2x_{2}x_{3}x_{6}+x_{2}x_{5}^{2}+x_{1}x_{4}x_{5}-x_{3}x_{4}%
^{2}+\frac{1}{2}x_{1}^{2}x_{6},\\
J_{2}  &  =x_{6},\\
J_{3}  &  =x_{7},\\
J_{4}  &  =x_{8}.
\end{align*}

\item  Semi-invariance conditions:%
\begin{align*}
\widehat{X}_{9}\left(  J_{1}\right)   &  =-2pJ_{1},\;\widehat{X}_{9}\left(
J_{2}\right)  =-2pJ_{2},\\
\widehat{X}_{9}\left(  J_{3}\right)   &  =-J_{3},\;\widehat{X}_{9}\left(
J_{4}\right)  =-J_{2}-pJ_{4}.
\end{align*}

\item  Invariants of $L_{9,19}^{p}$%
\begin{align*}
I_{1}  &  =\frac{J_{1}^{{}}}{J_{2}^{{}}},\;I_{2}=\frac{J_{3}^{2p}}{J_{2}^{{}}%
},\;I_{3}=-2p\frac{J_{4}^{{}}}{J_{2}}+\frac{1}{p}\ln J_{2},\;\left(
p\neq0\right)  ,\\
I_{1}  &  =J_{1},\;I_{2}=J_{2},\;I_{3}=J_{4}-J_{2}\ln J_{3},\;\left(
p=0\right)  .
\end{align*}
\end{itemize}

\newpage

$\mathbf{L}_{9,20}^{p}$

\begin{itemize}
\item  Levi decomposition: $L_{9,20}^{p}=\frak{sl}\left(  2,\mathbb{R}\right)
\overrightarrow{\oplus}_{R}\frak{g}_{6,21}$

\item  Describing representation: $R=D_{\frac{1}{2}}\oplus4D_{0}$

\item  Structure tensor:%
\[%
\begin{array}
[c]{llllll}%
C_{12}^{2}=2, & C_{13}^{3}=-2, & C_{23}^{1}=1, & C_{14}^{4}=1, & C_{15}%
^{5}=-1, & C_{25}^{4}=1,\\
C_{34}^{5}=1, & C_{49}^{4}=1, & C_{59}^{5}=1, & C_{69}^{6}=2p, & C_{78}%
^{6}=1, & C_{79}^{7}=p,\\
C_{89}^{7}=1, & C_{89}^{8}=p. &  &  &  & \\
&  &  &  &  & \\
&  &  &  &  &
\end{array}
\]

\item $\mathrm{codim}_{\frak{g}}\left[  \frak{g},\frak{g}\right]  =\left\{
\begin{tabular}
[c]{ll}%
$1,$ & $p\neq0$\\
$2,$ & $p=0$%
\end{tabular}
\right.  .$

\item  Rank $A\left(  \frak{g}\right)  =8$

\item  Conditions on invariants:
\[
\frac{\partial F}{\partial x_{i}}=0,\;i=7,8,9.
\]

\item  Codimension three subalgebra:%
\[
L\left\langle X_{1},..,X_{6}\right\rangle \simeq L_{5,1}\oplus L_{1}%
\]

\item  Invariants of subalgebra:%
\begin{align*}
J_{1}  &  =x_{3}x_{4}^{2}-x_{1}x_{4}x_{5}-x_{2}x_{5}^{2},\\
J_{2}  &  =x_{6}.
\end{align*}

\item  Semi-invariance conditions:%
\[
\widehat{X}_{9}\left(  J_{1}\right)  =-2J_{1},\;\widehat{X}_{9}\left(
J_{2}\right)  =-2pJ_{2}.
\]

\item  Invariants of $L_{9,20}^{p}$%
\begin{align*}
I_{1}  &  =\frac{J_{1}^{2p}}{J_{2}^{2}},\;\left(  p\neq0\right)  ,\\
I_{1}  &  =J_{2},\;\left(  p=0\right)  .
\end{align*}
\end{itemize}

\newpage

$\mathbf{L}_{9,21}^{p}$

\begin{itemize}
\item  Levi decomposition: $L_{9,21}^{p}=\frak{sl}\left(  2,\mathbb{R}\right)
\overrightarrow{\oplus}_{R}\frak{g}_{6,25}$

\item  Describing representation: $R=D_{\frac{1}{2}}\oplus4D_{0}$

\item  Structure tensor:%
\[%
\begin{array}
[c]{llllll}%
C_{12}^{2}=2, & C_{13}^{3}=-2, & C_{23}^{1}=1, & C_{14}^{4}=1, & C_{15}%
^{5}=-1, & C_{25}^{4}=1,\\
C_{34}^{5}=1, & C_{49}^{4}=1, & C_{59}^{5}=1, & C_{69}^{6}=2, & C_{79}%
^{7}=p, & C_{89}^{7}=1,\\
C_{89}^{8}=p, & C_{45}^{6}=1. &  &  &  & \\
&  &  &  &  & \\
&  &  &  &  &
\end{array}
\]

\item $\mathrm{codim}_{\frak{g}}\left[  \frak{g},\frak{g}\right]  =\left\{
\begin{tabular}
[c]{ll}%
$1,$ & $p\neq0$\\
$2,$ & $p=0$%
\end{tabular}
\right.  .$

\item  Rank $A\left(  \frak{g}\right)  =6$

\item  Conditions on invariants:
\[
\frac{\partial F}{\partial x_{9}}=0,
\]

\item  Codimension one subalgebra:%
\[
L\left\langle X_{1},..,X_{8}\right\rangle \simeq L_{6,2}\oplus2L_{1}%
\]

\item  Invariants of subalgebra:%
\begin{align*}
J_{1}  &  =2x_{2}x_{3}x_{6}+x_{2}x_{5}^{2}+x_{1}x_{4}x_{5}-x_{3}x_{4}%
^{2}+\frac{1}{2}x_{1}^{2}x_{6},\\
J_{2}  &  =x_{6},\\
J_{3}  &  =x_{7},\\
J_{4}  &  =x_{8}.
\end{align*}

\item  Semi-invariance conditions:%
\begin{align*}
\widehat{X}_{9}\left(  J_{1}\right)   &  =-2J_{1},\;\widehat{X}_{9}\left(
J_{2}\right)  =-2J_{2},\\
\widehat{X}_{9}\left(  J_{3}\right)   &  =-pJ_{3},\;\widehat{X}_{9}\left(
J_{4}\right)  =-J_{3}-pJ_{4}.
\end{align*}

\item  Invariants of $L_{9,21}^{p}$%
\[
I_{1}=\frac{J_{1}^{{}}}{J_{2}^{{}}},\;J_{2}=\frac{J_{3}^{2}}{J_{2}^{p}%
},\;J_{3}=\frac{\left(  -2J_{4}+J_{3}\ln\left(  J_{2}\right)  \right)  ^{2}%
}{J_{2}^{p}}.
\]
\end{itemize}

\newpage

$\mathbf{L}_{9,22}$

\begin{itemize}
\item  Levi decomposition: $L_{9,21}^{p}=\frak{sl}\left(  2,\mathbb{R}\right)
\overrightarrow{\oplus}_{R}\frak{g}_{6,26}$

\item  Describing representation: $R=D_{\frac{1}{2}}\oplus4D_{0}$

\item  Structure tensor:%
\[%
\begin{array}
[c]{llllll}%
C_{12}^{2}=2, & C_{13}^{3}=-2, & C_{23}^{1}=1, & C_{14}^{4}=1, & C_{15}%
^{5}=-1, & C_{25}^{4}=1,\\
C_{34}^{5}=1, & C_{45}^{6}=1, & C_{49}^{4}=1, & C_{59}^{5}=1, & C_{69}%
^{6}=2, & C_{79}^{6}=1,\\
C_{79}^{7}=2, & C_{89}^{7}=1, & C_{89}^{8}=2. &  &  & \\
&  &  &  &  & \\
&  &  &  &  &
\end{array}
\]

\item $\mathrm{codim}_{\frak{g}}\left[  \frak{g},\frak{g}\right]  =1.$

\item  Rank $A\left(  \frak{g}\right)  =6$

\item  Conditions on invariants:
\[
\frac{\partial F}{\partial x_{9}}=0,
\]

\item  Codimension one subalgebra:%
\[
L\left\langle X_{1},..,X_{8}\right\rangle \simeq L_{6,2}\oplus2L_{1}%
\]

\item  Invariants of subalgebra:%
\begin{align*}
J_{1}  &  =2x_{2}x_{3}x_{6}+x_{2}x_{5}^{2}+x_{1}x_{4}x_{5}-x_{3}x_{4}%
^{2}+\frac{1}{2}x_{1}^{2}x_{6},\\
J_{2}  &  =x_{6},\\
J_{3}  &  =x_{7},\\
J_{4}  &  =x_{8}.
\end{align*}

\item  Semi-invariance conditions:%
\begin{align*}
\widehat{X}_{9}\left(  J_{1}\right)   &  =-2J_{1},\;\widehat{X}_{9}\left(
J_{2}\right)  =-2J_{2},\\
\widehat{X}_{9}\left(  J_{3}\right)   &  =-J_{2}-2J_{3},\;\widehat{X}%
_{9}\left(  J_{4}\right)  =-J_{3}-2J_{4}.
\end{align*}

\item  Invariants of $L_{9,22}$%
\[
I_{1}=\frac{J_{1}^{{}}}{J_{2}^{{}}},\;I_{2}=\frac{-2J_{3}+J_{2}\ln\left(
J_{2}\right)  }{J_{2}},\;I_{3}=\frac{8J_{4}+J_{2}\ln\left(  J_{2}\right)
^{2}-4J_{3}\ln\left(  J_{2}\right)  }{J_{2}}.
\]
\end{itemize}

\newpage

$\mathbf{L}_{9,23}^{p,q}\;[p\neq0]$

\begin{itemize}
\item  Levi decomposition: $L_{9,23}^{p,q}=\frak{sl}\left(  2,\mathbb{R}%
\right)  \overrightarrow{\oplus}_{R}\frak{g}_{6,32}$

\item  Describing representation: $R=D_{\frac{1}{2}}\oplus4D_{0}$

\item  Structure tensor:%
\[%
\begin{array}
[c]{llllll}%
C_{12}^{2}=2, & C_{13}^{3}=-2, & C_{23}^{1}=1, & C_{14}^{4}=1, & C_{15}%
^{5}=-1, & C_{25}^{4}=1,\\
C_{34}^{5}=1, & C_{78}^{6}=1, & C_{49}^{4}=p, & C_{59}^{5}=p, & C_{69}%
^{6}=2q, & C_{79}^{7}=q,\\
C_{79}^{8}=-1, & C_{89}^{7}=1, & C_{89}^{8}=q. &  &  & \\
&  &  &  &  & \\
&  &  &  &  &
\end{array}
\]

\item $\mathrm{codim}_{\frak{g}}\left[  \frak{g},\frak{g}\right]  =1.$

\item  Rank $A\left(  \frak{g}\right)  =8$

\item  Conditions on invariants:
\[
\frac{\partial F}{\partial x_{j}}=0,\;j=7,8,9
\]

\item  Codimension one subalgebra:%
\[
L\left\langle X_{1},..,X_{8}\right\rangle \simeq L_{5,1}\oplus\frak{h}_{1}%
\]

\item  Invariants of subalgebra:%
\begin{align*}
J_{1}  &  =-x_{2}x_{5}^{2}-x_{1}x_{4}x_{5}+x_{3}x_{4}^{2},\\
J_{2}  &  =x_{6},
\end{align*}

\item  Semi-invariance conditions:%
\[
\widehat{X}_{9}\left(  J_{1}\right)  =-pJ_{1},\;\widehat{X}_{9}\left(
J_{2}\right)  =-qJ_{2}.
\]

\item  Invariants of $L_{9,23}^{p,q}$%
\begin{align*}
I_{1}  &  =\frac{J_{1}^{q}}{J_{2}^{p}},\;p\neq0\\
I_{1}  &  =J_{2},\;p=0.\;
\end{align*}
\end{itemize}

\newpage

\bigskip$\mathbf{L}_{9,24}^{p,q}\;[p\neq0,\;q\geq0]$

\begin{itemize}
\item  Levi decomposition: $L_{9,24}^{p,q}=\frak{sl}\left(  2,\mathbb{R}%
\right)  \overrightarrow{\oplus}_{R}\frak{g}_{6,35}$

\item  Describing representation: $R=D_{\frac{1}{2}}\oplus4D_{0}$

\item  Structure tensor:%
\[%
\begin{array}
[c]{llllll}%
C_{12}^{2}=2, & C_{13}^{3}=-2, & C_{23}^{1}=1, & C_{14}^{4}=1, & C_{15}%
^{5}=-1, & C_{25}^{4}=1,\\
C_{34}^{5}=1, & C_{45}^{6}=1, & C_{49}^{4}=p, & C_{59}^{5}=p, & C_{69}%
^{6}=2p, & C_{79}^{7}=q,\\
C_{79}^{8}=-1, & C_{89}^{7}=1, & C_{89}^{8}=q. &  &  & \\
&  &  &  &  & \\
&  &  &  &  &
\end{array}
\]

\item $\mathrm{codim}_{\frak{g}}\left[  \frak{g},\frak{g}\right]  =1.$

\item  Rank $A\left(  \frak{g}\right)  =6$

\item  Conditions on invariants:
\[
\frac{\partial F}{\partial x_{9}}=0,
\]

\item  Codimension one subalgebra:%
\[
L\left\langle X_{1},..,X_{8}\right\rangle \simeq L_{6,2}\oplus2L_{1}%
\]

\item  Invariants of subalgebra:%
\begin{align*}
J_{1}  &  =2x_{2}x_{3}x_{6}+x_{2}x_{5}^{2}+x_{1}x_{4}x_{5}-x_{3}x_{4}%
^{2}+\frac{1}{2}x_{1}^{2}x_{6},\\
J_{2}  &  =x_{6},\\
J_{3}  &  =x_{7},\\
J_{4}  &  =x_{8}.
\end{align*}

\item  Semi-invariance conditions:%
\begin{align*}
\widehat{X}_{9}\left(  J_{1}\right)   &  =-2pJ_{1},\;\widehat{X}_{9}\left(
J_{2}\right)  =-2pJ_{2},\\
\widehat{X}_{9}\left(  J_{3}\right)   &  =-qJ_{3}-J_{4},\;\widehat{X}%
_{9}\left(  J_{4}\right)  =-J_{3}-qJ_{4}.
\end{align*}

\item  Invariants of $L_{9,24}^{p,q}$%
\[
I_{1}=\frac{J_{1}^{{}}}{J_{2}^{{}}},\;I_{2}=\left(  J_{3}^{2}+J_{4}%
^{2}\right)  \left(  \frac{J_{3}-iJ_{4}}{J_{3}+iJ_{4}}\right)  ^{-iq}%
,\;I_{3}=J_{2}\exp\left(  2p\arctan\left(  J_{3}J_{4}^{-1}\right)  \right)  .
\]
\end{itemize}

\newpage

$\mathbf{L}_{9,25}^{p}\;[\,|p|\leq1\,]$

\begin{itemize}
\item  Levi decomposition: $L_{9,25}^{p}=\frak{sl}\left(  2,\mathbb{R}\right)
\overrightarrow{\oplus}_{R}\frak{g}_{6,82}$

\item  Describing representation: $R=D_{\frac{1}{2}}\oplus4D_{0}$

\item  Structure tensor:%
\[%
\begin{array}
[c]{llllll}%
C_{12}^{2}=2, & C_{13}^{3}=-2, & C_{23}^{1}=1, & C_{14}^{4}=1, & C_{15}%
^{5}=-1, & C_{25}^{4}=1,\\
C_{34}^{5}=1, & C_{45}^{6}=1, & C_{49}^{4}=p, & C_{59}^{5}=p, & C_{69}%
^{6}=2p, & C_{79}^{7}=1,\\
C_{89}^{8}=2p-1, & C_{78}^{6}=1. &  &  &  & \\
&  &  &  &  & \\
&  &  &  &  &
\end{array}
\]

\item $\mathrm{codim}_{\frak{g}}\left[  \frak{g},\frak{g}\right]  =\left\{
\begin{tabular}
[c]{ll}%
$1,$ & $p\neq\frac{1}{2}$\\
$2,$ & $p=\frac{1}{2}$%
\end{tabular}
\right.  .$

\item  Rank $A\left(  \frak{g}\right)  =\left\{
\begin{array}
[c]{cc}%
8 & \text{if }p\neq0\\
6 & \text{if }p=0
\end{array}
\right.  .$

\item  Conditions on invariants:
\[
\frac{\partial F}{\partial x_{j}}=0,\;j=7,8,9\;\text{if\ }p\neq0
\]

\item  Codimension one subalgebra for $p\neq0$:%
\[
L\left\langle X_{1},..,X_{8}\right\rangle \simeq L_{8,6}%
\]

\item  Invariants of subalgebra $\left(  p\neq0\right)  $:%
\begin{align*}
J_{1}  &  =2x_{2}x_{3}x_{6}+x_{2}x_{5}^{2}+x_{1}x_{4}x_{5}-x_{3}x_{4}%
^{2}+\frac{1}{2}x_{1}^{2}x_{6},\\
J_{2}  &  =x_{6}.
\end{align*}

\item  Semi-invariance conditions:%
\[
\widehat{X}_{9}\left(  J_{1}\right)  =-2pJ_{1},\;\widehat{X}_{9}\left(
J_{2}\right)  =-2pJ_{2}.
\]

\item  Invariants of $L_{9,25}^{p}$%
\begin{align*}
I_{1}  &  =\frac{J_{1}^{{}}}{J_{2}^{{}}},\;p\neq0,\\
I_{1}  &  =J_{1},\;I_{2}=J_{2},\;I_{3}=x_{6}x_{9}-x_{7}x_{8},\;p=0.
\end{align*}
\end{itemize}

\newpage

$\mathbf{L}_{9,26}^{\varepsilon}\;[\varepsilon=\pm1]$

\begin{itemize}
\item  Levi decomposition: $L_{9,26}^{\varepsilon}=\frak{sl}\left(
2,\mathbb{R}\right)  \overrightarrow{\oplus}_{R}\frak{g}_{6,85}$

\item  Describing representation: $R=D_{\frac{1}{2}}\oplus4D_{0}$

\item  Structure tensor:%
\[%
\begin{array}
[c]{llllll}%
C_{12}^{2}=2, & C_{13}^{3}=-2, & C_{23}^{1}=1, & C_{14}^{4}=1, & C_{15}%
^{5}=-1, & C_{25}^{4}=1,\\
C_{34}^{5}=1, & C_{45}^{6}=\varepsilon, & C_{49}^{4}=1, & C_{59}^{5}=1, &
C_{69}^{6}=2, & C_{79}^{7}=1,\\
C_{89}^{7}=1, & C_{89}^{8}=1, & C_{78}^{6}=1. &  &  & \\
&  &  &  &  & \\
&  &  &  &  &
\end{array}
\]

\item $\mathrm{codim}_{\frak{g}}\left[  \frak{g},\frak{g}\right]  =1.$

\item  Rank $A\left(  \frak{g}\right)  =1$

\item  Rank $A\left(  \frak{g}\right)  =8$

\item  Conditions on invariants:
\[
\frac{\partial F}{\partial x_{j}}=0,\;j=7,8,9.
\]

\item  Codimension one subalgebra:%
\[
L\left\langle X_{1},..,X_{8}\right\rangle \simeq L_{8,2}%
\]

\item  Invariants of subalgebra:%
\begin{align*}
J_{1}  &  =2\varepsilon x_{2}x_{3}x_{6}+x_{2}x_{5}^{2}+x_{1}x_{4}x_{5}%
-x_{3}x_{4}^{2}+\frac{\varepsilon}{2}x_{1}^{2}x_{6},\\
J_{2}  &  =x_{6}.
\end{align*}

\item  Semi-invariance conditions:%
\[
\widehat{X}_{9}\left(  J_{1}\right)  =-2J_{1},\;\widehat{X}_{9}\left(
J_{2}\right)  =-2J_{2}.
\]

\item  Invariants of $L_{9,26}^{\varepsilon}$%
\[
I_{1}=\frac{J_{1}^{{}}}{J_{2}^{{}}}.
\]
\end{itemize}

\newpage

$\mathbf{L}_{9,27}^{p}$

\begin{itemize}
\item  Levi decomposition: $L_{9,27}^{p}=\frak{sl}\left(  2,\mathbb{R}\right)
\overrightarrow{\oplus}_{R}\frak{g}_{6,89}$

\item  Describing representation: $R=D_{\frac{1}{2}}\oplus4D_{0}$

\item  Structure tensor:%
\[%
\begin{array}
[c]{llllll}%
C_{12}^{2}=2, & C_{13}^{3}=-2, & C_{23}^{1}=1, & C_{14}^{4}=1, & C_{15}%
^{5}=-1, & C_{25}^{4}=1,\\
C_{34}^{5}=1, & C_{45}^{6}=1, & C_{49}^{4}=p, & C_{59}^{5}=p, & C_{69}%
^{6}=2p, & C_{79}^{7}=p,\\
C_{79}^{8}=-1, & C_{89}^{7}=1, & C_{89}^{8}=p, & C_{78}^{6}=1. &  & \\
&  &  &  &  & \\
&  &  &  &  &
\end{array}
\]

\item $\mathrm{codim}_{\frak{g}}\left[  \frak{g},\frak{g}\right]  =1.$

\item  Rank $A\left(  \frak{g}\right)  =\left\{
\begin{array}
[c]{cc}%
8 & \text{if }p\neq0\\
6 & \text{if }p=0
\end{array}
\right.  $

\item  Conditions on invariants:
\[
\frac{\partial F}{\partial x_{j}}=0,\;j=7,8,9\text{ if }p\neq0.
\]

\item  Codimension one subalgebra:%
\[
L\left\langle X_{1},..,X_{8}\right\rangle \simeq L_{8,2}\text{ if }p\neq0.
\]

\item  Invariants of subalgebra:%
\begin{align*}
J_{1}  &  =2x_{2}x_{3}x_{6}+x_{2}x_{5}^{2}+x_{1}x_{4}x_{5}-x_{3}x_{4}%
^{2}+\frac{1}{2}x_{1}^{2}x_{6},\\
J_{2}  &  =x_{6}.
\end{align*}

\item  Semi-invariance conditions:%
\[
\widehat{X}_{9}\left(  J_{1}\right)  =-2pJ_{1},\;\widehat{X}_{9}\left(
J_{2}\right)  =-2pJ_{2}.
\]

\item  Invariants of $L_{9,27}^{p}$%
\begin{align*}
I_{1}  &  =\frac{J_{1}^{{}}}{J_{2}^{{}}},\;p\neq0\\
I_{1}  &  =J_{1},\;I_{2}=J_{2},\;I_{3}=2x_{6}x_{9}+x_{7}^{2}+x_{8}^{2},\;p=0.
\end{align*}
\end{itemize}

\newpage

$\mathbf{L}_{9,28}^{p,q}\;[pq\neq0]$

\begin{itemize}
\item  Levi decomposition: $L_{9,28}^{p,q}=\frak{sl}\left(  2,\mathbb{R}%
\right)  \overrightarrow{\oplus}_{R}N_{6,1}^{p,q,p,q}$

\item  Describing representation: $R=D_{\frac{1}{2}}\oplus4D_{0}$

\item  Structure tensor:%
\[%
\begin{array}
[c]{llllll}%
C_{12}^{2}=2, & C_{13}^{3}=-2, & C_{23}^{1}=1, & C_{14}^{4}=1, & C_{15}%
^{5}=-1, & C_{25}^{4}=1,\\
C_{34}^{5}=1, & C_{48}^{4}=p, & C_{49}^{4}=q, & C_{58}^{5}=p, & C_{59}%
^{5}=q, & C_{68}^{6}=1,\\
C_{79}^{7}=1. &  &  &  &  & \\
&  &  &  &  & \\
&  &  &  &  &
\end{array}
\]

\item $\mathrm{codim}_{\frak{g}}\left[  \frak{g},\frak{g}\right]  =2.$

\item  Rank $A\left(  \frak{g}\right)  =8$

\item  Conditions on invariants:
\[
\frac{\partial F}{\partial x_{j}}=0,\;j=8,9.
\]

\item  Codimension two subalgebra:%
\[
L\left\langle X_{1},..,X_{7}\right\rangle \simeq L_{5,1}\oplus2L_{1}%
\]

\item  Invariants of subalgebra:%
\begin{align*}
J_{1}  &  =-x_{2}x_{5}^{2}-x_{1}x_{4}x_{5}+x_{3}x_{4}^{2},\\
J_{2}  &  =x_{6},\\
J_{3}  &  =x_{7}.
\end{align*}

\item  Semi-invariance conditions:%
\begin{align*}
\widehat{X}_{8}\left(  J_{1}\right)   &  =-2pJ_{1},\;\widehat{X}_{8}\left(
J_{2}\right)  =-J_{2},\\
\widehat{X}_{9}\left(  J_{1}\right)   &  =-2qJ_{1},\;\widehat{X}_{9}\left(
J_{3}\right)  =-J_{3}.
\end{align*}

\item  Invariants of $L_{9,28}^{p,q}$%
\[
I_{1}=\frac{J_{1}^{{}}}{J_{2}^{2p}J_{3}^{2q}}.
\]
\end{itemize}

\newpage

$\mathbf{L}_{9,29}^{p}$

\begin{itemize}
\item  Levi decomposition: $L_{9,29}^{p}=\frak{sl}\left(  2,\mathbb{R}\right)
\overrightarrow{\oplus}_{R}N_{6,2}^{1,p,p}$

\item  Describing representation: $R=D_{\frac{1}{2}}\oplus4D_{0}$

\item  Structure tensor:%
\[%
\begin{array}
[c]{llllll}%
C_{12}^{2}=2, & C_{13}^{3}=-2, & C_{23}^{1}=1, & C_{14}^{4}=1, & C_{15}%
^{5}=-1, & C_{25}^{4}=1,\\
C_{34}^{5}=1, & C_{48}^{4}=1, & C_{49}^{4}=1, & C_{58}^{5}=1, & C_{59}%
^{5}=p, & C_{69}^{6}=1,\\
C_{78}^{6}=1, & C_{79}^{7}=1. &  &  &  & \\
&  &  &  &  & \\
&  &  &  &  &
\end{array}
\]

\item $\mathrm{codim}_{\frak{g}}\left[  \frak{g},\frak{g}\right]  =2.$

\item  Rank $A\left(  \frak{g}\right)  =8$

\item  Conditions on invariants:
\[
\frac{\partial F}{\partial x_{j}}=0,\;j=8,9.
\]

\item  Codimension two subalgebra:%
\[
L\left\langle X_{1},..,X_{7}\right\rangle \simeq L_{5,1}\oplus2L_{1}%
\]

\item  Invariants of subalgebra:%
\begin{align*}
J_{1}  &  =-x_{2}x_{5}^{2}-x_{1}x_{4}x_{5}+x_{3}x_{4}^{2},\\
J_{2}  &  =x_{6},\\
J_{3}  &  =x_{7}.
\end{align*}

\item  Semi-invariance conditions:%
\begin{align*}
\widehat{X}_{8}\left(  J_{1}\right)   &  =-2J_{1},\;\widehat{X}_{8}\left(
J_{3}\right)  =-J_{2},\\
\widehat{X}_{9}\left(  J_{1}\right)   &  =-2qJ_{1},\;\widehat{X}_{9}\left(
J_{2}\right)  =-J_{2},\;\widehat{X}_{9}\left(  J_{3}\right)  =-J_{3}.
\end{align*}

\item  Invariants of $L_{9,29}^{p}$%
\[
I_{1}=\ln\left(  \frac{J_{1}^{{}}}{J_{2}^{2q}}\right)  -2J_{3}J_{2}^{-1}.
\]
\end{itemize}

\newpage

$\mathbf{L}_{9,30}^{p,q}\;[p^{2}+q^{2}\neq0]$

\begin{itemize}
\item  Levi decomposition: $L_{9,30}^{p,q}=\frak{sl}\left(  2,\mathbb{R}%
\right)  \overrightarrow{\oplus}_{R}N_{6,13}^{p,q,p,q}$

\item  Describing representation: $R=D_{\frac{1}{2}}\oplus4D_{0}$

\item  Structure tensor:%
\[%
\begin{array}
[c]{llllll}%
C_{12}^{2}=2, & C_{13}^{3}=-2, & C_{23}^{1}=1, & C_{14}^{4}=1, & C_{15}%
^{5}=-1, & C_{25}^{4}=1,\\
C_{34}^{5}=1, & C_{48}^{4}=q, & C_{49}^{4}=p, & C_{58}^{5}=q, & C_{59}%
^{5}=p, & C_{68}^{6}=1,\\
C_{69}^{7}=-1, & C_{78}^{7}=1, & C_{79}^{6}=1. &  &  & \\
&  &  &  &  & \\
&  &  &  &  &
\end{array}
\]

\item $\mathrm{codim}_{\frak{g}}\left[  \frak{g},\frak{g}\right]  =2.$

\item  Rank $A\left(  \frak{g}\right)  =8$

\item  Conditions on invariants:
\[
\frac{\partial F}{\partial x_{j}}=0,\;j=8,9.
\]

\item  Codimension one subalgebra:%
\[
L\left\langle X_{1},..,X_{7}\right\rangle \simeq L_{5,1}\oplus2L_{1}%
\]

\item  Invariants of subalgebra:%
\begin{align*}
J_{1}  &  =-x_{2}x_{5}^{2}-x_{1}x_{4}x_{5}+x_{3}x_{4}^{2},\\
J_{2}  &  =x_{6},\\
J_{3}  &  =x_{7}.
\end{align*}

\item  Semi-invariance conditions:%
\begin{align*}
\widehat{X}_{8}\left(  J_{1}\right)   &  =-2qJ_{1},\;\widehat{X}_{8}\left(
J_{3}J_{2}^{-1}\right)  =0,\\
\widehat{X}_{9}\left(  J_{1}\right)   &  =-2pJ_{1},\;\widehat{X}_{9}\left(
J_{3}J_{2}^{-1}\right)  =-\left(  1+\left(  J_{3}J_{2}^{-1}\right)
^{2}\right)  .
\end{align*}

\item  Invariants of $L_{9,30}^{p,q}$%
\[
I_{1}=\ln\left(  \frac{J_{1}^{{}}}{\left(  J_{2}^{2}+J_{3}^{2}\right)  ^{q}%
}\right)  -2p\arctan\left(  J_{3}J_{2}^{-1}\right)  .
\]
\end{itemize}

\newpage

$\mathbf{L}_{9,31}^{{}}$

\begin{itemize}
\item  Levi decomposition: $L_{9,31}^{{}}=\frak{sl}\left(  2,\mathbb{R}%
\right)  \overrightarrow{\oplus}_{R}N_{6,20}^{1,0}$

\item  Describing representation: $R=D_{\frac{1}{2}}\oplus4D_{0}$

\item  Structure tensor:%
\[%
\begin{array}
[c]{llllll}%
C_{12}^{2}=2, & C_{13}^{3}=-2, & C_{23}^{1}=1, & C_{14}^{4}=1, & C_{15}%
^{5}=-1, & C_{25}^{4}=1,\\
C_{34}^{5}=1, & C_{48}^{4}=1, & C_{58}^{5}=1, & C_{69}^{6}=1, & C_{89}%
^{7}=1. & \\
&  &  &  &  & \\
&  &  &  &  &
\end{array}
\]

\item $\mathrm{codim}_{\frak{g}}\left[  \frak{g},\frak{g}\right]  =2.$

\item  Rank $A\left(  \frak{g}\right)  =8$

\item  Conditions on invariants:
\[
\frac{\partial F}{\partial x_{j}}=0,\;j\neq7.
\]

\item  Invariants of $L_{9,31}^{{}}$%
\[
I_{1}=x_{7}.
\]
\end{itemize}

\newpage

$\mathbf{L}_{9,32}^{p,q}\;[pq\neq0]$

\begin{itemize}
\item  Levi decomposition: $L_{9,32}^{p,q}=\frak{sl}\left(  2,\mathbb{R}%
\right)  \overrightarrow{\oplus}_{R}\frak{g}_{6,1}$

\item  Describing representation: $R=D_{1}\oplus3D_{0}$

\item  Structure tensor:%
\[%
\begin{array}
[c]{llllll}%
C_{12}^{2}=2, & C_{13}^{3}=-2, & C_{23}^{1}=1, & C_{14}^{4}=2, & C_{16}%
^{6}=-2, & C_{25}^{4}=2,\\
C_{26}^{5}=1, & C_{34}^{5}=1, & C_{35}^{6}=2, & C_{49}^{4}=1, & C_{59}%
^{5}=1, & C_{69}^{6}=1,\\
C_{79}^{7}=p, & C_{89}^{8}=q. &  &  &  & \\
&  &  &  &  & \\
&  &  &  &  &
\end{array}
\]

\item $\mathrm{codim}_{\frak{g}}\left[  \frak{g},\frak{g}\right]  =1.$

\item  Rank $A\left(  \frak{g}\right)  =6$

\item  Conditions on invariants:
\[
\frac{\partial F}{\partial x_{9}}=0.
\]

\item  Codimension one subalgebra:%
\[
L\left\langle X_{1},..,X_{8}\right\rangle \simeq L_{6,4}\oplus2L_{1}%
\]

\item  Invariants of subalgebra:%
\begin{align*}
J_{1}  &  =x_{1}x_{5}+2x_{2}x_{6}-x_{3}x_{4},\\
J_{2}  &  =x_{5}^{2}-4x_{4}x_{6},\\
J_{3}  &  =x_{7},\\
J_{4}  &  =x_{8}.
\end{align*}

\item  Semi-invariance conditions:%
\begin{align*}
\widehat{X}_{9}\left(  J_{1}\right)   &  =-J_{1},\;\widehat{X}_{8}\left(
J_{2}\right)  =-2J_{2},\\
\widehat{X}_{9}\left(  J_{3}\right)   &  =-pJ_{3},\;\widehat{X}_{9}\left(
J_{4}\right)  =-qJ_{4}.
\end{align*}

\item  Invariants of $L_{9,32}^{p,q}$%
\[
I_{1}=\frac{J_{1}^{2}}{J_{2}},\;I_{2}=\frac{J_{2}^{p}}{J_{3}^{2}}%
,\;I_{3}=\frac{J_{3}^{q}}{J_{4}^{p}}.
\]
\end{itemize}

\newpage

$\mathbf{L}_{9,33}^{p}$

\begin{itemize}
\item  Levi decomposition: $L_{9,33}^{p}=\frak{sl}\left(  2,\mathbb{R}\right)
\overrightarrow{\oplus}_{R}\frak{g}_{6,2}$

\item  Describing representation: $R=D_{1}\oplus3D_{0}$

\item  Structure tensor:%
\[%
\begin{array}
[c]{llllll}%
C_{12}^{2}=2, & C_{13}^{3}=-2, & C_{23}^{1}=1, & C_{14}^{4}=2, & C_{16}%
^{6}=-2, & C_{25}^{4}=2,\\
C_{26}^{5}=1, & C_{34}^{5}=1, & C_{35}^{6}=2, & C_{49}^{4}=1, & C_{59}%
^{5}=1, & C_{69}^{6}=1,\\
C_{79}^{7}=p, & C_{89}^{8}=p, & C_{89}^{7}=1. &  &  & \\
&  &  &  &  & \\
&  &  &  &  &
\end{array}
\]

\item $\mathrm{codim}_{\frak{g}}\left[  \frak{g},\frak{g}\right]  =\left\{
\begin{tabular}
[c]{ll}%
$1,$ & $p\neq0$\\
$2,$ & $p=0$%
\end{tabular}
\right.  .$

\item  Rank $A\left(  \frak{g}\right)  =6$

\item  Conditions on invariants:
\[
\frac{\partial F}{\partial x_{9}}=0.
\]

\item  Codimension one subalgebra:%
\[
L\left\langle X_{1},..,X_{8}\right\rangle \simeq L_{6,4}\oplus2L_{1}%
\]

\item  Invariants of subalgebra:%
\begin{align*}
J_{1}  &  =x_{1}x_{5}+2x_{2}x_{6}-x_{3}x_{4},\\
J_{2}  &  =x_{5}^{2}-4x_{4}x_{6},\\
J_{3}  &  =x_{7},\\
J_{4}  &  =x_{8}.
\end{align*}

\item  Semi-invariance conditions:%
\begin{align*}
\widehat{X}_{9}\left(  J_{1}\right)   &  =-J_{1},\;\widehat{X}_{8}\left(
J_{2}\right)  =-2J_{2},\\
\widehat{X}_{9}\left(  J_{3}\right)   &  =-pJ_{3},\;\widehat{X}_{9}\left(
J_{4}\right)  =-J_{3}-pJ_{4}.
\end{align*}

\item  Invariants of $L_{9,33}^{p}$%
\begin{align*}
I_{1}  &  =\frac{J_{1}^{2}}{J_{2}},\;I_{2}=\frac{J_{2}^{p}}{J_{3}^{2}}%
,\;I_{3}=\frac{pJ_{4}-J_{3}\ln\left(  J_{4}\right)  }{pJ_{3}},\;p\neq0\\
I_{1}  &  =\frac{J_{1}^{2}}{J_{2}},\;I_{2}=J_{3},\;I_{3}=\ln\left(
J_{2}\right)  -2\frac{J_{4}}{J_{3}},\;p=0.
\end{align*}
\end{itemize}

\newpage

$\mathbf{L}_{9,34}^{p,q}\;[p\neq0,\;q\geq0\,]$

\begin{itemize}
\item  Levi decomposition: $L_{9,34}^{p,q}=\frak{sl}\left(  2,\mathbb{R}%
\right)  \overrightarrow{\oplus}_{R}\frak{g}_{6,8}$

\item  Describing representation: $R=D_{1}\oplus3D_{0}$

\item  Structure tensor:%
\[%
\begin{array}
[c]{llllll}%
C_{12}^{2}=2, & C_{13}^{3}=-2, & C_{23}^{1}=1, & C_{14}^{4}=2, & C_{16}%
^{6}=-2, & C_{25}^{4}=2,\\
C_{26}^{5}=1, & C_{34}^{5}=1, & C_{35}^{6}=2, & C_{49}^{4}=p, & C_{59}%
^{5}=p, & C_{69}^{6}=p,\\
C_{79}^{7}=q, & C_{79}^{8}=-1, & C_{89}^{7}=1, & C_{89}^{8}=q. &  & \\
&  &  &  &  & \\
&  &  &  &  &
\end{array}
\]

\item $\mathrm{codim}_{\frak{g}}\left[  \frak{g},\frak{g}\right]  =1.$

\item  Rank $A\left(  \frak{g}\right)  =6$

\item  Conditions on invariants:
\[
\frac{\partial F}{\partial x_{9}}=0.
\]

\item  Codimension one subalgebra:%
\[
L\left\langle X_{1},..,X_{8}\right\rangle \simeq L_{6,4}\oplus2L_{1}%
\]

\item  Invariants of subalgebra:%
\begin{align*}
J_{1}  &  =x_{1}x_{5}+2x_{2}x_{6}-x_{3}x_{4},\\
J_{2}  &  =x_{5}^{2}-4x_{4}x_{6},\\
J_{3}  &  =x_{7},\\
J_{4}  &  =x_{8}.
\end{align*}

\item  Semi-invariance conditions:%
\begin{align*}
\widehat{X}_{9}\left(  J_{1}\right)   &  =-pJ_{1},\;\widehat{X}_{8}\left(
J_{2}\right)  =-2pJ_{2},\\
\widehat{X}_{9}\left(  J_{3}\right)   &  =-qJ_{3}-J_{4},\;\widehat{X}%
_{9}\left(  J_{4}\right)  =-J_{3}-qJ_{4}.
\end{align*}

\item  Invariants of $L_{9,34}^{p,q}$%
\[
I_{1}=\frac{J_{1}^{2}}{J_{2}},\;I_{2}=\frac{J_{2}^{q}}{\left(  J_{3}^{2}%
+J_{4}^{2}\right)  ^{p}},\;I_{3}=\left(  J_{3}^{2}+J_{4}^{2}\right)  \left(
\frac{J_{3}-iJ_{4}}{J_{3}+iJ_{4}}\right)  ^{-iq}.
\]
\end{itemize}

\newpage

$\mathbf{L}_{9,35}^{p}\;[p\neq0]$

\begin{itemize}
\item  Levi decomposition: $L_{9,35}^{p}=\frak{sl}\left(  2,\mathbb{R}\right)
\overrightarrow{\oplus}_{R}\frak{g}_{6,1}$

\item  Describing representation: $R=D_{\frac{3}{2}}\oplus2D_{0}$

\item  Structure tensor:%
\[%
\begin{array}
[c]{llllll}%
C_{12}^{2}=2, & C_{13}^{3}=-2, & C_{23}^{1}=1, & C_{14}^{4}=3, & C_{15}%
^{5}=1, & C_{16}^{6}=-1,\\
C_{17}^{7}=-3, & C_{25}^{4}=3, & C_{26}^{5}=2, & C_{27}^{6}=1, & C_{34}%
^{5}=1, & C_{35}^{6}=2,\\
C_{36}^{7}=3, & C_{49}^{4}=1, & C_{59}^{5}=1, & C_{69}^{6}=1, & C_{79}%
^{7}=1, & C_{89}^{8}=p.\\
&  &  &  &  & \\
&  &  &  &  &
\end{array}
\]

\item $\mathrm{codim}_{\frak{g}}\left[  \frak{g},\frak{g}\right]  =1.$

\item  Rank $A\left(  \frak{g}\right)  =8$

\item  Conditions on invariants:
\[
\frac{\partial F}{\partial x_{j}}=0,\;j=1,2,3,9.
\]

\item  Codimension one subalgebra:%
\[
L\left\langle X_{1},..,X_{8}\right\rangle \simeq L_{7,6}\oplus L_{1}%
\]

\item  Invariants of subalgebra:%
\begin{align*}
J_{1}  &  =27x_{4}^{2}x_{7}^{2}-18x_{4}x_{5}x_{6}x_{7}-x_{5}^{2}x_{6}%
^{2}+4\left(  x_{4}x_{6}^{3}+x_{5}^{3}x_{7}\right)  ,\\
J_{2}  &  =x_{8}.
\end{align*}

\item  Semi-invariance conditions:%
\[
\widehat{X}_{9}\left(  J_{1}\right)  =-4J_{1},\;\widehat{X}_{8}\left(
J_{2}\right)  =-qJ_{2}.
\]

\item  Invariants of $L_{9,35}^{p}$%
\[
I_{1}=\frac{J_{1}^{q}}{J_{2}^{4}}.
\]
\end{itemize}

\newpage

$\mathbf{L}_{9,36}$

\begin{itemize}
\item  Levi decomposition: $L_{9,36}=\frak{sl}\left(  2,\mathbb{R}\right)
\overrightarrow{\oplus}_{R}\frak{g}_{6,82}$

\item  Describing representation: $R=D_{\frac{3}{2}}\oplus2D_{0}$

\item  Structure tensor:%
\[%
\begin{array}
[c]{llllll}%
C_{12}^{2}=2, & C_{13}^{3}=-2, & C_{23}^{1}=1, & C_{14}^{4}=3, & C_{15}%
^{5}=1, & C_{16}^{6}=-1,\\
C_{17}^{7}=-3, & C_{25}^{4}=3, & C_{26}^{5}=2, & C_{27}^{6}=1, & C_{34}%
^{5}=1, & C_{35}^{6}=2,\\
C_{36}^{7}=3, & C_{49}^{4}=1, & C_{59}^{5}=1, & C_{69}^{6}=1, & C_{79}%
^{7}=1, & C_{89}^{8}=2,\\
C_{47}^{8}=1, & C_{56}^{8}=-3. &  &  &  & \\
&  &  &  &  &
\end{array}
\]

\item $\mathrm{codim}_{\frak{g}}\left[  \frak{g},\frak{g}\right]  =1.$

\item  Rank $A\left(  \frak{g}\right)  =8$

\item  Conditions on invariants:
\[
\frac{\partial F}{\partial x_{9}}=0.
\]

\item  Codimension one subalgebra:%
\[
L\left\langle X_{1},..,X_{8}\right\rangle \simeq L_{8,19}%
\]

\item  Invariants of subalgebra:%
\begin{align*}
J_{1}  &  =x_{8},\\
J_{2}  &  =\sqrt{D},
\end{align*}
where
\[
D:=\left|
\begin{array}
[c]{rrrrrrrr}%
0 & 2x_{2} & -2x_{3} & 3x_{4} & x_{5} & -x_{6} & -3x_{7} & x_{1}\\
-2x_{2} & 0 & x_{1} & 0 & 3x_{4} & 2x_{5} & x_{6} & x_{2}\\
2x_{3} & -x_{1} & 0 & x_{5} & 2x_{6} & 3x_{7} & 0 & x_{3}\\
-3x_{4} & 0 & -x_{5} & 0 & 0 & 0 & x_{8} & \frac{x_{4}}{2}\\
-x_{5} & -3x_{4} & -2x_{6} & 0 & 0 & -3x_{8} & 0 & \frac{x_{5}}{2}\\
x_{6} & -3x_{5} & -3x_{7} & 0 & 3x_{8} & 0 & 0 & \frac{x_{6}}{2}\\
3x_{7} & -x_{6} & 0 & -x_{8} & 0 & 0 & 0 & \frac{x_{7}}{2}\\
-x_{1} & -x_{2} & -x_{3} & -\frac{x_{4}}{2} & -\frac{x_{5}}{2} & -\frac{x_{6}%
}{2} & -\frac{x_{7}}{2} & 0
\end{array}
\right|
\]

\item  Semi-invariance conditions:%
\[
\widehat{X}_{9}\left(  J_{1}\right)  =-2J_{1},\;\widehat{X}_{8}\left(
J_{2}\right)  =-J_{2}.
\]

\item  Invariants of $L_{9,36}$%
\[
I_{1}=\frac{J_{2}}{J_{1}^{2}}.
\]
\end{itemize}

\newpage

$\mathbf{L}_{9,37}$

\begin{itemize}
\item  Levi decomposition: $L_{9,37}=\frak{sl}\left(  2,\mathbb{R}\right)
\overrightarrow{\oplus}_{R}\left(  2\frak{h}_{1}\right)  $

\item  Describing representation: $R=2D_{\frac{1}{2}}\oplus2D_{0}$

\item  Structure tensor:%
\[%
\begin{array}
[c]{llllll}%
C_{12}^{2}=2, & C_{13}^{3}=-2, & C_{23}^{1}=1, & C_{14}^{4}=1, & C_{15}%
^{5}=-1, & C_{16}^{6}=1,\\
C_{17}^{7}=-1, & C_{25}^{4}=1, & C_{27}^{6}=1, & C_{34}^{5}=1, & C_{36}%
^{7}=1, & C_{45}^{8}=1,\\
C_{67}^{9}=1. &  &  &  &  & \\
&  &  &  &  & \\
&  &  &  &  &
\end{array}
\]

\item $\mathrm{codim}_{\frak{g}}\left[  \frak{g},\frak{g}\right]  =0.$

\item  Rank $A\left(  \frak{g}\right)  =6$

\item  Invariants of $L_{9,37}$:%
\begin{align*}
I_{1}  &  =x_{8},\\
I_{2}  &  =x_{9},\\
I_{3}  &  =\sqrt{D},
\end{align*}
where
\[
D:=\left|
\begin{array}
[c]{rrrrrrrr}%
0 & 2x_{2} & -2x_{3} & x_{4} & -x_{5} & x_{6} & -x_{7} & x_{1}\\
-2x_{2} & 0 & x_{1} & 0 & x_{4} & 0 & x_{6} & x_{2}\\
2x_{3} & -x_{1} & 0 & x_{5} & 0 & x_{7} & 0 & x_{3}\\
-x_{4} & 0 & -x_{5} & 0 & x_{8} & 0 & 0 & \frac{x_{4}}{2}\\
x_{5} & -x_{4} & 0 & -x_{8} & 0 & 0 & 0 & \frac{x_{5}}{2}\\
-x_{6} & 0 & -x_{7} & 0 & 0 & 0 & x_{9} & \frac{x_{6}}{2}\\
x_{7} & -x_{6} & 0 & 0 & 0 & -x_{9} & 0 & \frac{x_{7}}{2}\\
-x_{1} & -x_{2} & -x_{3} & -\frac{x_{4}}{2} & -\frac{x_{5}}{2} & -\frac{x_{6}%
}{2} & -\frac{x_{7}}{2} & 0
\end{array}
\right|
\]
\end{itemize}

\newpage

$\mathbf{L}_{9,38}$

\begin{itemize}
\item  Levi decomposition: $L_{9,38}=\frak{sl}\left(  2,\mathbb{R}\right)
\overrightarrow{\oplus}_{R}\left(  2\mathcal{A}_{\frak{3,3}}\right)  $

\item  Describing representation: $R=2D_{\frac{1}{2}}\oplus2D_{0}$

\item  Structure tensor:%
\[%
\begin{array}
[c]{llllll}%
C_{12}^{2}=2, & C_{13}^{3}=-2, & C_{23}^{1}=1, & C_{14}^{4}=1, & C_{15}%
^{5}=-1, & C_{16}^{6}=1,\\
C_{17}^{7}=-1, & C_{25}^{4}=1, & C_{27}^{6}=1, & C_{34}^{5}=1, & C_{36}%
^{7}=1, & C_{48}^{8}=1,\\
C_{58}^{5}=1, & C_{69}^{6}=1, & C_{79}^{7}=1. &  &  & \\
&  &  &  &  & \\
&  &  &  &  &
\end{array}
\]

\item $\mathrm{codim}_{\frak{g}}\left[  \frak{g},\frak{g}\right]  =2.$

\item  Rank $A\left(  \frak{g}\right)  =8$

\item  Invariants of $L_{9,38}$%
\[
I_{1}=\frac{x_{4}x_{7}\left(  x_{8}-x_{9}+x_{1}\right)  -2x_{3}x_{4}%
x_{6}+2x_{2}x_{5}x_{7}+x_{5}x_{6}\left(  x_{9}-x_{8}+x_{1}\right)  }{\left(
x_{4}x_{7}-x_{5}x_{6}\right)  }.
\]
\end{itemize}

\newpage

$\mathbf{L}_{9,39}$

\begin{itemize}
\item  Levi decomposition: $L_{9,39}=\frak{sl}\left(  2,\mathbb{R}\right)
\overrightarrow{\oplus}_{R}\left(  \frak{h}_{1}\oplus\mathcal{A}_{3,3}\right)  $

\item  Describing representation: $R=2D_{\frac{1}{2}}\oplus2D_{0}$

\item  Structure tensor:%
\[%
\begin{array}
[c]{llllll}%
C_{12}^{2}=2, & C_{13}^{3}=-2, & C_{23}^{1}=1, & C_{14}^{4}=1, & C_{15}%
^{5}=-1, & C_{16}^{6}=1,\\
C_{17}^{7}=-1, & C_{25}^{4}=1, & C_{27}^{6}=1, & C_{34}^{5}=1, & C_{36}%
^{7}=1, & C_{49}^{4}=1,\\
C_{59}^{9}=1, & C_{67}^{8}=1. &  &  &  & \\
&  &  &  &  & \\
&  &  &  &  &
\end{array}
\]

\item $\mathrm{codim}_{\frak{g}}\left[  \frak{g},\frak{g}\right]  =1.$

\item  Rank $A\left(  \frak{g}\right)  =8$

\item  Conditions on invariants:
\[
\frac{\partial F}{\partial x_{j}}=0,\;j\neq8
\]

\item  Invariants of $L_{9,39}$%
\[
I_{1}=x_{8}.
\]
\end{itemize}

\newpage

$\mathbf{L}_{9,40}$

\begin{itemize}
\item  Levi decomposition: $L_{9,37}=\frak{sl}\left(  2,\mathbb{R}\right)
\overrightarrow{\oplus}_{R}\left(  \mathcal{A}_{6,3}\right)  $

\item  Describing representation: $R=2D_{\frac{1}{2}}\oplus2D_{0}$

\item  Structure tensor:%
\[%
\begin{array}
[c]{llllll}%
C_{12}^{2}=2, & C_{13}^{3}=-2, & C_{23}^{1}=1, & C_{14}^{4}=1, & C_{15}%
^{5}=-1, & C_{16}^{6}=1,\\
C_{17}^{7}=-1, & C_{25}^{4}=1, & C_{27}^{6}=1, & C_{34}^{5}=1, & C_{36}%
^{7}=1, & C_{67}^{8}=1,\\
C_{69}^{4}=1, & C_{79}^{5}=1. &  &  &  & \\
&  &  &  &  & \\
&  &  &  &  &
\end{array}
\]

\item $\mathrm{codim}_{\frak{g}}\left[  \frak{g},\frak{g}\right]  =1.$

\item  Rank $A\left(  \frak{g}\right)  =6$

\item  Invariants of $L_{9,40}$:%
\begin{align*}
I_{1}  &  =x_{8},\\
I_{2}  &  =x_{4}x_{7}-x_{5}x_{6}-x_{8}x_{9},\\
I_{3}  &  =\sqrt{D},
\end{align*}
where
\[
D:=\left|
\begin{array}
[c]{rrrrrrrr}%
0 & 2x_{2} & -2x_{3} & 3x_{4} & x_{5} & -x_{6} & -3x_{7} & x_{1}\\
-2x_{2} & 0 & x_{1} & 0 & 3x_{4} & 2x_{5} & x_{6} & x_{2}\\
2x_{3} & -x_{1} & 0 & x_{5} & 2x_{6} & 3x_{7} & 0 & x_{3}\\
-3x_{4} & 0 & -x_{5} & 0 & 0 & 0 & x_{8} & \frac{x_{4}}{2}\\
-x_{5} & -3x_{4} & -2x_{6} & 0 & 0 & -3x_{8} & 0 & \frac{x_{5}}{2}\\
x_{6} & -3x_{5} & -3x_{7} & 0 & 3x_{8} & 0 & 0 & \frac{x_{6}}{2}\\
3x_{7} & -x_{6} & 0 & -x_{8} & 0 & 0 & 0 & \frac{x_{7}}{2}\\
-x_{1} & -x_{2} & -x_{3} & -\frac{x_{4}}{2} & -\frac{x_{5}}{2} & -\frac{x_{6}%
}{2} & -\frac{x_{7}}{2} & 0
\end{array}
\right|
\]
\end{itemize}

\begin{remark}
Although here a fundamental system of invariants depends on all variables
$\left\{  x_{1},..,x_{9}\right\}  $, the functions $I_{1}$ and $I_{3}$ are
actually a fundamental system of invariants of the subalgebra $L\left\langle
X_{1},..,X_{8}\right\rangle \simeq\frak{sl}\left(  2,\mathbb{R}\right)
\overrightarrow{\oplus}_{2D_{\frac{1}{2}}\oplus D_{0}}\left(  \frak{h}%
_{1}\oplus2L_{1}\right)  $.
\end{remark}

\newpage

$\mathbf{L}_{9,41}$

\begin{itemize}
\item  Levi decomposition: $L_{9,41}=\frak{sl}\left(  2,\mathbb{R}\right)
\overrightarrow{\oplus}_{R}\mathcal{A}_{6,4}$

\item  Describing representation: $R=2D_{\frac{1}{2}}\oplus2D_{0}$

\item  Structure tensor:%
\[%
\begin{array}
[c]{llllll}%
C_{12}^{2}=2, & C_{13}^{3}=-2, & C_{23}^{1}=1, & C_{14}^{4}=1, & C_{15}%
^{5}=-1, & C_{16}^{6}=1,\\
C_{17}^{7}=-1, & C_{25}^{4}=1, & C_{27}^{6}=1, & C_{34}^{5}=1, & C_{36}%
^{7}=1, & C_{47}^{8}=1,\\
C_{56}^{8}=-1, & C_{67}^{9}=1. &  &  &  & \\
&  &  &  &  & \\
&  &  &  &  &
\end{array}
\]

\item $\mathrm{codim}_{\frak{g}}\left[  \frak{g},\frak{g}\right]  =0.$

\item  Rank $A\left(  \frak{g}\right)  =6$

\item  Invariants of $L_{9,41}$:%
\begin{align*}
I_{1}  &  =x_{8},\\
I_{2}  &  =x_{9},\\
I_{3}  &  =\sqrt{D},
\end{align*}
where
\[
D:=\left|
\begin{array}
[c]{rrrrrrrr}%
0 & 2x_{2} & -2x_{3} & x_{4} & -x_{5} & x_{6} & -x_{7} & x_{1}\\
-2x_{2} & 0 & x_{1} & 0 & x_{4} & 0 & x_{6} & x_{2}\\
2x_{3} & -x_{1} & 0 & x_{5} & 0 & x_{7} & 0 & x_{3}\\
-x_{4} & 0 & -x_{5} & 0 & 0 & 0 & x_{8} & \frac{x_{4}}{2}\\
x_{5} & -x_{4} & 0 & 0 & 0 & -x_{8} & 0 & \frac{x_{5}}{2}\\
-x_{6} & 0 & -x_{7} & 0 & x_{8} & 0 & x_{9} & \frac{x_{6}}{2}\\
x_{7} & -x_{6} & 0 & -x_{8} & 0 & -x_{9} & 0 & \frac{x_{7}}{2}\\
-x_{1} & -x_{2} & -x_{3} & -\frac{x_{4}}{2} & -\frac{x_{5}}{2} & -\frac{x_{6}%
}{2} & -\frac{x_{7}}{2} & 0
\end{array}
\right|  .
\]
\end{itemize}

\newpage

$\mathbf{L}_{9,42}$

\begin{itemize}
\item  Levi decomposition: $L_{9,42}=\frak{sl}\left(  2,\mathbb{R}\right)
\overrightarrow{\oplus}_{R}\mathcal{A}_{6,5}$

\item  Describing representation: $R=2D_{\frac{1}{2}}\oplus2D_{0}$

\item  Structure tensor:%
\[%
\begin{array}
[c]{llllll}%
C_{12}^{2}=2, & C_{13}^{3}=-2, & C_{23}^{1}=1, & C_{14}^{4}=1, & C_{15}%
^{5}=-1, & C_{16}^{6}=1,\\
C_{17}^{7}=-1, & C_{25}^{4}=1, & C_{27}^{6}=1, & C_{34}^{5}=1, & C_{36}%
^{7}=1, & C_{45}^{9}=-1,\\
C_{47}^{8}=1, & C_{56}^{8}=-1, & C_{67}^{9}=1. &  &  & \\
&  &  &  &  & \\
&  &  &  &  &
\end{array}
\]

\item $\mathrm{codim}_{\frak{g}}\left[  \frak{g},\frak{g}\right]  =0.$

\item  Rank $A\left(  \frak{g}\right)  =6$

\item  Invariants of $L_{9,42}$:%
\begin{align*}
I_{1}  &  =x_{8},\\
I_{2}  &  =x_{9},\\
I_{3}  &  =\sqrt{D},
\end{align*}
where
\[
D:=\left|
\begin{array}
[c]{rrrrrrrr}%
0 & 2x_{2} & -2x_{3} & x_{4} & -x_{5} & x_{6} & -x_{7} & x_{1}\\
-2x_{2} & 0 & x_{1} & 0 & x_{4} & 0 & x_{6} & x_{2}\\
2x_{3} & -x_{1} & 0 & x_{5} & 0 & x_{7} & 0 & x_{3}\\
-x_{4} & 0 & -x_{5} & 0 & -x_{9} & 0 & x_{8} & \frac{x_{4}}{2}\\
x_{5} & -x_{4} & 0 & x_{9} & 0 & -x_{8} & 0 & \frac{x_{5}}{2}\\
-x_{6} & 0 & -x_{7} & 0 & x_{8} & 0 & x_{9} & \frac{x_{6}}{2}\\
x_{7} & -x_{6} & 0 & -x_{8} & 0 & -x_{9} & 0 & \frac{x_{7}}{2}\\
-x_{1} & -x_{2} & -x_{3} & -\frac{x_{4}}{2} & -\frac{x_{5}}{2} & -\frac{x_{6}%
}{2} & -\frac{x_{7}}{2} & 0
\end{array}
\right|  .
\]
\end{itemize}

\newpage

$\mathbf{L}_{9,43}^{p,q}\;[q\neq0,\;|p|\leq1\,]$

\begin{itemize}
\item  Levi decomposition: $L_{9,43}^{p,q}=\frak{sl}\left(  2,\mathbb{R}%
\right)  \overrightarrow{\oplus}_{R}\frak{g}_{6,1}$

\item  Describing representation: $R=2D_{\frac{1}{2}}\oplus2D_{0}$

\item  Structure tensor:%
\[%
\begin{array}
[c]{llllll}%
C_{12}^{2}=2, & C_{13}^{3}=-2, & C_{23}^{1}=1, & C_{14}^{4}=1, & C_{15}%
^{5}=-1, & C_{16}^{6}=1,\\
C_{17}^{7}=-1, & C_{25}^{4}=1, & C_{27}^{6}=1, & C_{34}^{5}=1, & C_{36}%
^{7}=1, & C_{49}^{4}=1,\\
C_{59}^{5}=1, & C_{69}^{6}=p, & C_{79}^{7}=p, & C_{89}^{8}=q. &  & \\
&  &  &  &  & \\
&  &  &  &  &
\end{array}
\]

\item $\mathrm{codim}_{\frak{g}}\left[  \frak{g},\frak{g}\right]  =1.$

\item  Rank $A\left(  \frak{g}\right)  =8$

\item  Conditions on invariants:%
\begin{align*}
\frac{\partial F}{\partial x_{j}}  &  =0,\;j=1,2,3,9\;\text{if\ }p+1\neq0\\
\frac{\partial F}{\partial x_{j}}  &  =0,\;j=1,2,3,8,9\text{ if }p+1=0
\end{align*}

\item  Codimension one subalgebra%
\[
L\left\langle X_{1},..,X_{8}\right\rangle \simeq L_{7,7}\oplus L_{1}%
\]

\item  Invariants of subalgebra:%
\begin{align*}
J_{1}  &  =x_{4}x_{7}-x_{5}x_{6}\\
J_{2}  &  =x_{8}%
\end{align*}

\item  Semi-invariance conditions:%
\begin{align*}
\widehat{X}_{9}\left(  J_{1}\right)   &  =-\left(  1+p\right)  J_{1},\\
\widehat{X}_{9}\left(  J_{2}\right)   &  =-qJ_{2}.
\end{align*}

\item  Invariants of $L_{9,43}^{p,q}$:%
\[
I_{1}=\frac{J_{1}^{q}}{J_{2}^{1+p}}.
\]
\end{itemize}

\newpage

$\mathbf{L}_{9,44}^{p}$

\begin{itemize}
\item  Levi decomposition: $L_{9,44}^{p}=\frak{sl}\left(  2,\mathbb{R}\right)
\overrightarrow{\oplus}_{R}\frak{g}_{6,6}$

\item  Describing representation: $R=2D_{\frac{1}{2}}\oplus2D_{0}$

\item  Structure tensor:%
\[%
\begin{array}
[c]{llllll}%
C_{12}^{2}=2, & C_{13}^{3}=-2, & C_{23}^{1}=1, & C_{14}^{4}=1, & C_{15}%
^{5}=-1, & C_{16}^{6}=1,\\
C_{17}^{7}=-1, & C_{25}^{4}=1, & C_{27}^{6}=1, & C_{34}^{5}=1, & C_{36}%
^{7}=1, & C_{49}^{4}=p,\\
C_{59}^{5}=p, & C_{69}^{4}=1, & C_{69}^{6}=p, & C_{79}^{5}=1, & C_{79}%
^{7}=p, & C_{89}^{8}=1.\\
&  &  &  &  & \\
&  &  &  &  &
\end{array}
\]

\item $\mathrm{codim}_{\frak{g}}\left[  \frak{g},\frak{g}\right]  =1.$

\item  Rank $A\left(  \frak{g}\right)  =8$

\item  Conditions on invariants:%
\[
\frac{\partial F}{\partial x_{j}}=0,\;j=1,2,3,9
\]

\item  Codimension one subalgebra%
\[
L\left\langle X_{1},..,X_{8}\right\rangle \simeq L_{7,7}\oplus L_{1}%
\]

\item  Invariants of subalgebra:%
\begin{align*}
J_{1}  &  =x_{4}x_{7}-x_{5}x_{6}\\
J_{2}  &  =x_{8}%
\end{align*}

\item  Semi-invariance conditions:%
\begin{align*}
\widehat{X}_{9}\left(  J_{1}\right)   &  =-2pJ_{1},\\
\widehat{X}_{9}\left(  J_{2}\right)   &  =-J_{2}.
\end{align*}

\item  Invariants of $L_{9,44}^{p}$:%
\[
I_{1}=\frac{J_{1}}{J_{2}^{2p}}.
\]
\end{itemize}

\newpage

$\mathbf{L}_{9,45}^{p,q}\;[q\neq0]$

\begin{itemize}
\item  Levi decomposition: $L_{9,45}^{p,q}=\frak{sl}\left(  2,\mathbb{R}%
\right)  \overrightarrow{\oplus}_{R}\frak{g}_{6,11}$

\item  Describing representation: $R=2D_{\frac{1}{2}}\oplus2D_{0}$

\item  Structure tensor:%
\[%
\begin{array}
[c]{llllll}%
C_{12}^{2}=2, & C_{13}^{3}=-2, & C_{23}^{1}=1, & C_{14}^{4}=1, & C_{15}%
^{5}=-1, & C_{16}^{6}=1,\\
C_{17}^{7}=-1, & C_{25}^{4}=1, & C_{27}^{6}=1, & C_{34}^{5}=1, & C_{36}%
^{7}=1, & C_{49}^{4}=p,\\
C_{49}^{6}=-1, & C_{59}^{5}=p, & C_{59}^{7}=-1, & C_{69}^{4}=1, & C_{69}%
^{6}=p, & C_{79}^{5}=1,\\
C_{79}^{7}=p, & C_{89}^{8}=q. &  &  &  & \\
&  &  &  &  &
\end{array}
\]

\item $\mathrm{codim}_{\frak{g}}\left[  \frak{g},\frak{g}\right]  =1.$

\item  Rank $A\left(  \frak{g}\right)  =8$

\item  Conditions on invariants:%
\[
\frac{\partial F}{\partial x_{j}}=0,\;j=1,2,3,9
\]

\item  Codimension one subalgebra%
\[
L\left\langle X_{1},..,X_{8}\right\rangle \simeq L_{7,7}\oplus L_{1}%
\]

\item  Invariants of subalgebra:%
\begin{align*}
J_{1}  &  =x_{4}x_{7}-x_{5}x_{6}\\
J_{2}  &  =x_{8}%
\end{align*}

\item  Semi-invariance conditions:%
\begin{align*}
\widehat{X}_{9}\left(  J_{1}\right)   &  =-2pJ_{1},\\
\widehat{X}_{9}\left(  J_{2}\right)   &  =-qJ_{2}.
\end{align*}

\item  Invariants of $L_{9,45}^{p,q}$:%
\[
I_{1}=\frac{J_{1}^{q}}{J_{2}^{2p}}.
\]
\end{itemize}

\newpage

$\mathbf{L}_{9,46}^{p}$

\begin{itemize}
\item  Levi decomposition: $L_{9,46}^{p}=\frak{sl}\left(  2,\mathbb{R}\right)
\overrightarrow{\oplus}_{R}\frak{g}_{6,13}$

\item  Describing representation: $R=2D_{\frac{1}{2}}\oplus2D_{0}$

\item  Structure tensor:%
\[%
\begin{array}
[c]{llllll}%
C_{12}^{2}=2, & C_{13}^{3}=-2, & C_{23}^{1}=1, & C_{14}^{4}=1, & C_{15}%
^{5}=-1, & C_{16}^{6}=1,\\
C_{17}^{7}=-1, & C_{25}^{4}=1, & C_{27}^{6}=1, & C_{34}^{5}=1, & C_{36}%
^{7}=1, & C_{49}^{4}=1,\\
C_{45}^{8}=1, & C_{59}^{5}=1, & C_{69}^{6}=p, & C_{79}^{7}=p, & C_{89}%
^{8}=2. & \\
&  &  &  &  & \\
&  &  &  &  &
\end{array}
\]

\item $\mathrm{codim}_{\frak{g}}\left[  \frak{g},\frak{g}\right]  =1.$

\item  Rank $A\left(  \frak{g}\right)  =8$

\item  Conditions on invariants:%
\[
\frac{\partial F}{\partial x_{9}}=0.
\]

\item  Codimension one subalgebra%
\[
L\left\langle X_{1},..,X_{8}\right\rangle \simeq L_{8,13}^{\epsilon=-1}%
\]

\item  Invariants of subalgebra:%
\begin{align*}
J_{1}  &  =x_{8}\\
J_{2}  &  =\sqrt{D},
\end{align*}
where
\[
D:=\left|
\begin{array}
[c]{rrrrrrrr}%
0 & 2x_{2} & -2x_{3} & x_{4} & -x_{5} & x_{6} & -x_{7} & x_{1}\\
-2x_{2} & 0 & x_{1} & 0 & x_{4} & 0 & x_{6} & x_{2}\\
2x_{3} & -x_{1} & 0 & x_{5} & 0 & x_{7} & 0 & x_{3}\\
-x_{4} & 0 & -x_{5} & 0 & x_{8} & 0 & 0 & \frac{x_{4}}{2}\\
x_{5} & -x_{4} & 0 & -x_{8} & 0 & 0 & 0 & \frac{x_{5}}{2}\\
-x_{6} & 0 & -x_{7} & 0 & 0 & 0 & 0 & \frac{x_{6}}{2}\\
x_{7} & -x_{6} & 0 & 0 & 0 & 0 & 0 & \frac{x_{7}}{2}\\
-x_{1} & -x_{2} & -x_{3} & -\frac{x_{4}}{2} & -\frac{x_{5}}{2} & -\frac{x_{6}%
}{2} & -\frac{x_{7}}{2} & 0
\end{array}
\right|  .
\]

\item  Semi-invariance conditions:%
\begin{align*}
\widehat{X}_{9}\left(  J_{1}\right)   &  =-2J_{1},\\
\widehat{X}_{9}\left(  J_{2}\right)   &  =-2\left(  p+1\right)  J_{2}.
\end{align*}

\item  Invariants of $L_{9,46}^{p}$:%
\[
I_{1}=\frac{J_{2}}{J_{2}^{1+p}}.
\]
\end{itemize}

\newpage

$\mathbf{L}_{9,47}$

\begin{itemize}
\item  Levi decomposition: $L_{9,47}=\frak{sl}\left(  2,\mathbb{R}\right)
\overrightarrow{\oplus}_{R}\frak{g}_{6,15}$

\item  Describing representation: $R=2D_{\frac{1}{2}}\oplus2D_{0}$

\item  Structure tensor:%
\[%
\begin{array}
[c]{llllll}%
C_{12}^{2}=2, & C_{13}^{3}=-2, & C_{23}^{1}=1, & C_{14}^{4}=1, & C_{15}%
^{5}=-1, & C_{16}^{6}=1,\\
C_{17}^{7}=-1, & C_{25}^{4}=1, & C_{27}^{6}=1, & C_{34}^{5}=1, & C_{36}%
^{7}=1, & C_{49}^{4}=1,\\
C_{67}^{8}=1, & C_{59}^{5}=1, & C_{69}^{4}=1, & C_{69}^{6}=1, & C_{79}%
^{5}=1, & C_{79}^{7}=1,\\
C_{89}^{8}=2. &  &  &  &  & \\
&  &  &  &  &
\end{array}
\]

\item $\mathrm{codim}_{\frak{g}}\left[  \frak{g},\frak{g}\right]  =1.$

\item  Rank $A\left(  \frak{g}\right)  =8$

\item  Conditions on invariants:%
\[
\frac{\partial F}{\partial x_{9}}=0.
\]

\item  Codimension one subalgebra%
\[
L\left\langle X_{1},..,X_{8}\right\rangle \simeq L_{8,13}^{\epsilon=-1}%
\]

\item  Invariants of subalgebra:%
\begin{align*}
J_{1}  &  =x_{8}\\
J_{2}  &  =\sqrt{D},
\end{align*}
where
\[
D:=\left|
\begin{array}
[c]{rrrrrrrr}%
0 & 2x_{2} & -2x_{3} & x_{4} & -x_{5} & x_{6} & -x_{7} & x_{1}\\
-2x_{2} & 0 & x_{1} & 0 & x_{4} & 0 & x_{6} & x_{2}\\
2x_{3} & -x_{1} & 0 & x_{5} & 0 & x_{7} & 0 & x_{3}\\
-x_{4} & 0 & -x_{5} & 0 & 0 & 0 & 0 & \frac{x_{4}}{2}\\
x_{5} & -x_{4} & 0 & 0 & 0 & 0 & 0 & \frac{x_{5}}{2}\\
-x_{6} & 0 & -x_{7} & 0 & 0 & 0 & x_{8} & \frac{x_{6}}{2}\\
x_{7} & -x_{6} & 0 & 0 & 0 & -x_{8} & 0 & \frac{x_{7}}{2}\\
-x_{1} & -x_{2} & -x_{3} & -\frac{x_{4}}{2} & -\frac{x_{5}}{2} & -\frac{x_{6}%
}{2} & -\frac{x_{7}}{2} & 0
\end{array}
\right|  .
\]

\item  Semi-invariance conditions:%
\begin{align*}
\widehat{X}_{9}\left(  J_{1}\right)   &  =-2J_{1},\\
\widehat{X}_{9}\left(  J_{2}\right)   &  =-4J_{2}.
\end{align*}

\item  Invariants of $L_{9,47}$:%
\[
I_{1}=\frac{J_{2}}{J_{2}^{2}}.
\]
\end{itemize}

\newpage

$\mathbf{L}_{9,48}$

\begin{itemize}
\item  Levi decomposition: $L_{9,48}=\frak{sl}\left(  2,\mathbb{R}\right)
\overrightarrow{\oplus}_{R}\frak{g}_{6,53}$

\item  Describing representation: $R=2D_{\frac{1}{2}}\oplus2D_{0}$

\item  Structure tensor:%
\[%
\begin{array}
[c]{llllll}%
C_{12}^{2}=2, & C_{13}^{3}=-2, & C_{23}^{1}=1, & C_{14}^{4}=1, & C_{15}%
^{5}=-1, & C_{16}^{6}=1,\\
C_{17}^{7}=-1, & C_{25}^{4}=1, & C_{27}^{6}=1, & C_{34}^{5}=1, & C_{36}%
^{7}=1, & C_{68}^{4}=1,\\
C_{69}^{9}=1, & C_{78}^{5}=1, & C_{79}^{7}=1, & C_{89}^{8}=-1. &  & \\
&  &  &  &  & \\
&  &  &  &  &
\end{array}
\]

\item $\mathrm{codim}_{\frak{g}}\left[  \frak{g},\frak{g}\right]  =1.$

\item  Rank $A\left(  \frak{g}\right)  =8$

\item  Conditions on invariants:%
\[
\frac{\partial F}{\partial x_{9}}=0.
\]

\item  Codimension one subalgebra%
\[
L\left\langle X_{1},..,X_{8}\right\rangle \simeq L_{8,14}%
\]

\item  Invariants of subalgebra:%
\begin{align*}
J_{1}  &  =x_{1}x_{4}x_{5}+x_{5}x_{6}x_{8}-x_{4}x_{7}x_{8}+x_{2}x_{5}%
^{2}-x_{3}x_{4}^{2}\\
J_{2}  &  =x_{4}x_{7}-x_{5}x_{6}.
\end{align*}

\item  Semi-invariance conditions:%
\begin{align*}
\widehat{X}_{9}\left(  J_{1}\right)   &  =0,\\
\widehat{X}_{9}\left(  J_{2}\right)   &  =-J_{2}.
\end{align*}

\item  Invariants of $L_{9,48}$:%
\[
I_{1}=J_{1}.
\]
\end{itemize}

\newpage

$\mathbf{L}_{9,49}^{p}$

\begin{itemize}
\item  Levi decomposition: $L_{9,49}^{p}=\frak{sl}\left(  2,\mathbb{R}\right)
\overrightarrow{\oplus}_{R}\frak{g}_{6,53}$

\item  Describing representation: $R=2D_{\frac{1}{2}}\oplus2D_{0}$

\item  Structure tensor:%
\[%
\begin{array}
[c]{llllll}%
C_{12}^{2}=2, & C_{13}^{3}=-2, & C_{23}^{1}=1, & C_{14}^{4}=1, & C_{15}%
^{5}=-1, & C_{16}^{6}=1,\\
C_{17}^{7}=-1, & C_{25}^{4}=1, & C_{27}^{6}=1, & C_{34}^{5}=1, & C_{36}%
^{7}=1, & C_{49}^{4}=1,\\
C_{59}^{5}=1, & C_{68}^{4}=1, & C_{69}^{6}=1-p, & C_{78}^{5}=1, & C_{79}%
^{9}=1-p, & C_{89}^{8}=p.\\
&  &  &  &  & \\
&  &  &  &  &
\end{array}
\]

\item $\mathrm{codim}_{\frak{g}}\left[  \frak{g},\frak{g}\right]  =\left\{
\begin{tabular}
[c]{ll}%
$1,$ & $p\neq0$\\
$2,$ & $p=0$%
\end{tabular}
\right.  .$

\item  Rank $A\left(  \frak{g}\right)  =8$

\item  Conditions on invariants:%
\[
\frac{\partial F}{\partial x_{9}}=0.
\]

\item  Codimension one subalgebra%
\[
L\left\langle X_{1},..,X_{8}\right\rangle \simeq L_{8,14}%
\]

\item  Invariants of subalgebra:%
\begin{align*}
J_{1}  &  =x_{1}x_{4}x_{5}+x_{5}x_{6}x_{8}-x_{4}x_{7}x_{8}+x_{2}x_{5}%
^{2}-x_{3}x_{4}^{2}\\
J_{2}  &  =x_{4}x_{7}-x_{5}x_{6}.
\end{align*}

\item  Semi-invariance conditions:%
\begin{align*}
\widehat{X}_{9}\left(  J_{1}\right)   &  =-2J_{1},\\
\widehat{X}_{9}\left(  J_{2}\right)   &  =\left(  p-2\right)  J_{2}.
\end{align*}

\item  Invariants of $L_{9,49}^{p}$:%
\[
I_{1}=J_{1}^{p-2}J_{2}^{2}.
\]
\end{itemize}

\newpage

$\mathbf{L}_{9,50}$

\begin{itemize}
\item  Levi decomposition: $L_{9,50}=\frak{sl}\left(  2,\mathbb{R}\right)
\overrightarrow{\oplus}_{R}\frak{g}_{6,76}$

\item  Describing representation: $R=2D_{\frac{1}{2}}\oplus2D_{0}$

\item  Structure tensor:%
\[%
\begin{array}
[c]{llllll}%
C_{12}^{2}=2, & C_{13}^{3}=-2, & C_{23}^{1}=1, & C_{14}^{4}=1, & C_{15}%
^{5}=-1, & C_{16}^{6}=1,\\
C_{17}^{7}=-1, & C_{25}^{4}=1, & C_{27}^{6}=1, & C_{34}^{5}=1, & C_{36}%
^{7}=1, & C_{49}^{4}=3,\\
C_{59}^{5}=3, & C_{67}^{8}=1, & C_{68}^{4}=1, & C_{69}^{6}=1, & C_{78}%
^{5}=1, & C_{79}^{7}=1,\\
C_{89}^{8}=2. &  &  &  &  & \\
&  &  &  &  &
\end{array}
\]

\item $\mathrm{codim}_{\frak{g}}\left[  \frak{g},\frak{g}\right]  =1.$

\item  Rank $A\left(  \frak{g}\right)  =8$

\item  Conditions on invariants:%
\[
\frac{\partial F}{\partial x_{9}}=0.
\]

\item  Codimension one subalgebra%
\[
L\left\langle X_{1},..,X_{8}\right\rangle \simeq L_{8,15}%
\]

\item  Invariants of subalgebra:%
\begin{align*}
J_{1}  &  =x_{1}x_{4}x_{5}+x_{5}x_{6}x_{8}-x_{4}x_{7}x_{8}+x_{2}x_{5}%
^{2}-x_{3}x_{4}^{2}+\frac{1}{3}x_{8}^{3}\\
J_{2}  &  =x_{4}x_{7}-x_{5}x_{6}-\frac{1}{2}x_{8}^{2}.
\end{align*}

\item  Semi-invariance conditions:%
\begin{align*}
\widehat{X}_{9}\left(  J_{1}\right)   &  =-6J_{1},\\
\widehat{X}_{9}\left(  J_{2}\right)   &  =-4J_{2}.
\end{align*}

\item  Invariants of $L_{9,50}$:%
\[
I_{1}=\frac{J_{1}^{2}}{J_{2}^{3}}.
\]
\end{itemize}

\newpage

$\mathbf{L}_{9,51}^{p}\;[\,|p|\leq1\,]$

\begin{itemize}
\item  Levi decomposition: $L_{9,51}^{p}=\frak{sl}\left(  2,\mathbb{R}\right)
\overrightarrow{\oplus}_{R}\frak{g}_{6,82}$

\item  Describing representation: $R=2D_{\frac{1}{2}}\oplus2D_{0}$

\item  Structure tensor:%
\[%
\begin{array}
[c]{llllll}%
C_{12}^{2}=2, & C_{13}^{3}=-2, & C_{23}^{1}=1, & C_{14}^{4}=1, & C_{15}%
^{5}=-1, & C_{16}^{6}=1,\\
C_{17}^{7}=-1, & C_{25}^{4}=1, & C_{27}^{6}=1, & C_{34}^{5}=1, & C_{36}%
^{7}=1, & C_{49}^{4}=1,\\
C_{47}^{8}=1, & C_{56}^{8}=-1, & C_{59}^{5}=1, & C_{69}^{6}=p, & C_{79}%
^{7}=p, & C_{89}^{8}=1+p.\\
&  &  &  &  & \\
&  &  &  &  &
\end{array}
\]

\item $\mathrm{codim}_{\frak{g}}\left[  \frak{g},\frak{g}\right]  =1.$

\item  Rank $A\left(  \frak{g}\right)  =\left\{
\begin{array}
[c]{cc}%
8 & \text{if }p+1\neq0\\
6 & \text{if }p+1=0
\end{array}
\right.  $

\item  Conditions on invariants:%
\[
\frac{\partial F}{\partial x_{9}}=0\text{\ if }p+1\neq0.
\]

\item  Codimension one (and three if $p+1=0$) subalgebras:%
\begin{align*}
L\left\langle X_{1},..,X_{8}\right\rangle  &  \simeq L_{8,13}\\
L\left\langle X_{4},..,X_{9}\right\rangle  &  \simeq\frak{g}_{6,82}\text{ \ if
}p+1=0
\end{align*}

\item  Invariants of subalgebra:%
\begin{align*}
J_{1}  &  =\left(  x_{1}^{2}+4x_{2}x_{3}\right)  +2x_{1}x_{8}\left(
x_{4}x_{7}+x_{5}x_{6}\right)  -2x_{4}x_{5}x_{6}x_{7}+\\
&  +4x_{8}\left(  x_{2}x_{5}x_{7}-x_{3}x_{4}x_{6}\right)  +x_{4}^{2}x_{7}%
^{2}+x_{5}^{2}x_{6}^{2},\\
J_{2}  &  =x_{8},\\
J_{3}  &  =x_{4}x_{7}-x_{5}x_{6}-x_{8}x_{9}\text{ if }p+1=0.
\end{align*}

\item  Semi-invariance conditions:%
\begin{align*}
\widehat{X}_{9}\left(  J_{1}\right)   &  =-2\left(  1+p\right)  J_{1},\\
\widehat{X}_{9}\left(  J_{2}\right)   &  =-\left(  1+p\right)  J_{2},\\
\widehat{X}_{9}\left(  J_{3}\right)   &  =0,\;\left(  p+1=0\right)  .
\end{align*}

\item  Invariants of $L_{9,51}^{p}$:%
\begin{align*}
I_{1}  &  =\frac{J_{1}}{J_{2}^{2}}\text{\ if }p+1\neq0,\\
I_{1}  &  =J_{1},\;I_{2}=J_{2},\;\\
I_{3}  &  =x_{4}x_{7}-x_{5}x_{6}-x_{8}x_{9}\;\text{if }p+1=0
\end{align*}
\end{itemize}

\newpage

$\mathbf{L}_{9,52}$

\begin{itemize}
\item  Levi decomposition: $L_{9,52}=\frak{sl}\left(  2,\mathbb{R}\right)
\overrightarrow{\oplus}_{R}\frak{g}_{6,82}$

\item  Describing representation: $R=2D_{\frac{1}{2}}\oplus2D_{0}$

\item  Structure tensor:%
\[%
\begin{array}
[c]{llllll}%
C_{12}^{2}=2, & C_{13}^{3}=-2, & C_{23}^{1}=1, & C_{14}^{4}=1, & C_{15}%
^{5}=-1, & C_{16}^{6}=1,\\
C_{17}^{7}=-1, & C_{25}^{4}=1, & C_{27}^{6}=1, & C_{34}^{5}=1, & C_{36}%
^{7}=1, & C_{45}^{8}=1,\\
C_{49}^{4}=1, & C_{59}^{5}=1, & C_{67}^{8}=1, & C_{69}^{6}=1, & C_{79}%
^{7}=1, & C_{89}^{8}=2.\\
&  &  &  &  & \\
&  &  &  &  &
\end{array}
\]

\item $\mathrm{codim}_{\frak{g}}\left[  \frak{g},\frak{g}\right]  =1.$

\item  Rank $A\left(  \frak{g}\right)  =8$

\item  Conditions on invariants:%
\[
\frac{\partial F}{\partial x_{9}}=0.
\]

\item  Codimension one subalgebra%
\[
L\left\langle X_{1},..,X_{8}\right\rangle \simeq L_{8,13}%
\]

\item  Invariants of subalgebra:%
\begin{align*}
J_{1}  &  =\left(  x_{1}^{2}+4x_{2}x_{3}\right)  +2x_{1}x_{8}\left(
x_{4}x_{7}+x_{5}x_{6}\right)  -2x_{4}x_{5}x_{6}x_{7}+\\
&  +4x_{8}\left(  x_{2}x_{5}x_{7}-x_{3}x_{4}x_{6}\right)  +x_{4}^{2}x_{7}%
^{2}+x_{5}^{2}x_{6}^{2},\\
J_{2}  &  =x_{8}.
\end{align*}

\item  Semi-invariance conditions:%
\begin{align*}
\widehat{X}_{9}\left(  J_{1}\right)   &  =-4J_{1},\\
\widehat{X}_{9}\left(  J_{2}\right)   &  =-2J_{2}.
\end{align*}

\item  Invariants of $L_{9,52}$:%
\[
I_{1}=\frac{J_{1}}{J_{2}^{2}}.
\]
\end{itemize}

\newpage

$\mathbf{L}_{9,53}^{p}\;[p\geq0]$

\begin{itemize}
\item  Levi decomposition: $L_{9,53}^{p}=\frak{sl}\left(  2,\mathbb{R}\right)
\overrightarrow{\oplus}_{R}\frak{g}_{6,92}$

\item  Describing representation: $R=2D_{\frac{1}{2}}\oplus2D_{0}$

\item  Structure tensor:%
\[%
\begin{array}
[c]{llllll}%
C_{12}^{2}=2, & C_{13}^{3}=-2, & C_{23}^{1}=1, & C_{14}^{4}=1, & C_{15}%
^{5}=-1, & C_{16}^{6}=1,\\
C_{17}^{7}=-1, & C_{25}^{4}=1, & C_{27}^{6}=1, & C_{34}^{5}=1, & C_{36}%
^{7}=1, & C_{45}^{8}=1,\\
C_{49}^{4}=p, & C_{49}^{6}=-1, & C_{59}^{5}=p, & C_{59}^{7}=-1, & C_{67}%
^{8}=1, & C_{69}^{4}=1,\\
C_{69}^{6}=p, & C_{79}^{5}=1, & C_{79}^{7}=p, & C_{89}^{8}=2p. &  & \\
&  &  &  &  &
\end{array}
\]

\item $\mathrm{codim}_{\frak{g}}\left[  \frak{g},\frak{g}\right]  =1.$

\item  Rank $A\left(  \frak{g}\right)  =8$

\item  Conditions on invariants:%
\[
\frac{\partial F}{\partial x_{9}}=0.
\]

\item  Codimension one subalgebra%
\begin{align*}
L\left\langle X_{1},..,X_{8}\right\rangle  &  \simeq L_{8,13}\\
L\left\langle X_{4},..,X_{9}\right\rangle  &  \simeq\frak{g}_{6,92}\text{ if
}p=0.
\end{align*}

\item  Invariants of subalgebra:%
\begin{align*}
J_{1}  &  =\left(  x_{1}^{2}+4x_{2}x_{3}\right)  +2x_{1}x_{8}\left(
x_{4}x_{7}+x_{5}x_{6}\right)  -2x_{4}x_{5}x_{6}x_{7}+\\
&  +4x_{8}\left(  x_{2}x_{5}x_{7}-x_{3}x_{4}x_{6}\right)  +x_{4}^{2}x_{7}%
^{2}+x_{5}^{2}x_{6}^{2},\\
J_{2}  &  =x_{8},\\
J_{3}  &  =x_{4}x_{7}-x_{5}x_{6}-x_{8}x_{9}\text{ if }p=0.
\end{align*}

\item  Semi-invariance conditions:%
\begin{align*}
\widehat{X}_{9}\left(  J_{1}\right)   &  =-4pJ_{1},\\
\widehat{X}_{9}\left(  J_{2}\right)   &  =-2pJ_{2},\\
\widehat{X}_{9}\left(  J_{3}\right)   &  =0\text{ if }p=0.
\end{align*}

\item  Invariants of $L_{9,53}^{p}$:%
\begin{align*}
I_{1}  &  =\frac{J_{1}}{J_{2}^{2}}\text{ if }p\neq0,\;\\
I_{1}  &  =J_{1},I_{2}=J_{2},I_{3}=J_{3}\text{ if }p=0.
\end{align*}
\end{itemize}

\newpage

$\mathbf{L}_{9,54}$

\begin{itemize}
\item  Levi decomposition: $L_{9,54}=\frak{sl}\left(  2,\mathbb{R}\right)
\overrightarrow{\oplus}_{R}\mathcal{N}_{6,18}^{0,1,1}$

\item  Describing representation: $R=2D_{\frac{1}{2}}\oplus2D_{0}$

\item  Structure tensor:%
\[%
\begin{array}
[c]{llllll}%
C_{12}^{2}=2, & C_{13}^{3}=-2, & C_{23}^{1}=1, & C_{14}^{4}=1, & C_{15}%
^{5}=-1, & C_{16}^{6}=1,\\
C_{17}^{7}=-1, & C_{25}^{4}=1, & C_{27}^{6}=1, & C_{34}^{5}=1, & C_{36}%
^{7}=1, & C_{48}^{8}=1,\\
C_{49}^{6}=1, & C_{58}^{5}=1, & C_{59}^{7}=1, & C_{68}^{6}=1, & C_{69}%
^{4}=-1, & C_{78}^{8}=1,\\
C_{79}^{5}=-1. &  &  &  &  & \\
&  &  &  &  &
\end{array}
\]

\item $\mathrm{codim}_{\frak{g}}\left[  \frak{g},\frak{g}\right]  =2.$

\item  Rank $A\left(  \frak{g}\right)  =8$

\item  Conditions on invariants:%
\[
\frac{\partial F}{\partial x_{8}}=0.
\]

\item  Codimension one subalgebra%
\[
L\left\langle X_{1},..X_{7},X_{9}\right\rangle \simeq L_{8,18}^{0}%
\]

\item  Invariants of subalgebra:%
\begin{align*}
J_{1}  &  =-2x_{1}\left(  x_{6}x_{7}+x_{4}x_{5}\right)  -2x_{9}\left(
x_{4}x_{7}-x_{5}x_{6}\right)  +2x_{3}\left(  x_{4}^{2}+x_{6}^{2}\right)  +\\
&  -2x_{2}\left(  x_{5}^{2}+x_{7}^{2}\right)  ,\\
J_{2}  &  =x_{4}x_{7}-x_{5}x_{6}.
\end{align*}

\item  Semi-invariance conditions:%
\begin{align*}
\widehat{X}_{9}\left(  J_{1}\right)   &  =-2J_{1},\\
\widehat{X}_{9}\left(  J_{2}\right)   &  =-2J_{2}.
\end{align*}

\item  Invariants of $L_{9,54}$:%
\[
I_{1}=\frac{J_{1}}{J_{2}}.
\]
\end{itemize}

\newpage

$\mathbf{L}_{9,55}$

\begin{itemize}
\item  Levi decomposition: $L_{9,55}=\frak{sl}\left(  2,\mathbb{R}\right)
\overrightarrow{\oplus}_{R}\frak{g}_{6,1}$

\item  Describing representation: $R=D_{2}\oplus D_{0}$

\item  Structure tensor:%
\[%
\begin{array}
[c]{llllll}%
C_{12}^{2}=2, & C_{13}^{3}=-2, & C_{23}^{1}=1, & C_{14}^{4}=4, & C_{15}%
^{5}=2, & C_{17}^{7}=-2,\\
C_{18}^{8}=-4, & C_{25}^{4}=4, & C_{26}^{5}=3, & C_{27}^{6}=2, & C_{28}%
^{7}=1, & C_{34}^{5}=1,\\
C_{35}^{6}=2, & C_{36}^{7}=3, & C_{37}^{8}=4, & C_{49}^{4}=1, & C_{59}%
^{5}=1, & C_{69}^{6}=1,\\
C_{79}^{7}=1, & C_{89}^{8}=1. &  &  &  & \\
&  &  &  &  &
\end{array}
\]

\item $\mathrm{codim}_{\frak{g}}\left[  \frak{g},\frak{g}\right]  =1.$

\item  Rank $A\left(  \frak{g}\right)  =8$

\item  Conditions on invariants:%
\[
\frac{\partial F}{\partial x_{j}}=0,\;j=1,2,3,9.
\]

\item  Codimension one subalgebra%
\[
L\left\langle X_{1},..X_{8}\right\rangle \simeq L_{8,21}%
\]

\item  Invariants of subalgebra:%
\begin{align*}
J_{1}  &  =3x_{5}x_{7}-x_{6}^{2}-12x_{4}x_{8},\\
J_{2}  &  =-2x_{6}^{3}+9x_{5}x_{6}x_{7}+72x_{4}x_{6}x_{8}-27\left(  x_{4}%
x_{7}^{2}+x_{5}^{2}x_{8}\right)  .
\end{align*}

\item  Semi-invariance conditions:%
\begin{align*}
\widehat{X}_{9}\left(  J_{1}\right)   &  =-2J_{1},\\
\widehat{X}_{9}\left(  J_{2}\right)   &  =-3J_{2}.
\end{align*}

\item  Invariants of $L_{9,55}$:%
\[
I_{1}=\frac{J_{1}^{3}}{J_{2}^{2}}.
\]
\end{itemize}

\newpage

$\mathbf{L}_{9,56}^{p}$

\begin{itemize}
\item  Levi decomposition: $L_{9,56}^{p}=\left\{
\begin{array}
[c]{cc}%
\frak{sl}\left(  2,\mathbb{R}\right)  \overrightarrow{\oplus}_{R}%
\frak{g}_{6,1}, & p\neq0\\
\frak{sl}\left(  2,\mathbb{R}\right)  \overrightarrow{\oplus}_{R}\left(
A_{4,5}^{1,1}\oplus2L_{1}\right)  , & p=0.
\end{array}
\right.  $

\item  Describing representation: $R=D_{1}\oplus D_{\frac{1}{2}}\oplus D_{0}$

\item  Structure tensor:%
\[%
\begin{array}
[c]{llllll}%
C_{12}^{2}=2, & C_{13}^{3}=-2, & C_{23}^{1}=1, & C_{14}^{4}=2, & C_{16}%
^{6}=-2, & C_{17}^{7}=1,\\
C_{18}^{8}=-1, & C_{25}^{4}=2, & C_{26}^{5}=1, & C_{28}^{7}=1, & C_{34}%
^{5}=1, & C_{35}^{6}=2,\\
C_{37}^{8}=1, & C_{49}^{4}=1, & C_{59}^{5}=1, & C_{69}^{6}=1, & C_{79}%
^{7}=p, & C_{89}^{8}=p.\\
&  &  &  &  & \\
&  &  &  &  &
\end{array}
\]

\item $\mathrm{codim}_{\frak{g}}\left[  \frak{g},\frak{g}\right]  =1.$

\item  Rank $A\left(  \frak{g}\right)  =8$

\item  Conditions on invariants:%
\[
\frac{\partial F}{\partial x_{j}}=0,\;j=1,2,3,9.
\]

\item  Codimension one subalgebra%
\[
L\left\langle X_{1},..X_{8}\right\rangle \simeq L_{8,22}%
\]

\item  Invariants of subalgebra:%
\begin{align*}
J_{1}  &  =-4x_{4}x_{6}+x_{5}^{2},\\
J_{2}  &  =x_{5}x_{7}x_{8}-x_{4}x_{8}^{2}-x_{6}x_{7}^{2}.
\end{align*}

\item  Semi-invariance conditions:%
\begin{align*}
\widehat{X}_{9}\left(  J_{1}\right)   &  =-2J_{1},\\
\widehat{X}_{9}\left(  J_{2}\right)   &  =-\left(  1+2p\right)  J_{2}.
\end{align*}

\item  Invariants of $L_{9,56}^{p}$:%
\[
I_{1}=\frac{J_{1}^{2p+1}}{J_{2}^{2}}.
\]
\end{itemize}

\newpage

$\mathbf{L}_{9,57}$

\begin{itemize}
\item  Levi decomposition: $L_{9,57}=\frak{sl}\left(  2,\mathbb{R}\right)
\overrightarrow{\oplus}_{R}\left(  3L_{1}\oplus A_{3,3}\right)  $

\item  Describing representation: $R=D_{1}\oplus D_{\frac{1}{2}}\oplus D_{0}$

\item  Structure tensor:%
\[%
\begin{array}
[c]{llllll}%
C_{12}^{2}=2, & C_{13}^{3}=-2, & C_{23}^{1}=1, & C_{14}^{4}=2, & C_{16}%
^{6}=-2, & C_{17}^{7}=1,\\
C_{18}^{8}=-1, & C_{25}^{4}=2, & C_{26}^{5}=1, & C_{28}^{7}=1, & C_{34}%
^{5}=1, & C_{35}^{6}=2,\\
C_{37}^{8}=1, & C_{79}^{7}=1, & C_{89}^{8}=1. &  &  & \\
&  &  &  &  & \\
&  &  &  &  &
\end{array}
\]

\item $\mathrm{codim}_{\frak{g}}\left[  \frak{g},\frak{g}\right]  =1.$

\item  Rank $A\left(  \frak{g}\right)  =8$

\item  Conditions on invariants:%
\[
\frac{\partial F}{\partial x_{j}}=0,\;j=1,2,3,7,8,9.
\]

\item  Codimension one subalgebra%
\[
L\left\langle X_{1},..,X_{8}\right\rangle \simeq L_{8,22}%
\]

\item  Invariants of subalgebra:%
\begin{align*}
J_{1}  &  =-4x_{4}x_{6}+x_{5}^{2},\\
J_{2}  &  =x_{5}x_{7}x_{8}-x_{4}x_{8}^{2}-x_{6}x_{7}^{2}.
\end{align*}

\item  Semi-invariance conditions:%
\begin{align*}
\widehat{X}_{9}\left(  J_{1}\right)   &  =0,\\
\widehat{X}_{9}\left(  J_{2}\right)   &  =-2J_{2}.
\end{align*}

\item  Invariants of $L_{9,57}$:%
\[
I_{1}=J_{1}.
\]
\end{itemize}

\newpage

$\mathbf{L}_{9,58}$

\begin{itemize}
\item  Levi decomposition: $L_{9,58}=\frak{sl}\left(  2,\mathbb{R}\right)
\overrightarrow{\oplus}_{R}\left(  \frak{h}_{1}\oplus3L_{1}\right)  $

\item  Describing representation: $R=D_{1}\oplus D_{\frac{1}{2}}\oplus D_{0}$

\item  Structure tensor:%
\[%
\begin{array}
[c]{llllll}%
C_{12}^{2}=2, & C_{13}^{3}=-2, & C_{23}^{1}=1, & C_{14}^{4}=2, & C_{16}%
^{6}=-2, & C_{17}^{7}=1,\\
C_{18}^{8}=-1, & C_{25}^{4}=2, & C_{26}^{5}=1, & C_{28}^{7}=1, & C_{34}%
^{5}=1, & C_{35}^{6}=2,\\
C_{37}^{8}=1, & C_{78}^{9}=1. &  &  &  & \\
&  &  &  &  & \\
&  &  &  &  &
\end{array}
\]

\item $\mathrm{codim}_{\frak{g}}\left[  \frak{g},\frak{g}\right]  =0.$

\item  Rank $A\left(  \frak{g}\right)  =6$

\item  Invariants of $L_{9,58}$:%
\begin{align*}
I_{1}  &  =x_{9},\\
I_{2}  &  =x_{5}^{2}-4x_{4}x_{6},\\
I_{3}  &  =x_{1}x_{5}x_{9}-x_{4}x_{8}^{2}-x_{6}x_{7}^{2}+x_{5}x_{7}%
x_{8}+2x_{9}\left(  x_{2}x_{6}-x_{3}x_{4}\right)  .
\end{align*}
\end{itemize}

\newpage

$\mathbf{L}_{9,59}$

\begin{itemize}
\item  Levi decomposition: $L_{9,59}=\frak{sl}\left(  2,\mathbb{R}\right)
\overrightarrow{\oplus}_{R}6L_{1}$

\item  Describing representation: $R=D_{\frac{5}{2}}$

\item  Structure tensor:%
\[%
\begin{array}
[c]{llllll}%
C_{12}^{2}=2, & C_{13}^{3}=-2, & C_{23}^{1}=1, & C_{14}^{4}=5, & C_{15}%
^{5}=3, & C_{16}^{6}=1,\\
C_{17}^{7}=-1, & C_{18}^{8}=-3, & C_{19}^{9}=-5, & C_{25}^{4}=5, & C_{26}%
^{5}=4, & C_{27}^{6}=3,\\
C_{28}^{7}=2, & C_{29}^{8}=1, & C_{34}^{5}=1, & C_{35}^{6}=2, & C_{36}%
^{7}=3, & C_{37}^{8}=4.\\
C_{38}^{9}=5. &  &  &  &  & \\
&  &  &  &  &
\end{array}
\]

\item $\mathrm{codim}_{\frak{g}}\left[  \frak{g},\frak{g}\right]  =0.$

\item  Rank $A\left(  \frak{g}\right)  =6$

\item  Conditions on invariants:%
\[
\frac{\partial F}{\partial x_{j}}=0,\;j=1,2,3.
\]

\item  Invariants of $L_{9,59}$:%
\begin{align*}
I_{1}  &  =625x_{4}^{2}x_{9}^{2}-5x_{4}\left(  50x_{5}x_{8}x_{9}-5x_{6}%
x_{7}x_{9}+3x_{7}^{2}x_{8}-8x_{6}x_{8}^{2}\right)  +x_{5}^{2}\left(
40x_{7}x_{9}+9x_{8}^{2}\right)  +\\
&  -x_{5}\left(  15x_{6}^{2}x_{9}+19x_{6}x_{7}x_{8}-6x_{7}^{3}\right)
+2x_{6}^{2}\left(  3x_{6}x_{8}-x_{7}^{2}\right)  ,
\end{align*}%

\begin{align*}
I_{2}  &  =-3375x_{4}x_{9}^{2}x_{6}^{5}-6x_{7}^{2}x_{8}x_{6}^{5}-45x_{5}%
x_{8}x_{9}x_{6}^{5}+x_{7}^{4}x_{6}^{4}-380x_{4}x_{8}^{3}x_{6}^{4}-57x_{5}%
x_{7}x_{8}^{2}x_{6}^{4}\\
&  +900x_{5}^{2}x_{9}^{2}x_{6}^{4}+15x_{5}x_{7}^{2}x_{9}x_{6}^{4}%
+2325x_{4}x_{7}x_{8}x_{9}x_{6}^{4}+40x_{4}x_{7}^{2}x_{8}^{2}x_{6}%
^{3}+19500x_{4}x_{5}x_{7}x_{9}^{2}x_{6}^{3}\\
&  +37x_{5}x_{7}^{3}x_{8}x_{6}^{3}-525x_{4}x_{7}^{3}x_{9}x_{6}^{3}%
-1800x_{4}x_{5}x_{8}^{2}x_{9}x_{6}^{3}-300x_{5}^{2}x_{7}x_{8}x_{9}x_{6}%
^{3}-6x_{5}x_{7}^{5}x_{6}^{2}\\
&  +4400x_{4}^{2}x_{8}^{4}x_{6}^{2}+2120x_{4}x_{5}x_{7}x_{8}^{3}x_{6}%
^{2}-75000x_{4}^{2}x_{5}x_{9}^{3}x_{6}^{2}+50x_{5}^{2}x_{7}^{2}x_{8}^{2}%
x_{6}^{2}-4800x_{5}^{3}x_{7}x_{9}^{2}x_{6}^{2}\\
&  -30000x_{4}x_{5}^{2}x_{8}x_{9}^{2}x_{6}^{2}+15x_{4}x_{7}^{4}x_{8}x_{6}%
^{2}+40x_{5}^{2}x_{7}^{3}x_{9}x_{6}^{2}+120x_{5}^{3}x_{8}^{2}x_{9}x_{6}%
^{2}-17000x_{4}^{2}x_{7}x_{8}^{2}x_{9}x_{6}^{2}\\
&  -11000x_{4}x_{5}x_{7}^{2}x_{8}x_{9}x_{6}^{2}-4320x_{4}x_{5}^{2}x_{8}%
^{4}x_{6}-4800x_{4}^{2}x_{7}^{2}x_{8}^{3}x_{6}+50000x_{4}x_{5}^{3}x_{9}%
^{3}x_{6}\\
&  +125000x_{4}^{3}x_{7}x_{9}^{3}x_{6}-17000x_{4}x_{5}^{2}x_{7}^{2}x_{9}%
^{2}x_{6}-50000x_{4}^{3}x_{8}^{2}x_{9}^{2}x_{6}+6000x_{5}^{4}x_{8}x_{9}%
^{2}x_{6}\\
&  +55000x_{4}^{2}x_{5}x_{7}x_{8}x_{9}^{2}x_{6}-57x_{5}^{2}x_{7}^{4}x_{8}%
x_{6}+2325x_{4}x_{5}x_{7}^{4}x_{9}x_{6}-10000x_{4}^{2}x_{5}x_{8}^{3}x_{9}%
x_{6}\\
&  +2120x_{5}^{3}x_{7}^{2}x_{8}x_{9}x_{6}+9x_{5}^{2}x_{7}^{6}-8000x_{4}%
^{3}x_{8}^{5}+864x_{5}^{4}x_{8}^{4}+6000x_{4}^{2}x_{5}x_{7}x_{8}^{4}%
+120x_{4}x_{5}^{2}x_{7}^{2}x_{8}^{3}\\
&  +900x_{4}^{2}x_{7}^{4}x_{8}^{2}+152x_{5}^{3}x_{7}^{3}x_{8}^{2}%
+30000x_{4}^{2}x_{5}x_{7}^{3}x_{9}^{2}+4400x_{5}^{4}x_{7}^{2}x_{9}%
^{2}+20000x_{4}^{2}x_{5}^{2}x_{8}^{2}x_{9}^{2}\\
&  -75000x_{4}^{3}x_{7}^{2}x_{8}x_{9}^{2}-10000x_{4}x_{5}^{3}x_{7}x_{8}%
x_{9}^{2}-45x_{4}x_{5}x_{7}^{5}x_{8}-3375x_{4}^{2}x_{7}^{5}x_{9}-380x_{5}%
^{3}x_{7}^{4}x_{9}\\
&  -30000x_{4}^{2}x_{5}x_{7}^{2}x_{8}^{2}x_{9}-4320x_{5}^{4}x_{7}x_{8}%
^{2}x_{9}-1800x_{4}x_{5}^{2}x_{7}^{3}x_{8}x_{9}+9x_{8}^{2}x_{6}^{6}%
+152x_{5}^{2}x_{8}^{3}x_{6}^{3}\\
&  +30000x_{4}^{2}x_{8}x_{9}^{2}x_{6}^{3}-26250x_{4}^{2}x_{7}^{2}x_{9}%
^{2}x_{6}^{2}-648x_{5}^{3}x_{7}x_{8}^{3}x_{6}-300x_{4}x_{5}x_{7}^{3}x_{8}%
^{2}x_{6}\\
&  +26600x_{4}x_{5}^{2}x_{7}x_{8}^{2}x_{9}x_{6}+19500x_{4}^{2}x_{7}^{3}%
x_{8}x_{9}x_{6}-8000x_{5}^{5}x_{9}^{3}-50000x_{4}^{2}x_{5}^{2}x_{7}x_{9}^{3}\\
&  +50000x_{4}^{3}x_{7}x_{8}^{3}x_{9}%
\end{align*}%
\begin{align*}
I_{12}  &  =1992x_{8}^{3}x_{6}^{9}-702x_{7}x_{8}x_{9}x_{6}^{9}-1875x_{7}%
^{2}x_{8}^{2}x_{6}^{8}-14040x_{5}x_{7}x_{9}^{2}x_{6}^{8}+21600x_{4}x_{8}%
x_{9}^{2}x_{6}^{8}\\
&  +1053x_{9}^{2}x_{6}^{10}+156x_{7}^{3}x_{9}x_{6}^{8}-13536x_{5}x_{8}%
^{2}x_{9}x_{6}^{8}+52608x_{4}x_{8}^{4}x_{6}^{7}-19392x_{5}x_{7}x_{8}^{3}%
x_{6}^{7}\\
&  -54000x_{4}x_{5}x_{9}^{3}x_{6}^{7}++4500x_{4}x_{7}^{2}x_{9}^{2}x_{6}%
^{7}+42480x_{5}^{2}x_{8}x_{9}^{2}x_{6}^{7}+8160x_{4}x_{7}x_{8}^{2}x_{9}%
x_{6}^{7}\\
&  +17916x_{5}x_{7}^{2}x_{8}x_{9}x_{6}^{7}+612x_{7}^{4}x_{8}x_{6}^{7}%
-68x_{7}^{6}x_{6}^{6}+6240x_{5}^{2}x_{8}^{4}x_{6}^{6}-55344x_{4}x_{7}^{2}%
x_{8}^{3}x_{6}^{6}\\
&  -26400x_{5}^{3}x_{9}^{3}x_{6}^{6}+90000x_{4}^{2}x_{7}x_{9}^{3}x_{6}%
^{6}+17604x_{5}x_{7}^{3}x_{8}^{2}x_{6}^{6}+60720x_{5}^{2}x_{7}^{2}x_{9}%
^{2}x_{6}^{6}\\
&  +1464000x_{4}^{2}x_{8}^{2}x_{9}^{2}x_{6}^{6}-318000x_{4}x_{5}x_{7}%
x_{8}x_{9}^{2}x_{6}^{6}-3402x_{5}x_{7}^{4}x_{9}x_{6}^{6}-542400x_{4}x_{5}%
x_{8}^{3}x_{9}x_{6}^{6}\\
&  +116064x_{5}^{2}x_{7}x_{8}^{2}x_{9}x_{6}^{6}-14580x_{4}x_{7}^{3}x_{8}%
x_{9}x_{6}^{6}+221952x_{4}^{2}x_{8}^{5}x_{6}^{5}-331008x_{4}x_{5}x_{7}%
x_{8}^{4}x_{6}^{5}\\
&  -7500000x_{4}^{3}x_{9}^{4}x_{6}^{5}+62832x_{5}^{2}x_{7}^{2}x_{8}^{3}%
x_{6}^{5}+636000x_{4}x_{5}^{2}x_{7}x_{9}^{3}x_{6}^{5}-7500000x_{4}^{2}%
x_{5}x_{8}x_{9}^{3}x_{6}^{5}\\
&  +21552x_{4}x_{7}^{4}x_{8}^{2}x_{6}^{5}-10800x_{4}x_{5}x_{7}^{3}x_{9}%
^{2}x_{6}^{5}+1620000x_{4}x_{5}^{2}x_{8}^{2}x_{9}^{2}x_{6}^{5}-756000x_{4}%
^{2}x_{7}^{2}x_{8}x_{9}^{2}x_{6}^{5}\\
&  -370080x_{5}^{3}x_{7}x_{8}x_{9}^{2}x_{6}^{5}-5610x_{5}x_{7}^{5}x_{8}%
x_{6}^{5}+3078x_{4}x_{7}^{5}x_{9}x_{6}^{5}-7776x_{5}^{3}x_{8}^{3}x_{9}%
x_{6}^{5}+612x_{5}x_{7}^{7}x_{6}^{4}\\
&  -72000x_{4}^{2}x_{7}x_{8}^{3}x_{9}x_{6}^{5}+416880x_{4}x_{5}x_{7}^{2}%
x_{8}^{2}x_{9}x_{6}^{5}-120816x_{5}^{2}x_{7}^{3}x_{8}x_{9}x_{6}^{5}%
-47328x_{5}^{3}x_{7}x_{8}^{4}x_{6}^{4}\\
&  +220032x_{4}x_{5}^{2}x_{8}^{5}x_{6}^{4}-82080x_{4}^{2}x_{7}^{2}x_{8}%
^{4}x_{6}^{4}+15000000x_{4}^{2}x_{5}^{2}x_{9}^{4}x_{6}^{4}+325680x_{4}%
x_{5}x_{7}^{3}x_{8}^{3}x_{6}^{4}\\
&  +1080000x_{4}^{2}x_{5}x_{7}^{2}x_{9}^{3}x_{6}^{4}+211200x_{5}^{4}x_{7}%
x_{9}^{3}x_{6}^{4}-624000x_{4}x_{5}^{3}x_{8}x_{9}^{3}x_{6}^{4}-2880x_{5}%
^{4}x_{8}^{2}x_{9}^{2}x_{6}^{4}\\
&  +22500000x_{4}^{3}x_{7}x_{8}x_{9}^{3}x_{6}^{4}-56556x_{5}^{2}x_{7}^{4}%
x_{8}^{2}x_{6}^{4}+3298080x_{4}x_{5}^{2}x_{7}x_{8}^{3}x_{9}x_{6}^{4}%
+156x_{4}x_{7}^{8}x_{6}^{3}\\
&  +130500x_{4}^{2}x_{7}^{4}x_{9}^{2}x_{6}^{4}-79680x_{5}^{3}x_{7}^{3}%
x_{9}^{2}x_{6}^{4}+1680000x_{4}^{3}x_{8}^{3}x_{9}^{2}x_{6}^{4}-9612000x_{4}%
^{2}x_{5}x_{7}x_{8}^{2}x_{9}^{2}x_{6}^{4}\\
&  +799200x_{4}x_{5}^{2}x_{7}^{2}x_{8}x_{9}^{2}x_{6}^{4}-3402x_{4}x_{7}%
^{6}x_{8}x_{6}^{4}+21552x_{5}^{2}x_{7}^{5}x_{9}x_{6}^{4}-2630400x_{4}^{2}%
x_{5}x_{8}^{4}x_{9}x_{6}^{4}\\
&  -15600x_{4}^{2}x_{7}^{3}x_{8}^{2}x_{9}x_{6}^{4}-333120x_{5}^{3}x_{7}%
^{2}x_{8}^{2}x_{9}x_{6}^{4}-7320x_{4}x_{5}x_{7}^{4}x_{8}x_{9}x_{6}%
^{4}+563200x_{4}^{3}x_{8}^{6}x_{6}^{3}\\
&  -3840x_{5}^{4}x_{8}^{5}x_{6}^{3}-1336320x_{4}^{2}x_{5}x_{7}x_{8}^{5}%
x_{6}^{3}+305280x_{4}x_{5}^{2}x_{7}^{2}x_{8}^{4}x_{6}^{3}-3360000x_{4}%
x_{5}^{4}x_{9}^{4}x_{6}^{3}\\
&  -79680x_{4}^{2}x_{7}^{4}x_{8}^{3}x_{6}^{3}-57600000x_{4}^{3}x_{5}x_{8}%
^{2}x_{9}^{3}x_{6}^{3}+21504000x_{4}^{2}x_{5}^{2}x_{8}^{3}x_{9}^{2}x_{6}%
^{3}+1992x_{5}^{3}x_{7}^{9}\\
&  -60544x_{5}^{3}x_{7}^{3}x_{8}^{3}x_{6}^{3}-5600000x_{4}^{3}x_{7}^{3}%
x_{9}^{3}x_{6}^{3}-2616000x_{4}x_{5}^{3}x_{7}^{2}x_{9}^{3}x_{6}^{3}%
-19200x_{5}^{5}x_{8}x_{9}^{3}x_{6}^{3}\\
&  +33120000x_{4}^{2}x_{5}^{2}x_{7}x_{8}x_{9}^{3}x_{6}^{3}-120816x_{4}%
x_{5}x_{7}^{5}x_{8}^{2}x_{6}^{3}-15600x_{4}x_{5}^{2}x_{7}^{4}x_{9}^{2}%
x_{6}^{3}\\
&  -4920000x_{4}^{3}x_{7}^{2}x_{8}^{2}x_{9}^{2}x_{6}^{3}-10156800x_{4}%
x_{5}^{3}x_{7}x_{8}^{2}x_{9}^{2}x_{6}^{3}+4704000x_{4}^{2}x_{5}x_{7}^{3}%
x_{8}x_{9}^{2}x_{6}^{3}\\
&  +1129920x_{5}^{4}x_{7}^{2}x_{8}x_{9}^{2}x_{6}^{3}+144000000x_{4}^{3}%
x_{5}^{2}x_{8}x_{9}^{4}x_{6}^{2}-401664x_{4}x_{5}^{3}x_{7}x_{8}^{5}x_{6}^{2}\\
&  +17604x_{5}^{2}x_{7}^{6}x_{8}x_{6}^{3}-14580x_{4}x_{5}x_{7}^{6}x_{9}%
x_{6}^{3}-2238720x_{4}x_{5}^{3}x_{8}^{4}x_{9}x_{6}^{3}+3264000x_{4}^{3}%
x_{7}x_{8}^{4}x_{9}x_{6}^{3}\\
&  +1454400x_{4}^{2}x_{5}x_{7}^{2}x_{8}^{3}x_{9}x_{6}^{3}+53760x_{5}^{4}%
x_{7}x_{8}^{3}x_{9}x_{6}^{3}-2747520x_{4}x_{5}^{2}x_{7}^{3}x_{8}^{2}x_{9}%
x_{6}^{3}\\
&  +325680x_{5}^{3}x_{7}^{4}x_{8}x_{9}x_{6}^{3}-1875x_{5}^{2}x_{7}^{8}%
x_{6}^{2}+913920x_{4}^{2}x_{5}^{2}x_{8}^{6}x_{6}^{2}-556800x_{4}^{3}x_{7}%
^{2}x_{8}^{5}x_{6}^{2}\\
&  +1129920x_{4}^{2}x_{5}x_{7}^{3}x_{8}^{4}x_{6}^{2}+123648x_{5}^{4}x_{7}%
^{2}x_{8}^{4}x_{6}^{2}+288000x_{5}^{6}x_{9}^{4}x_{6}^{2}+6240x_{5}^{4}%
x_{7}^{6}x_{8}^{2}\\
&  -333120x_{4}x_{5}^{2}x_{7}^{4}x_{8}^{3}x_{6}^{2}-4920000x_{4}^{2}x_{5}%
^{2}x_{7}^{3}x_{9}^{3}x_{6}^{2}-556800x_{5}^{5}x_{7}^{2}x_{9}^{3}x_{6}^{2}\\
&  -47040000x_{4}^{2}x_{5}^{3}x_{8}^{2}x_{9}^{3}x_{6}^{2}-57600000x_{4}%
^{2}x_{5}^{3}x_{7}x_{9}^{4}x_{6}^{2}+60720x_{4}^{2}x_{7}^{6}x_{8}^{2}x_{6}%
^{2}\\
&  -10800x_{4}^{2}x_{7}^{5}x_{8}x_{9}x_{6}^{3}+96000000x_{4}^{4}x_{7}x_{8}%
^{2}x_{9}^{3}x_{6}^{2}-76800000x_{4}^{3}x_{5}x_{7}^{2}x_{8}x_{9}^{3}x_{6}%
^{2}\\
&  +7968000x_{4}x_{5}^{4}x_{7}x_{8}x_{9}^{3}x_{6}^{2}-23040000x_{4}^{2}%
x_{5}^{3}x_{7}^{2}x_{8}x_{9}^{3}x_{6}+1053x_{4}^{2}x_{7}^{10}-3840x_{5}%
^{5}x_{7}^{3}x_{8}^{4}%
\end{align*}
\end{itemize}%

\begin{align*}
& \\
&  +62832x_{5}^{3}x_{7}^{5}x_{8}^{2}x_{6}^{2}-756000x_{4}^{2}x_{5}x_{7}%
^{5}x_{9}^{2}x_{6}^{2}-82080x_{5}^{4}x_{7}^{4}x_{9}^{2}x_{6}^{2}%
+21600000x_{4}^{4}x_{8}^{4}x_{9}^{2}x_{6}^{2}\\
&  +5299200x_{4}x_{5}^{4}x_{8}^{3}x_{9}^{2}x_{6}^{2}+1080000x_{4}^{3}x_{7}%
^{4}x_{8}x_{9}^{2}x_{6}^{2}+95040000x_{4}^{3}x_{5}^{2}x_{8}^{4}x_{9}^{2}%
x_{6}-702x_{4}x_{5}x_{7}^{9}x_{6}\\
&  -14040x_{4}^{2}x_{7}^{8}x_{8}x_{6}-23040000x_{4}^{3}x_{5}x_{7}x_{8}%
^{3}x_{9}^{2}x_{6}^{2}+12960000x_{4}^{2}x_{5}^{2}x_{7}^{2}x_{8}^{2}x_{9}%
^{2}x_{6}^{2}\\
&  +34560x_{5}^{5}x_{7}x_{8}^{2}x_{9}^{2}x_{6}^{2}+1454400x_{4}x_{5}^{3}%
x_{7}^{3}x_{8}x_{9}^{2}x_{6}^{2}+17916x_{4}x_{5}x_{7}^{7}x_{8}x_{6}%
^{2}+4500x_{4}^{2}x_{7}^{7}x_{9}x_{6}^{2}\\
&  -55344x_{5}^{3}x_{7}^{6}x_{9}x_{6}^{2}-14976000x_{4}^{3}x_{5}x_{8}^{5}%
x_{9}x_{6}^{2}+211968x_{5}^{5}x_{8}^{4}x_{9}x_{6}^{2}+10886400x_{4}^{2}%
x_{5}^{2}x_{7}x_{8}^{4}x_{9}x_{6}^{2}\\
&  -2616000x_{4}^{3}x_{7}^{3}x_{8}^{3}x_{9}x_{6}^{2}-2603520x_{4}x_{5}%
^{3}x_{7}^{2}x_{8}^{3}x_{9}x_{6}^{2}+799200x_{4}^{2}x_{5}x_{7}^{4}x_{8}%
^{2}x_{9}x_{6}^{2}\\
&  +305280x_{5}^{4}x_{7}^{3}x_{8}^{2}x_{9}x_{6}^{2}+416880x_{4}x_{5}^{2}%
x_{7}^{5}x_{8}x_{9}x_{6}^{2}-55296x_{5}^{5}x_{7}x_{8}^{5}x_{6}-19200x_{4}%
^{3}x_{5}x_{7}^{3}x_{8}^{5}\\
&  -768000x_{4}^{4}x_{8}^{7}x_{6}-27648x_{4}x_{5}^{4}x_{8}^{6}x_{6}%
-307200x_{4}^{3}x_{5}x_{7}x_{8}^{6}x_{6}+34560x_{4}^{2}x_{5}^{2}x_{7}^{2}%
x_{8}^{5}x_{6}\\
&  -120000000x_{4}^{3}x_{5}^{3}x_{9}^{5}x_{6}+211200x_{4}^{3}x_{7}^{4}%
x_{8}^{4}x_{6}+53760x_{4}x_{5}^{3}x_{7}^{3}x_{8}^{4}x_{6}+96000000x_{4}%
^{3}x_{5}^{2}x_{7}^{2}x_{9}^{4}x_{6}\\
&  +600000000x_{4}^{5}x_{8}^{2}x_{9}^{4}x_{6}+10560000x_{4}x_{5}^{5}x_{7}%
x_{9}^{4}x_{6}+33600000x_{4}^{2}x_{5}^{4}x_{8}x_{9}^{4}x_{6}\\
&  -480000000x_{4}^{4}x_{5}x_{7}x_{8}x_{9}^{4}x_{6}-401664x_{5}^{5}x_{7}%
^{2}x_{8}^{3}x_{9}x_{6}+33120000x_{4}^{3}x_{5}x_{7}^{3}x_{8}^{2}x_{9}^{2}%
x_{6}\\
&  -370080x_{4}^{2}x_{5}x_{7}^{5}x_{8}^{3}x_{6}-47328x_{5}^{4}x_{7}^{4}%
x_{8}^{3}x_{6}+22500000x_{4}^{3}x_{5}x_{7}^{4}x_{9}^{3}x_{6}+3264000x_{4}%
x_{5}^{4}x_{7}^{3}x_{9}^{3}x_{6}\\
&  -408000000x_{4}^{4}x_{5}x_{8}^{3}x_{9}^{3}x_{6}-1728000x_{4}x_{5}^{5}%
x_{8}^{2}x_{9}^{3}x_{6}+273600000x_{4}^{3}x_{5}^{2}x_{7}x_{8}^{2}x_{9}%
^{3}x_{6}\\
&  -307200x_{5}^{6}x_{7}x_{8}x_{9}^{3}x_{6}+116064x_{4}x_{5}^{2}x_{7}^{6}%
x_{8}^{2}x_{6}+90000x_{4}^{3}x_{7}^{6}x_{9}^{2}x_{6}-72000x_{4}x_{5}^{3}%
x_{7}^{5}x_{9}^{2}x_{6}\\
&  -806400x_{5}^{6}x_{8}^{3}x_{9}^{2}x_{6}-57600000x_{4}^{4}x_{7}^{2}x_{8}%
^{3}x_{9}^{2}x_{6}-74304000x_{4}^{2}x_{5}^{3}x_{7}x_{8}^{3}x_{9}^{2}%
x_{6}-2880x_{4}^{2}x_{5}^{2}x_{7}^{4}x_{8}^{4}\\
&  +10886400x_{4}x_{5}^{4}x_{7}^{2}x_{8}^{2}x_{9}^{2}x_{6}-9612000x_{4}%
^{2}x_{5}^{2}x_{7}^{4}x_{8}x_{9}^{2}x_{6}-1336320x_{5}^{5}x_{7}^{3}x_{8}%
x_{9}^{2}x_{6}\\
&  -19392x_{5}^{3}x_{7}^{7}x_{8}x_{6}+8160x_{4}x_{5}^{2}x_{7}^{7}x_{9}%
x_{6}-4339200x_{4}^{2}x_{5}^{3}x_{8}^{5}x_{9}x_{6}+10560000x_{4}^{4}x_{7}%
x_{8}^{5}x_{9}x_{6}\\
&  +7968000x_{4}^{3}x_{5}x_{7}^{2}x_{8}^{4}x_{9}x_{6}+3655680x_{4}x_{5}%
^{4}x_{7}x_{8}^{4}x_{9}x_{6}-10156800x_{4}^{2}x_{5}^{2}x_{7}^{3}x_{8}^{3}%
x_{9}x_{6}\\
&  +636000x_{4}^{3}x_{7}^{5}x_{8}^{2}x_{9}x_{6}+3298080x_{4}x_{5}^{3}x_{7}%
^{4}x_{8}^{2}x_{9}x_{6}-318000x_{4}^{2}x_{5}x_{7}^{6}x_{8}x_{9}x_{6}%
-331008x_{5}^{4}x_{7}^{5}x_{8}x_{9}x_{6}\\
&  +1075200x_{4}^{3}x_{5}^{2}x_{8}^{7}+20736x_{5}^{6}x_{8}^{6}+288000x_{4}%
^{4}x_{7}^{2}x_{8}^{6}-806400x_{4}^{2}x_{5}^{3}x_{7}x_{8}^{6}+2500000000x_{4}%
^{6}x_{9}^{6}\\
&  +211968x_{4}x_{5}^{4}x_{7}^{2}x_{8}^{5}+19200000x_{4}^{2}x_{5}^{5}x_{9}%
^{5}+600000000x_{4}^{4}x_{5}^{2}x_{7}x_{9}^{5}-3000000000x_{4}^{5}x_{5}%
x_{8}x_{9}^{5}\\
&  +21600000x_{4}^{2}x_{5}^{4}x_{7}^{2}x_{9}^{4}+1260000000x_{4}^{4}x_{5}%
^{2}x_{8}^{2}x_{9}^{4}-768000x_{5}^{7}x_{7}x_{9}^{4}-7680000x_{4}x_{5}%
^{6}x_{8}x_{9}^{4}\\
&  -408000000x_{4}^{3}x_{5}^{3}x_{7}x_{8}x_{9}^{4}-26400x_{4}^{3}x_{7}%
^{6}x_{8}^{3}-7776x_{4}x_{5}^{3}x_{7}^{5}x_{8}^{3}-7500000x_{4}^{4}x_{7}%
^{5}x_{9}^{3}\\
&  +1680000x_{4}^{2}x_{5}^{3}x_{7}^{4}x_{9}^{3}+144000000x_{4}^{4}x_{5}%
x_{7}^{2}x_{8}^{2}x_{9}^{3}+33600000x_{4}^{4}x_{5}x_{7}x_{8}^{4}x_{9}%
^{2}-54000x_{4}^{3}x_{7}^{7}x_{8}x_{9}\\
&  +563200x_{5}^{6}x_{7}^{3}x_{9}^{3}-227200000x_{4}^{3}x_{5}^{3}x_{8}%
^{3}x_{9}^{3}-120000000x_{4}^{5}x_{7}x_{8}^{3}x_{9}^{3}+1075200x_{5}^{7}%
x_{8}^{2}x_{9}^{3}\\
&  +95040000x_{4}^{2}x_{5}^{4}x_{7}x_{8}^{2}x_{9}^{3}-57600000x_{4}^{3}%
x_{5}^{2}x_{7}^{3}x_{8}x_{9}^{3}-14976000x_{4}x_{5}^{5}x_{7}^{2}x_{8}x_{9}%
^{3}+42480x_{4}^{2}x_{5}x_{7}^{7}x_{8}^{2}\\
&  +1464000x_{4}^{2}x_{5}^{2}x_{7}^{6}x_{9}^{2}+221952x_{5}^{5}x_{7}^{5}%
x_{9}^{2}+19200000x_{4}^{5}x_{8}^{5}x_{9}^{2}+15072000x_{4}^{2}x_{5}^{4}%
x_{8}^{4}x_{9}^{2}\\
&  -47040000x_{4}^{3}x_{5}^{2}x_{7}^{2}x_{8}^{3}x_{9}^{2}-4339200x_{4}%
x_{5}^{5}x_{7}x_{8}^{3}x_{9}^{2}+15000000x_{4}^{4}x_{7}^{4}x_{8}^{2}x_{9}%
^{2}+21504000x_{4}^{2}x_{5}^{3}x_{7}^{3}x_{8}^{2}x_{9}^{2}\\
&  +52608x_{5}^{4}x_{7}^{7}x_{9}+913920x_{5}^{6}x_{7}^{2}x_{8}^{2}x_{9}%
^{2}-7500000x_{4}^{3}x_{5}x_{7}^{5}x_{8}x_{9}^{2}-2630400x_{4}x_{5}^{4}%
x_{7}^{4}x_{8}x_{9}^{2}\\
&  -13536x_{4}x_{5}^{2}x_{7}^{8}x_{8}+21600x_{4}^{2}x_{5}x_{7}^{8}%
x_{9}-7680000x_{4}^{4}x_{5}x_{8}^{6}x_{9}-511488x_{4}x_{5}^{5}x_{8}^{5}%
x_{9}-1728000x_{4}^{3}x_{5}^{2}x_{7}x_{8}^{5}x_{9}\\
&  -3360000x_{4}^{4}x_{7}^{3}x_{8}^{4}x_{9}+5299200x_{4}^{2}x_{5}^{3}x_{7}%
^{2}x_{8}^{4}x_{9}-27648x_{5}^{6}x_{7}x_{8}^{4}x_{9}-624000x_{4}^{3}x_{5}%
x_{7}^{4}x_{8}^{3}x_{9}\\
&  -2238720x_{4}x_{5}^{4}x_{7}^{3}x_{8}^{3}x_{9}+1620000x_{4}^{2}x_{5}%
^{2}x_{7}^{5}x_{8}^{2}x_{9}+220032x_{5}^{5}x_{7}^{4}x_{8}^{2}x_{9}%
-542400x_{4}x_{5}^{3}x_{7}^{6}x_{8}x_{9}.
\end{align*}

\newpage

$\mathbf{L}_{9,60}$

\begin{itemize}
\item  Levi decomposition: $L_{9,60}=\frak{sl}\left(  2,\mathbb{R}\right)
\overrightarrow{\oplus}_{R}6L_{1}$

\item  Describing representation: $R=D_{\frac{3}{2}}\oplus D_{\frac{1}{2}}$

\item  Structure tensor:%
\[%
\begin{array}
[c]{llllll}%
C_{12}^{2}=2, & C_{13}^{3}=-2, & C_{23}^{1}=1, & C_{14}^{4}=3, & C_{15}%
^{5}=1, & C_{16}^{6}=-1,\\
C_{17}^{7}=-3, & C_{18}^{8}=1, & C_{19}^{9}=-1, & C_{25}^{4}=3, & C_{26}%
^{5}=3, & C_{27}^{6}=1,\\
C_{29}^{8}=1, & C_{34}^{5}=1, & C_{35}^{6}=2, & C_{36}^{7}=3, & C_{37}%
^{9}=1. & \\
&  &  &  &  & \\
&  &  &  &  &
\end{array}
\]

\item $\mathrm{codim}_{\frak{g}}\left[  \frak{g},\frak{g}\right]  =0.$

\item  Rank $A\left(  \frak{g}\right)  =6$

\item  Conditions on invariants:%
\[
\frac{\partial F}{\partial x_{j}}=0,\;j=1,2,3.
\]

\item  Invariants of $L_{9,60}$:%
\begin{align*}
I_{1}  &  =27x_{4}^{2}x_{7}^{2}-x_{5}^{2}x_{6}^{2}+4\left(  x_{4}x_{6}%
^{3}+x_{5}^{3}x_{7}\right)  -18x_{4}x_{5}x_{6}x_{7},\\
I_{2}  &  =18x_{4}x_{7}x_{8}x_{9}-6x_{4}x_{6}x_{9}^{2}+2x_{5}^{2}x_{9}%
^{2}-2x_{5}x_{6}x_{8}x_{9}-6x_{5}x_{7}x_{8}^{2}+2x_{6}^{2}x_{8}^{2},\\
I_{3}  &  =x_{4}x_{9}^{3}-x_{7}x_{8}^{3}+x_{6}x_{8}^{2}x_{9}-x_{5}x_{8}%
x_{9}^{2}.
\end{align*}
\end{itemize}

\newpage

$\mathbf{L}_{9,61}$

\begin{itemize}
\item  Levi decomposition: $L_{9,61}=\frak{sl}\left(  2,\mathbb{R}\right)
\overrightarrow{\oplus}_{R}6L_{1}$

\item  Describing representation: $R=2D_{1}$

\item  Structure tensor:%
\[%
\begin{array}
[c]{llllll}%
C_{12}^{2}=2, & C_{13}^{3}=-2, & C_{23}^{1}=1, & C_{14}^{4}=2, & C_{16}%
^{6}=-2, & C_{17}^{7}=2,\\
C_{19}^{9}=-2, & C_{25}^{4}=2, & C_{26}^{5}=1, & C_{28}^{7}=2, & C_{29}%
^{8}=1, & C_{34}^{5}=1,\\
C_{35}^{6}=2, & C_{37}^{8}=1, & C_{38}^{9}=2. &  &  & \\
&  &  &  &  & \\
&  &  &  &  &
\end{array}
\]

\item $\mathrm{codim}_{\frak{g}}\left[  \frak{g},\frak{g}\right]  =0.$

\item  Rank $A\left(  \frak{g}\right)  =6$

\item  Conditions on invariants:%
\[
\frac{\partial F}{\partial x_{j}}=0,\;j=1,2,3.
\]

\item  Codimension three subalgebras:%
\[
L\left\langle X_{1},..X_{6}\right\rangle \simeq L\left\langle X_{1}%
,X_{2},X_{3},X_{7},X_{8},X_{9}\right\rangle \simeq L_{6,4}%
\]

\item  Invariants of $L_{9,61}$:%
\begin{align*}
J_{1}  &  =x_{4}x_{6}-\frac{1}{4}x_{5}^{2},\\
J_{2}  &  =x_{7}x_{9}-\frac{1}{4}x_{8}^{2},\\
I_{3}  &  =2x_{4}x_{9}-x_{5}x_{8}+2x_{6}x_{7}.
\end{align*}
\end{itemize}

\newpage

$\mathbf{L}_{9,62}$

\begin{itemize}
\item  Levi decomposition: $L_{9,62}=\frak{sl}\left(  2,\mathbb{R}\right)
\overrightarrow{\oplus}_{R}A_{6,3}$

\item  Describing representation: $R=2D_{1}$

\item  Structure tensor:%
\[%
\begin{array}
[c]{llllll}%
C_{12}^{2}=2, & C_{13}^{3}=-2, & C_{23}^{1}=1, & C_{14}^{4}=2, & C_{16}%
^{6}=-2, & C_{17}^{7}=2,\\
C_{19}^{9}=-2, & C_{25}^{4}=2, & C_{26}^{5}=1, & C_{28}^{7}=2, & C_{29}%
^{8}=1, & C_{34}^{5}=1,\\
C_{35}^{6}=2, & C_{37}^{8}=1, & C_{38}^{9}=2, & C_{45}^{7}=2, & C_{46}%
^{8}=1, & C_{56}^{9}=2.\\
&  &  &  &  & \\
&  &  &  &  &
\end{array}
\]

\item $\mathrm{codim}_{\frak{g}}\left[  \frak{g},\frak{g}\right]  =0.$

\item  Rank $A\left(  \frak{g}\right)  =6$

\item  Invariants of $L_{9,62}$:%
\begin{align*}
J_{1}  &  =2x_{4}x_{9}-x_{5}x_{8}+2x_{6}x_{7},\\
J_{2}  &  =x_{7}x_{9}-\frac{1}{4}x_{8}^{2},\\
I_{3}  &  =-2x_{1}x_{8}-4x_{2}x_{9}+4x_{3}x_{7}+x_{5}^{2}-4x_{4}x_{6}.
\end{align*}
\end{itemize}

\newpage

$\mathbf{L}_{9,63}$

\begin{itemize}
\item  Levi decomposition: $L_{9,63}=\frak{sl}\left(  2,\mathbb{R}\right)
\overrightarrow{\oplus}_{R}6L_{1}$

\item  Describing representation: $R=3D_{\frac{1}{2}}$

\item  Structure tensor:%
\[%
\begin{array}
[c]{llllll}%
C_{12}^{2}=2, & C_{13}^{3}=-2, & C_{23}^{1}=1, & C_{14}^{4}=1, & C_{15}%
^{5}=-1, & C_{16}^{6}=1,\\
C_{17}^{7}=-1, & C_{18}^{8}=1, & C_{19}^{9}=-1, & C_{25}^{4}=1, & C_{27}%
^{6}=1, & C_{29}^{8}=1,\\
C_{34}^{5}=1, & C_{36}^{7}=1, & C_{38}^{9}=1. &  &  & \\
&  &  &  &  & \\
&  &  &  &  &
\end{array}
\]

\item $\mathrm{codim}_{\frak{g}}\left[  \frak{g},\frak{g}\right]  =0.$

\item  Rank $A\left(  \frak{g}\right)  =8$

\item  Conditions on invariants:%
\[
\frac{\partial F}{\partial x_{j}}=0,\;j=1,2,3.
\]

\item  Invariants of $L_{9,63}$:%
\begin{align*}
I_{1}  &  =x_{4}x_{7}-x_{5}x_{6},\\
I_{2}  &  =x_{4}x_{9}-x_{5}x_{8},\\
I_{3}  &  =x_{6}x_{9}-x_{7}x_{8}.
\end{align*}
\end{itemize}

\newpage

\subsubsection*{Acknowledgement} The author benefited during the preparation of the paper from
fruitful discussions with V. M. Boyko, to whom he expresses his
gratitude.

\end{document}